\documentclass[preprint]{elsarticle}
\usepackage{color}
\usepackage{hhline}
\usepackage{mathrsfs}
\usepackage{graphicx}
\usepackage{dcolumn}
\usepackage{bm}
\usepackage{multirow}
\usepackage{booktabs}
\usepackage{afterpage}
\usepackage{amsmath}

\usepackage{algorithm}
\usepackage{algorithmicx}
\usepackage{algpseudocode}

\usepackage{longtable}

\journal{Computer Physics Communications}

\begin{document}

\begin{frontmatter}

\title{TC++: First-principles calculation code for solids using the transcorrelated method}

\author[1,2]{Masayuki Ochi}
\ead{ochi@phys.sci.osaka-u.ac.jp}

\affiliation[1]{organization={Forefront Research Center, Osaka University},
addressline={1-1 Machikaneyama-cho, Toyonaka},
postcode={560-0043},
city={Osaka},
country={Japan}}
\affiliation[2]{organization={Department of Physics, Osaka University},
addressline={1-1 Machikaneyama-cho, Toyonaka},
postcode={560-0043},
city={Osaka},
country={Japan}}

\begin{abstract}
TC\verb!++! is a free/libre open-source software of the transcorrelated (TC) method for first-principles calculation of solids.
Here, the TC method is one of the promising wave-function theories that can be applied to periodic systems with reasonable computational cost and satisfactory accuracy.
We present our implementation of TC\verb!++! including a detailed description of the divergence correction technique applied to the TC effective interactions.
We also present the way to use TC\verb!++! and some results of application to simple periodic systems: bulk silicon and homogeneous electron gas.
\end{abstract}


\end{frontmatter}

{\bf PROGRAM SUMMARY/NEW VERSION PROGRAM SUMMARY}

\begin{small}
\noindent
{\em Program Title:} TC\verb!++! \\
{\em CPC Library link to program files:} (to be added by Technical Editor) \\
{\em Developer's respository link:} https://github.com/masaochi/TC \\
{\em Licensing provisions:} MIT \\
{\em Programming language:} C\verb!++! and partially Fortran90\\
 {\em External routines/libraries:}  Boost, Eigen, FFTW \\
{\em Nature of problem:}
First-principles calculation of many-electron systems with the periodic boundary condition.
TC\verb!++! gives physical quantities such as total energy, orbital energies, and magnetic moment.
In this method, one-electron orbitals in the Slater determinant are optimized in the existence of the Jastrow correlation factor. \\ 
{\em Solution method:} Hartree-Fock and (biorthogonal) transcorrelated methods using a plane-wave basis set with the periodic boundary condition. Norm-conserving pseudopotential without partial core correction provided as a UPF file is used to represent core electrons. \\ 
\end{small}

\section{Introduction}
\label{Intro}

Accurate electronic-structure calculation of materials has been a long-standing problem in materials science.
Density functional theory (DFT) is one of the most successful theories for this purpose, which enables efficient first-principles calculation with satisfactory accuracy in many cases.
However, several problems in accuracy are known for standard approximations in DFT, such as difficulties in describing electronic structure of strongly correlated systems.
Since it is difficult to systematically improve accuracy of the exchange-correlation energy functional in DFT, another theoretical framework has also gathered attention: wave function theory (WFT).

In WFT, many-body wave functions are explicitly handled, which requires expensive computational cost while its accuracy can be systematically improved.
While WFT has been developed mainly in quantum chemistry, it has now been applied to several solids.
One representative example is quantum Monte Carlo (QMC) methods~\cite{QMC} including variational Monte Carlo (VMC) and diffusion Monte Carlo (DMC), where many-body integration such as the expectation value of the physical quantities is performed with the Monte Carlo technique. Other famous methods in WFT are the Hartree-Fock (HF) method and post-HF methods such as M{\o}ller-Plesset (MP) perturbation theory and coupled-cluster (CC) theory, where the Slater determinant consisting of the HF orbitals is used as a starting point of approximation.
More recently, full-configuration-interaction (FCI) QMC method~\cite{FCI1, FCI2, FCI3} has been paid much attention, where linear combination of Slater determinants is optimized with a Monte Carlo technique.
For small systems, accurate description of electronic structure using WFT is well-established.
On the other hand, for solid-state calculations, many electrons require much demanding computation, which hinders efficient and accurate calculation for many systems.

Transcorrelated (TC) method~\cite{BoysHandy,Handy,Ten-no1,Ten-no2,Umezawa} is one of the promising WFTs, which can be applied to calculations of homogeneous electron gas~\cite{elgas_Armour, Umezawa_elgas, Sakuma, Luo_elgas, FCI_TC_elgas, perturbation_elgas} and solids~\cite{Sakuma, TCaccel, TCjfo, TCPW, TCZnO} with efficient computational cost and reasonable accuracy.
In the TC method, many-body Hamiltonian is similarity-transformed by the Jastrow correlation factor, by which electron correlation effects are partially incorporated into Hamiltonian. 
For example, the similarity-transformed Hamiltonian, called TC Hamiltonian, includes an effective Coulomb interaction without divergence at the electron coalescence point for a singlet spin pair when the cusp condition~\cite{cusp,cusp2} is imposed to the Jastrow factor.
In the single-determinant TC method, the HF approximation is applied to the TC Hamiltonian, namely, the one-electron orbitals optimized for the TC Hamiltonian are obtained.
Because of this construction, several post-HF methods can be applied to the TC Hamiltonian in a straightforward manner.
For example, the TC method was combined with the coupled-cluster theory~\cite{Ten-no3, TCCC2021}, M{\o}ller-Plesset (MP) perturbation theory~\cite{Ten-no1,Ten-no2},
and configuration interaction (CI) theory~\cite{Umezawa_CIS, LuoVTC, Luo_multiconf, Giner_He, SCI} for atomic and molecular systems.
The combination of the TC method with the post-HF methods was also reported for solid-state calculations: calculation of optical absorption spectra by TC-CI singles~\cite{TCCIS}, and TC-MP2 calculation for simple solids~\cite{TCMP2}.
Several QMC methods using the Jastrow factor can also be combined with the TC method~\cite{Umezawa, Umezawa_beta, LuoTC, TCatoms_HFJastrow, TCatoms_oneparam, TCatom}.
In particular, similarity-transformed FCIQMC, the combination of the TC and FCIQMC methods, has recently been much attention~\cite{FCI_TC_elgas, FCI_canoTC_DMRG, FCI_canoTC, FCI_Bedimer, FCI_largeCI, FCI_Hubbard, FCI_1dgas, FCI_cold}.
In addition, several important developments of the TC method have been recently reported.
Canonical TC method~\cite{CanonicalTC, CanonicalTC_qCCSD} is an important development for treating the non-Hermiticity of similarity-transformed Hamiltonian in the TC method.
TC method has also been applied to model systems~\cite{TCHubbard, FCI_Hubbard, TCDMRG, LieAlgebra}, quantum gas~\cite{FCI_1dgas}, and cold-atom systems~\cite{FCI_cold}.
Using the TC method in the context of quantum simulation is an interesting new direction of studies~\cite{McArdle,quantum_simulation}.
Construction of the exchange-correlation functional of DFT using the TC method is also an intriguing attempt~\cite{Umezawa_Exc,Umezawa_Exc2}.

As seen in the previous paragraph, the number of studies for solid-state TC calculation~\cite{Sakuma, TCaccel, TCjfo, TCPW, TCZnO, TCCIS, TCMP2} is relatively limited compared with that for molecular systems.
One problem in solid-state calculation is that we should handle very complicated interaction terms of the similarity-transformed Hamiltonian, which shows divergent behavior in reciprocal space.
Thus, it is important to publish computational code of the TC method for solids and present how to resolve these numerical difficulties.
In this paper, we present our implementation of TC\verb!++!, a free/libre open-source software of the single-determinant TC method for first-principles calculation of solids,
which was recently published in github~\cite{github_TC}.
This paper is organized as follows.
In Chapter~\ref{algo}, we present computational algorithm of the single-determinant TC method for solid-state calculations as implemented in our code.
The way to use TC\verb!++! is shown in Chapter~\ref{howto}. We present some results of application to simple systems in Chapter~\ref{results}.
This paper is summarized in Chapter~\ref{summary}.
Since our computational code is focused on the single-determinant version of the TC method, we simply use `the TC method` to represent the single-determinant TC method hereafter in this paper.

\section{Algorithm}
\label{algo}

\subsection{Transcorrelated method}

For an $N$-electron system under an external potential $v_{\mathrm{ext}}({\bm r})$,
Hamiltonian $\mathcal{H}$ reads
\begin{equation}
\mathcal{H}=\sum_{i=1}^{N} \left( -\frac{1}{2}\nabla_i ^2 + v_{\mathrm{ext}}({\bm r}_i) \right) +
\sum_{i=1}^{N}\sum_{j>i}^N \frac{1}{|{\bm r}_i-{\bm r}_j|},\label{eq:Hamil}
\end{equation}
where $x=({\bm r},\sigma)$ denotes a set of spatial and spin coordinates associated with an electron. 
The many-body wave function $\Psi$ can be factorized as $\Psi=F\Phi$ where
\begin{equation}
F=\mathrm{exp}(-\sum_{i,j(>i)}^N u(x_i,x_j)), \label{eq:Jastrowfactor}
\end{equation}
is the Jastrow factor and $\Phi \equiv \Psi/F$. 
We assume the Jastrow function $u(x_i,x_j)$ to be symmetric, i.e., $u(x_i,x_j)=u(x_j,x_i)$, without loss of generality.
By introducing a similarity-transformed Hamiltonian,
\begin{equation}
\mathcal{H}_{\mathrm{TC}} \equiv F^{-1}\mathcal{H}F,
\end{equation}
the Schr{\"o}dinger equation is rewritten as,
\begin{equation}
\mathcal{H}\Psi = E\Psi \Leftrightarrow \mathcal{H}_{\mathrm{TC}}\Phi = E \Phi \label{eq:simtr}.
\end{equation}
In this way, electron correlation effects described by the Jastrow factor are incorporated into
the similarity-transformed Hamiltonian $\mathcal{H}_{\mathrm{TC}}$, which is called the TC Hamiltonian.
TC Hamiltonian can be explicitly written as,
\begin{gather}
\mathcal{H}_{\mathrm{TC}}=\sum_{i=1}^{N} \left( -\frac{1}{2}\nabla_i ^2 + v_{\mathrm{ext}}({\bm r}_i) \right) +
\sum_{i=1}^{N}\sum_{j>i}^N v_{\mathrm{2body}}(x_1,x_2)  \notag \\
- \sum_{i=1}^{N}\sum_{j>i}^N \sum_{k>j}^N v_{\mathrm{3body}}(x_1,x_2,x_3),
\end{gather}
where $v_{\mathrm{2body}}(x_1,x_2)$ and $v_{\mathrm{3body}}(x_1,x_2,x_3)$ are the effective interactions defined as,
\begin{gather}
v_{\mathrm{2body}}(x_1,x_2)\notag\\
\equiv \frac{1}{|{\bm r}_1-{\bm r}_2|}+\frac{1}{2}\bigg[ \nabla_1^2 u(x_1,x_2)+\nabla_2^2 u(x_1,x_2)\notag \\
-(\nabla_1 u(x_1,x_2))^2-(\nabla_2 u(x_1,x_2))^2\bigg] \notag \\
+ \nabla_1 u(x_1,x_2)\cdot \nabla_1 + \nabla_2 u(x_1,x_2)\cdot \nabla_2, \label{eq:V2body}
\end{gather}
and
\begin{gather}
v_{\mathrm{3body}}(x_1,x_2,x_3)\notag\\
\equiv\nabla_1 u(x_1,x_2)\cdot \nabla_1 u(x_1,x_3) 
+ \nabla_2 u(x_2,x_1) \cdot \nabla_2 u(x_2,x_3) \notag \\
+ \nabla_3 u(x_3,x_1) \cdot \nabla_3 u(x_3,x_2).  \label{eq:V3body}
\end{gather}

By applying the single-Slater-determinant (i.e., Hartree--Fock) approximation to TC Hamiltonian,
$\Phi$ can be written as $\Phi=\mathrm{det}[ \phi_i({\bm r}_j) ]$ consisting of one-electron orbitals $\phi({\bm r})$.
In this paper, we assume one-electron orbitals are assigned as spin-up or spin-down: to say, we do not consider a spinor orbital where up- and down-components are hybridized.
The following one-body self-consistent-field (SCF) equation for one-electron orbitals can be derived (see, e.g., \cite{Umezawa}):
\begin{align}
\left( -\frac{1}{2}\nabla_1^2 +v_{\mathrm{ext}}({\bm r}_1) \right) \phi_i ({\bm r}_1)\notag \\
+ \sum_{j=1}^N
\int \mathrm{d}{\bm r}_2\  \phi_j^*({\bm r}_2) v_{\mathrm{2body}}(x_1,x_2)
\mathrm{det} \left[
\begin{array}{rrr}
\phi_i({\bm r}_1) & \phi_i({\bm r}_2) \\
\phi_j({\bm r}_1) & \phi_j({\bm r}_2) \\
\end{array} \right] \notag \\
- \sum_{j=1}^N \sum_{k>j}^N
\int \mathrm{d}{\bm r}_2 \mathrm{d}{\bm r}_3\  \phi_j^*({\bm r}_2)\phi_k^*({\bm r}_3)v_{\mathrm{3body}}(x_1,x_2,x_3)  \notag \\
\times 
\mathrm{det} \left[
\begin{array}{rrr}
\phi_i({\bm r}_1) & \phi_i({\bm r}_2) &  \phi_i({\bm r}_3) \\
\phi_j({\bm r}_1) & \phi_j({\bm r}_2) & \phi_j({\bm r}_3) \\
\phi_k({\bm r}_1) & \phi_k({\bm r}_2) & \phi_k({\bm r}_3)
\end{array} \right]
= \sum_{j=1}^N \epsilon_{ij} \phi_j({\bm r}_1), \label{eq:SCF}
\end{align}
where the orthonormal condition, $\langle \phi_i | \phi_j \rangle = \delta_{i,j}$, is imposed.
The TC one-electron orbitals $\phi_i({\bm r})$ are optimized by solving Eq.~(\ref{eq:SCF}).
This procedure costs just the same order as the HF method thanks to an efficient algorithm of the TC method~\cite{TCaccel}.
Note that this equation comes down to the HF equation when $u=0$.
In TC\verb!++!, one-electron orbitals are expanded with a plane-wave basis set, and their coefficients are determined by an iterative diagonalization scheme.
In particular, we adopt the block-Davidson method~\cite{Davidson1, Davidson2}, a detail of which are presented in our previous study~\cite{TCPW} and shall be presented in Sec.~\ref{sec:diagonalization}.
The total energy,
\begin{equation}
E = \frac{\langle \Psi | \mathcal{H} | \Psi \rangle}{\langle \Psi  | \Psi \rangle} \label{eq:totE},
\end{equation}
is often approximated by the TC pseudoenergy,
\begin{equation}
E_{\mathrm{TC}} = \mathrm{Re} \bigg[ \frac{\langle \Phi | \mathcal{H}_{\mathrm{TC}} | \Phi \rangle}{\langle \Phi  | \Phi \rangle} \bigg],\label{eq:EtotTC}
\end{equation}
where these two quantities coincide when $\Phi$ is the exact eigenstate of $\mathcal{H}_{\mathrm{TC}}$. 
An important advantage for using the TC pseudoenergy is that $E_{\mathrm{TC}}$ requires only nine-dimensional (three-body) integration.

We also describe the biorthogonal formulation of the TC method, called the biorthogonal TC (BITC) method.
In the BITC method, we use left and right Slater determinants consisting of different one-electron orbitals: $X=\mathrm{det}[\chi_i({\bm r}_j)]$ and $\Phi=\mathrm{det}[\phi_i({\bm r}_j)]$, respectively, with the biorthogonal condition $\langle \chi_i | \phi_j \rangle = \delta_{i,j}$ and the normalization condition $\langle \phi_i | \phi_i \rangle = 1$. Then a one-body SCF equation becomes slightly different from Eq.~(\ref{eq:SCF}): $\phi^*({\bm r})$ are replaced with $\chi^*({\bm r})$ and the right-hand side of the SCF equation can be diagonal, i.e., $\epsilon_{ij} = 0\ (i\neq j)$:
\begin{align}
\left( -\frac{1}{2}\nabla_1^2 +v_{\mathrm{ext}}({\bm r}_1) \right) \phi_i ({\bm r}_1)\notag \\
+ \sum_{j=1}^N
\int \mathrm{d}{\bm r}_2\  \chi_j^*({\bm r}_2) v_{\mathrm{2body}}(x_1,x_2)
\mathrm{det} \left[
\begin{array}{rrr}
\phi_i({\bm r}_1) & \phi_i({\bm r}_2) \\
\phi_j({\bm r}_1) & \phi_j({\bm r}_2) \\
\end{array} \right] \notag \\
- \sum_{j=1}^N \sum_{k>j}^N
\int \mathrm{d}{\bm r}_2 \mathrm{d}{\bm r}_3\  \chi_j^*({\bm r}_2)\chi_k^*({\bm r}_3)v_{\mathrm{3body}}(x_1,x_2,x_3)  \notag \\
\times 
\mathrm{det} \left[
\begin{array}{rrr}
\phi_i({\bm r}_1) & \phi_i({\bm r}_2) &  \phi_i({\bm r}_3) \\
\phi_j({\bm r}_1) & \phi_j({\bm r}_2) & \phi_j({\bm r}_3) \\
\phi_k({\bm r}_1) & \phi_k({\bm r}_2) & \phi_k({\bm r}_3)
\end{array} \right]
= \epsilon_{ii} \phi_i({\bm r}_1). \label{eq:BITCSCF}
\end{align}
 When diagonalizing this one-body SCF equation, we get $\chi$ and $\phi$ as the left and right eigenstates.
This procedure is equivalent to that we also impose the one-body SCF equation for the left orbitals $\chi$:
\begin{align}
\left( -\frac{1}{2}\nabla_1^2 +v_{\mathrm{ext}}({\bm r}_1) \right) \chi_i ({\bm r}_1)\notag \\
+ \sum_{j=1}^N
\int \mathrm{d}{\bm r}_2\  \phi_j^*({\bm r}_2) v^{\dag}_{\mathrm{2body}}(x_1,x_2)
\mathrm{det} \left[
\begin{array}{rrr}
\chi_i({\bm r}_1) & \chi_i({\bm r}_2) \\
\chi_j({\bm r}_1) & \chi_j({\bm r}_2) \\
\end{array} \right] \notag \\
- \sum_{j=1}^N \sum_{k>j}^N
\int \mathrm{d}{\bm r}_2 \mathrm{d}{\bm r}_3\  \phi_j^*({\bm r}_2)\phi_k^*({\bm r}_3)v_{\mathrm{3body}}(x_1,x_2,x_3)  \notag \\
\times 
\mathrm{det} \left[
\begin{array}{rrr}
\chi_i({\bm r}_1) & \chi_i({\bm r}_2) &  \chi_i({\bm r}_3) \\
\chi_j({\bm r}_1) & \chi_j({\bm r}_2) & \chi_j({\bm r}_3) \\
\chi_k({\bm r}_1) & \chi_k({\bm r}_2) & \chi_k({\bm r}_3)
\end{array} \right]
= \epsilon_{ii}^* \chi_i({\bm r}_1). \label{eq:BITCSCFconj}
\end{align}
The BITC pseudoenergy is defined as,
\begin{equation}
E_{\mathrm{BITC}} = \mathrm{Re} \bigg[  \frac{\langle X | \mathcal{H}_{\mathrm{TC}} | \Phi \rangle}{\langle X  | \Phi \rangle} \bigg].\label{eq:EtotBiTC}
\end{equation}

We can use many kinds of the Jastrow function $u$.
At present, TC\verb!++! supports the following simple Jastrow function~\cite{Sakuma,QMCreview,Ceperley,CeperleyAlder}:
\begin{equation}
u(x, x')
= u_{\sigma, \sigma'}(|{\bm r} - {\bm r'}|)
= \frac{A_{\sigma, \sigma'}}{|{\bm r} - {\bm r'}|}\left( 1 - \mathrm{e}^{-|{\bm r} - {\bm r'}|/C_{\sigma, \sigma'}} \right) \label{eq:Jastrow},
\end{equation}
where
\begin{equation}
C_{\sigma, \sigma'} = \sqrt{2A_{\sigma, \sigma'}}\ (\sigma=\sigma'),\ \ \sqrt{A_{\sigma, \sigma'}}\ (\sigma\neq\sigma').\label{eq:cusp}
\end{equation}
This relation is derived by the cusp condition~\cite{cusp, cusp2}.
Note that this Jastrow function is actually spin-contaminated and does not satisfy the (true) cusp condition. This deficiency can be avoided by constructing the Jastrow factor with the permutation operator, but this procedure introduces non-terminating series of interaction in the TC Hamiltonian~\cite{Ten-nocusp}. Thus, we adopt this approximate cusp condition here. Fortunately, a VMC study reported that an effect of spin contamination on accuracy of the wave function and its energy is small~\cite{cuspUmrigar}.
The parameter $A$ is often determined by
the long-range asymptotic behavior of the Jastrow function determined by the random-phase approximation of homogeneous electron gas~\cite{BohmPines}:
\begin{equation}
A = \sqrt{\frac{\Omega}{4\pi N_{\mathrm{unit}}}} \label{eq:A_normalize},
\end{equation}
where $\Omega$ and $N_{\mathrm{unit}}$ are the unit-cell volume and the number of electrons therein, respectively.
In TC\verb!++!, one can use different values for the parameter $A$ while keeping the cusp condition by imposing Eq.~(\ref{eq:cusp}), as adopted in \cite{TCjfo}.

\subsection{Fourier transform and convolution formula~\label{sec:Fourier}}

Before describing details of implementation, we define some notations in our paper.
Fourier transformation of the periodic function $f({\bm r})$ satisfying $f({\bm r} + {\bm R}) = f({\bm r})$ for an arbitrary lattice vector ${\bm R}$, is denoted as $\tilde{f}({\bm G})$ in this paper:
\begin{equation}
f({\bm r}) = \sum_{{\bm G}} {\tilde{f}} ({\bm G}) \mathrm{e}^{i{\bm G}\cdot{\bm r}}, \ \ 
\tilde{f}({\bm G})= \frac{1}{\Omega} \int_{\Omega} \mathrm{d}{\bm r}\  f ({\bm r}) \mathrm{e}^{-i{\bm G}\cdot{\bm r}},
\end{equation}
where the integration is performed in the unit cell, which yields the factor of $1/\Omega$.
We also represent it as $\tilde{f}({\bm G})=$ FT[$f({\bm x})$] or $f({\bm x}) =$ FT$^{-1}$[$\tilde{f}({\bm G})$], both of which are calculated using the Fast-Fourier-Transform technique.
On the other hand, Fourier transformation of the non-periodic function $g({\bm r})$ such as the Coulomb potential is defined as
\begin{equation}
g({\bm r})= \frac{1}{(2\pi)^3} \int \mathrm{d}{\bm G}\  \tilde{g} ({\bm G}) \mathrm{e}^{i{\bm G}\cdot{\bm r}},\ \ 
\tilde{g}({\bm G})= \int \mathrm{d}{\bm r}\  g ({\bm r}) \mathrm{e}^{-i{\bm G}\cdot{\bm r}},
\end{equation}
in this paper, where the integration is performed in the infinitely large region.
For example, it is well known that the Fourier transform of $1/r$ is $4\pi/G^2$.

The following integration often takes place in electronic-structure calculation:
\begin{equation}
\int \mathrm{d}{\bm r'}\ g({\bm r}-{\bm r'})f({\bm r'}) \mathrm{e}^{i{\bm k}\cdot {\bm r'}}
\end{equation}
where $f$ is periodic with respect to the lattice-vector translation while $g$ and $\mathrm{e}^{i{\bm k}\cdot {\bm r'}}$ are not.
We can derive the convolution formula even in this case:
\begin{align}
&\int \mathrm{d}{\bm r'}\ g({\bm r}-{\bm r'})f({\bm r'}) \mathrm{e}^{i{\bm k}\cdot {\bm r'}} \\
&= \frac{1}{(2\pi)^3}  \sum_{\bm G} \int \mathrm{d}{\bm G'} \mathrm{d}{\bm r'}\ \tilde{g}({\bm G'}) \tilde{f}({\bm G})
\mathrm{e}^{i{\bm G'}\cdot({\bm r}-{\bm r'})} \mathrm{e}^{i({\bm k} +{\bm G})\cdot{\bm r'}} \\
&=  \sum_{\bm G} \tilde{g}({\bm k}+{\bm G}) \tilde{f}({\bm G})
 \mathrm{e}^{i({\bm k}+{\bm G})\cdot{\bm r}}, \label{eq:convolution}
\end{align}
as is often used in many calculation codes.

\subsection{One-electron orbitals}

The left and right one-electron orbitals in the BITC method can be written as
\begin{align}
\phi_{j}({\bm r}) &= \frac{\mathrm{e}^{i {\bm k}\cdot {\bm r}}}{\sqrt{N_{\bm k}}} \phi_{\mathrm{periodic}, j}({\bm r})
=   \frac{\mathrm{e}^{i {\bm k}\cdot {\bm r}}}{\sqrt{N_{\bm k}}}  \sum_{{\bm G}} \tilde{\phi}_{\mathrm{periodic},j}({\bm G}) \mathrm{e}^{i {\bm G}\cdot {\bm r}},\\
\chi_{j}({\bm r}) &= \frac{\mathrm{e}^{i {\bm k}\cdot {\bm r}}}{\sqrt{N_{\bm k}}}  \chi_{\mathrm{periodic}, j}({\bm r})
=  \frac{\mathrm{e}^{i {\bm k}\cdot {\bm r}}}{\sqrt{N_{\bm k}}} \sum_{{\bm G}} \tilde{\chi}_{\mathrm{periodic}, j}({\bm G}) \mathrm{e}^{i {\bm G}\cdot {\bm r}},
\end{align}
where the orbital index $i$ denotes a pair of spin, ${\bm k}$-vector, and band indices: $j = (\sigma, {\bm k},\mu_j)$. $N_{\bm k}$ is the number of ${\bm k}$ points.
The functions $\phi_{\mathrm{periodic}, j}$ and $\chi_{\mathrm{periodic}, j}$ are periodic with respect to arbitrary translation compatible with the unit cell.
$\tilde{\phi}_{\mathrm{periodic},j}$ and $\tilde{\chi}_{\mathrm{periodic},j}$ are the Fourier transform of $\phi_{\mathrm{periodic}, j}$ and $\chi_{\mathrm{periodic}, j}$, respectively, as defined in Sec.~\ref{sec:Fourier}.
Normalization of the one-electron orbital is imposed as follows:
\begin{equation}
\langle \phi_j | \phi_j \rangle = \int_{\mathrm{super}} \mathrm{d}{\bm r}\ |\phi_{j}({\bm r})|^2 =  \int_{\Omega} \mathrm{d}{\bm r}\ |\phi_{\mathrm{periodic}, j}({\bm r})|^2 = 1,
\end{equation}
where the integration denoted as $\int_{\mathrm{super}}$ is performed in the supercell with a volume of $N_{\bm k}\Omega$.

Here we explain how we impose the orthonormal condition on one-electron orbitals in the TC method, while eigenvectors of a non-Hermitian operator are not orthogonal each other.
This method is described in \cite{Umezawa}.
First, we rewrite the left-hand side of the SCF equation, Eq.~(\ref{eq:SCF}), as $\hat{h}\phi_i({\bm r})$.
By diagonalizing the non-Hermitian operator $\hat{h}$, we get one-electron orbitals that are not orthogonal each other.
Then, we apply the Gram--Schmidt orthonormalization for these orbitals.
By this procedure, we can get the orthonormalized orbitals $\phi_i$ while the eigenvalue matrix $\epsilon_{ij}$ becomes non-diagonal owing to the Gram--Schmidt orthonormalization.
We note that the diagonal element $\epsilon_{ii}$ is unchanged by the Gram--Schmidt orthonormalization as proven in \cite{Umezawa}.
Another simpler proof is shown in Appendix A.
The real part of $\epsilon_{ii}$ can be regarded as an one-electron energy on the basis of the Koopmans' theorem, which was also proven in \cite{Umezawa}.
This allows ones to depict the effective band dispersion, which is an important advantage of the TC method.

For the BITC method, the situation is rather simple: ones just diagonalize the SCF operator $\hat{h}$ and set left and right eigenvectors as $\chi$ and $\phi$.
The biorthogonal condition is automatically satisfied because
\begin{equation}
\epsilon_{ii} \langle \chi_i | \phi_j \rangle = \langle \chi_i | \hat{h} | \phi_j \rangle = \epsilon_{jj} \langle \chi_i | \phi_j \rangle \ \ \Rightarrow\ \ 
\langle \chi_i | \phi_j \rangle = 0\ \ (\epsilon_{ii} \neq \epsilon_{jj}).
\end{equation}
Note that the biorthogonal condition can be easily applied also to eigenvectors with degenerate eigenvalues.

\subsection{Diagonalization~\label{sec:diagonalization}}

By rewriting the left-hand side of the SCF equation, Eq.~(\ref{eq:SCF}), as $\hat{h}\phi_i({\bm r})$, the SCF equation can be regarded as the eigenvalue problem of the non-Hermitian operator $\hat{h}$. We adopt the block-Davidson method~\cite{Davidson1, Davidson2} for solving the eigenvalue problem, a detail of which are presented in our previous study~\cite{TCPW}.

In the block Davidson algorithm, we begin with the initial trial vectors \{$v_1^{(1)}, v_2^{(1)}, \dots, v_{n_{\mathrm{bands}}}^{(1)}$\} and estimated eigenvalues \{$\epsilon_1^{(1)}, \epsilon_2^{(1)}, \dots, \epsilon_{n_{\mathrm{bands}}}^{(1)}$\} for $\hat{h}[\phi]$, where $n_{\mathrm{bands}}$ is the number of bands considered at each ${\bm k}$-point. Here, we consider the diagonalization problem at each ${\bm k}$-point while orbitals are updated for all the ${\bm k}$-points simultaneously. The trial vectors are used as basis functions to represent the cell-periodic part of one-electron orbitals, $\phi_{\mathrm{periodic}}$, and satisfy the orthonormal condition, $\langle v_i^{(1)} | v_j^{(1)} \rangle = \delta_{i,j}$.
The initial trial vectors $v^{(1)}$ and eigenvalues $\epsilon^{(1)}$ are extracted from Quantum ESPRESSO (e.g., DFT or HF results).

To obtain $v_{j}^{(1)}$ ($pn_{\mathrm{bands}}+1 \leq j \leq (p+1)n_{\mathrm{bands}}$) from  $v_{j}^{(1)}$ ($(p-1)n_{\mathrm{bands}}+1 \leq j \leq pn_{\mathrm{bands}}$) 
($p = 1,2,\dots, p_{\mathrm{max}}$), we calculate
\begin{equation}
v_{pn_{\mathrm{bands}}+i}^{(1)}= \hat{P} (\hat{h}^{(1)} - \epsilon^{(1)}_i)v_{(p-1)n_{\mathrm{bands}}+i}^{(1)}\ \ \ (i=1,2,\dots,n_{\mathrm{bands}}), \label{eq:blockDavid}
\end{equation}
where we used the preconditioner $\hat{P}$ proposed by Payne {\it et} {\it al}.~\cite{Payne_precon}
\begin{equation}
\hat{P}({\bm G}) = \frac{27+18x+12x^2+8x^3}{27+18x+12x^2+8x^3+16x^4},
\end{equation}
where
\begin{equation}
x = \frac{1}{2}|{\bm k}+{\bm G}|^2 \left( \sum_{{\bm G}'} \frac{1}{2}|{\bm k}+{\bm G}'|^2 \tilde{v}_{(p-1)n_{\mathrm{bands}}+i}^{(1)} ({\bm G}') \right)^{-1}.
\end{equation}
This operator $\hat{P}$ acts on the Fourier transform of $(\hat{h}^{(1)} - \epsilon^{(1)}_i)v^{(1)}$ in reciprocal space to suppress high-frequency components of plane waves.
After that, we perform the Gram-Schmidt orthonormalization for the new trial vectors $v_{j}^{(1)}$ ($pn_{\mathrm{bands}}+1 \leq j \leq (p+1)n_{\mathrm{bands}}$) so that all $v_{j}^{(1)}$ obtained so far (i.e., $1 \leq j \leq (p+1)n_{\mathrm{bands}}$) are orthonormalized.
By repeating these processes until all $v_j^{(1)}$ and $\hat{h}^{(1)} v_j^{(1)}$ are obtained for $1 \leq j \leq (p_{\mathrm{max}}+1)n_{\mathrm{bands}}$, we can construct a subspace Hamiltonian $\langle v^{(1)}_i | \hat{h}^{(1)} | v^{(1)}_j\rangle$.
The subspace dimension is $ (p_{\mathrm{max}}+1)n_{\mathrm{bands}}$.
Note that the subspace dimension is sometimes smaller than this value in real calculation, because we exclude a trial vector that is linearly dependent on other trial vectors or has a very small norm ($<10^{-8}$ in the present implementation) before Gram-Schmidt orthonormalization.

By diagonalizing the subspace Hamiltonian, we can get a better estimate of the eigenvectors and the eigenvalues of $\hat{h}$.
By using them, we update orbitals included in $\hat{h}^{(1)}$ and call it $\hat{h}^{(2)}$.
The eigenvectors that have the lowest $n_{\mathrm{bands}}$ eigenvalues are used as new trial vectors \{$v_1^{(2)}, v_2^{(2)}, \dots, v_{n_{\mathrm{bands}}}^{(2)}$\} after Gram-Schmidt orthonormalization, and these eigenvalues are also used as a new estimate \{$\epsilon_1^{(2)}, \epsilon_2^{(2)}, \dots, \epsilon_{n_{\mathrm{bands}}}^{(2)}$\}. 
Starting from these trial vectors and estimated eigenvalues, other trial vectors are constructed using Eq.~(\ref{eq:blockDavid}), namely,
\begin{equation}
v_{pn_{\mathrm{bands}}+i}^{(2)}= \hat{P} (\hat{h}^{(2)} - \epsilon^{(2)}_i)v_{(p-1)n_{\mathrm{bands}}+i}^{(2)}\ \ \ (i=1,2,\dots,n_{\mathrm{bands}}),
\end{equation}
for $p = 1,2,\dots, p_{\mathrm{max}}$.
This self-consistent procedure is continued until the total energy and the charge density are sufficiently converged.
For band-structure calculation, we instead check convergence of the summation over the band eigenvalues.

We did not update orbitals in $\hat{h}$ for every loop in our old implementation shown in Ref.~\cite{TCPW}.
However, we update them in every loop of subspace diagonalization in our present implementation for efficient computation.
For some systems, there can be a chance that the former way is more efficient.
In addition, we implemented the linear mixing of the electron density or the density matrix in this self-consistent loop.
The latter option is given because our TC-SCF Hamiltonian is not determined only by the electron density.
For the linear mixing of the (spin) density~\cite{memo_density} , the density is replaced as
\begin{align}
&n_{\sigma}({\bm r}) = \sum_{{\bm k},\mu} \chi_i^{\mathrm{new,*}}({\bm r}) \phi^{\mathrm{new}}_i ({\bm r}) f^{\mathrm{new}}_i,\ \ \ (i =  (\sigma, {\bm k},\mu)) \\
&\to \beta \sum_{{\bm k},\mu} \chi_i^{\mathrm{new,*}}({\bm r}) \phi^{\mathrm{new}}_i ({\bm r}) f_i^{\mathrm{new}} + (1-\beta) \sum_{{\bm k},\mu} \chi_k^{\mathrm{old,*}}({\bm r}) \phi^{\mathrm{old}}_k ({\bm r}) f_i^{\mathrm{old}},
\end{align}
where $\beta$ is a mixing ratio and new (old) quantities in the self-consistent loop are shown with a superscript `new' (`old'). On the other hand, the linear mixing of the density matrix is given as
\begin{align}
&\sum_{{\bm k},\mu} \chi_i^{\mathrm{new,*}}({\bm r'}) \phi^{\mathrm{new}}_i ({\bm r}) f^{\mathrm{new}}_i \\
&\to  \beta \sum_{{\bm k},\mu} \chi_i^{\mathrm{new,*}} ({\bm r'}) \phi^{\mathrm{new}}_i ({\bm r}) f^{\mathrm{new}}_i +  (1-\beta) \sum_{{\bm k},\mu} \chi_i^{\mathrm{old,*}} ({\bm r'}) \phi^{\mathrm{old}}_i ({\bm r}) f^{\mathrm{old}}_i.
\end{align}
We implemented the density-matrix mixing just by replacing orbitals and fillings $\{\chi^{\mathrm{new}}_i, \phi^{\mathrm{new}}_i, f^{\mathrm{new}}_i\}$ with $\{ \{\chi^{\mathrm{new}}_i, \phi^{\mathrm{new}}_i, \beta f^{\mathrm{new}}_i\}, \{\chi^{\mathrm{old}}_i, \phi^{\mathrm{old}}_i, (1-\beta) f^{\mathrm{old}}_i\} \}$. To say, old orbitals are considered with a fictitious band index with a rescaled band filling.

For the BITC method, both the left and right eigenvectors are expanded with the trial vectors $v_i$ in our algorithm.
We have another choice for diagonalization; we can use left and right trial vectors separately, as described in Ref.~\cite{Hirao_Nakatsuji}.
At present, we do not implement this alternative algorithm since it did not improve convergence at least by our implementation.
While more sophisticate implementation might resolve this problem, this is a future issue for improving convergence of BITC calculations.

\subsection{Divergence correction in reciprocal space\label{sec:divcorr}}

Both the Coulomb potential and the Jastrow function used in this study, Eq.~(\ref{eq:Jastrow}), asymptotically behave as $\propto 1/|{\bm r}-{\bm r'}|$ for a long electron-electron distance, which yields a singularity of $\propto 1/q^2\ (q\to 0)$ in reciprocal space after Fourier transformation.
The presence of the singularity means that we should integrate a rapidly varying function in reciprocal space, which makes the convergence with respect to the number of ${\bm k}$-points much worse. To say, the problem is that the integral
\begin{equation}
I = \frac{\Omega}{(2\pi)^3} \int_{\mathrm{1st\ BZ}} \mathrm{d}{\bm q}\ \sum_{\bm{G}} \frac{1}{|{\bm q} + {\bm G}|^2} f({\bm q}+{\bm G}) \label{eq:divcorr_prob},
\end{equation}
is difficult to approximate by a finite ${\bm k}$-point sampling:
\begin{equation}
S = \frac{1}{N_{\bm k}} \sum_{\substack{{\bm q}, {\bm G}\\ ({\bm q} + {\bm G}\ne {\bm 0})}} \frac{1}{|{\bm q} + {\bm G}|^2} f({\bm q}+{\bm G}), \label{eq:divcorr_summation}
\end{equation}
where the summation over ${\bm q}$ is performed within the first Brillouin zone.
Here we assume that $f({\bm q})$ is a slowly varying function and the integral, Eq.~(\ref{eq:divcorr_prob}), itself is well-defined.
While $I = S$ holds at the limit of $N_{\bm k}\to \infty$, the diverging behavior of the integrand makes it difficult to achieve good convergence with a small number of ${\bm k}$-point.

Gygi and Baldereschi proposed a way to alleviate this problem~\cite{Gygi}.
By using an auxiliary function $A_{\mathrm{aux}}({\bm q})$ having a singularity of $1/q^2\ (q\to 0)$ in reciprocal space,
the difference $I-S$ can be well approximated as
\begin{equation}
I-S \simeq \bigg[  
 \frac{\Omega}{(2\pi)^3} \int_{\mathrm{1st\ BZ}} \mathrm{d}{\bm q}\ \sum_{\bm{G}}\ A_{\mathrm{aux}}({\bm q}+{\bm G}) -
 \frac{1}{N_{\bm k}}\sum_{\substack{{\bm q}, {\bm G}\\ ({\bm q} + {\bm G}\ne {\bm 0})}}  A_{\mathrm{aux}}({\bm q}+{\bm G}) 
 \bigg] f({\bm 0}). \label{eq:correction}
\end{equation}
Therefore, if one can evaluate the right-hand side of Eq.~(\ref{eq:correction}), this quantity can be a good correction to be added to $S$ for approximating $I$.
For this purpose, we adopt the auxiliary function proposed in Ref.~\cite{div_corr1}:
\begin{equation}
A_{\mathrm{aux}}({\bm q}) = \frac{\mathrm{e}^{-\alpha q^2}}{q^2}, \label{eq:aux}
\end{equation}
integration of which can be analytically evaluated as
\begin{equation}
\int_{\mathrm{1st\ BZ}} \mathrm{d}{\bm q}\ \sum_{\bm{G}} A_{\mathrm{aux}}({\bm q}+{\bm G}) = \int_{\mathrm{whole\ BZ}} \mathrm{d}{\bm q}\ A_{\mathrm{aux}}({\bm q})
= 2\pi\sqrt{\frac{\pi}{\alpha}}.
\end{equation}
For the finite ${\bm k}$-point sampling, ${\bm q}={\bm 0}$ is not necessarily included in $\sum_{{\bm q}}$. 
When ${\bm q}={\bm 0}$ is included in $\sum_{\bm q}$ of Eqs.~(\ref{eq:divcorr_summation}) and (\ref{eq:correction}), 
while the ${\bm q}+{\bm G}={\bm 0}$ term, $A_{\mathrm{aux}}({\bm 0})$, having an infinite value is excluded from it, this equation is a bit modified as follows:
\begin{align}
&I-S \\
&\simeq \bigg[  
 \frac{\Omega}{(2\pi)^3} \int_{\mathrm{1st\ BZ}} \mathrm{d}{\bm q}\ \sum_{\bm{G}}\ A_{\mathrm{aux}}({\bm q}+{\bm G}) -
 \frac{1}{N_{\bm k}} \bigg( -\alpha + \sum_{\substack{{\bm q}, {\bm G}\\ ({\bm q} + {\bm G}\ne {\bm 0})}}  A_{\mathrm{aux}}({\bm q}+{\bm G}) \bigg)
 \bigg] f({\bm 0}) \\
 &= \bigg[ \frac{\Omega}{4\sqrt{\pi^{3} \alpha}} -
 \frac{1}{N_{\bm k}}\bigg( -\alpha + \sum_{\substack{{\bm q}, {\bm G}\\ ({\bm q} + {\bm G}\ne {\bm 0})}}  A_{\mathrm{aux}}({\bm q}+{\bm G}) \bigg)
 \bigg] f({\bm 0}) \label{eq:correction2}
\end{align}
because $\lim_{{\bm q}\to {\bm 0}} (A_{\mathrm{aux}}({\bm q})-1/q^2) = -\alpha$ for the auxiliary function, Eq.~(\ref{eq:aux}).

In the TC method, we should handle another type of divergence like
\begin{align}
{\bm I}' &= \frac{\Omega}{(2\pi)^3} \int_{\mathrm{1st\ BZ}} \mathrm{d}{\bm q}\ \sum_{\bm{G}} \frac{{\bm q}+{\bm G}}{|{\bm q} + {\bm G}|^2} f({\bm q}+{\bm G}),\\
{\bm S}' &=  \frac{1}{N_{\bm k}}\sum_{\substack{{\bm q}, {\bm G}\\ ({\bm q} + {\bm G}\ne {\bm 0})}} \frac{{\bm q}+{\bm G}}{|{\bm q} + {\bm G}|^2} f({\bm q}+{\bm G}),
\end{align}
for considering $\nabla u$. The correction term for it is
\begin{align}
&{\bm I}' -{\bm S}'  \\
&\simeq \bigg[  
 \frac{\Omega}{(2\pi)^3} \int_{\mathrm{1st\ BZ}} \mathrm{d}{\bm q}\ \sum_{\bm{G}}\ ({\bm q}+{\bm G})A_{\mathrm{aux}}({\bm q}+{\bm G}) -
 \frac{1}{N_{\bm k}}\sum_{\substack{{\bm q}, {\bm G}\\ ({\bm q} + {\bm G}\ne {\bm 0})}}  ({\bm q}+{\bm G})A_{\mathrm{aux}}({\bm q}+{\bm G}) 
 \bigg] f({\bm 0})\\
 &= \bigg[ - \frac{1}{N_{\bm k}}\sum_{\substack{{\bm q}, {\bm G}\\ ({\bm q} + {\bm G}\ne {\bm 0})}}  ({\bm q}+{\bm G})A_{\mathrm{aux}}({\bm q}+{\bm G}) 
 \bigg] f({\bm 0}), \label{eq:correction_nabla}
\end{align}
because $\int_{\mathrm{whole\ BZ}} \mathrm{d}{\bm q}\ {\bm q}A_{\mathrm{aux}}({\bm q})={\bm 0}$.

Implementation of the divergence correction for each term in the TC-SCF equation shall be presented later in this paper.

\subsection{Details of implementation}

From here on, we focus on $\hat{h}\phi$ in the BITC method for simplicity in this paper.
However, that for the TC method can be obtained by simply replacing $\chi^*({\bm r})$ to $\phi^*({\bm r})$.
We use the following notation to represent integration:
\begin{align}
&\langle *, q_1, q_2 | \nabla_1 u_{12} \nabla_1 u_{13}| q_1, q_2, j\rangle \equiv \notag \\
& \int \mathrm{d}{\bm r}_2 \mathrm{d}{\bm r}_3\ \chi^*_{q_1} ({\bm r}_2) \chi^*_{q_2} ({\bm r}_3)
\nabla_1 u_{\sigma_1, \sigma_2}(|{\bm r}_1 - {\bm r}_2|) \nabla_1 u_{\sigma_1, \sigma}(|{\bm r}_1 - {\bm r}_3|) \notag \\
 &\ \ \ \ \ \ \ \ \phi_{q_1} ({\bm r}_1) \phi_{q_2} ({\bm r}_2) \phi_j ({\bm r}_3) \label{eq:example_notation_3body}
\end{align}
where each one-electron orbital is specified by a set of spin, ${\bm k}$-vector, and band indices:
\begin{equation}
q_1 =  (\sigma_1, {\bm q}_1,\mu_1),\ \  q_2 =  (\sigma_2, {\bm q}_2,\mu_2),\ \   j = (\sigma, {\bm k},\mu_j).
\end{equation}
In our notation, $*$ in $\langle *, q_1, q_2 | \nabla_1 u_{12} \nabla_1 u_{13}| q_1, q_2, j\rangle$ means that there is no bra orbital for ${\bm x}_1$ in the integrand, and integration over ${\bm x}_1$ is not performed.
Because bra and ket orbitals with the same variable should have the same spin indices (e.g., $\chi^*_{q_2} ({\bm x}_3)$ and $\phi_j ({\bm x}_3)$ in Eq.~(\ref{eq:example_notation_3body})), $\sigma = \sigma_1 = \sigma_2$ is imposed for the above integral.
If one considers another term, $\langle *, q_1, q_2 | \nabla_1 u_{12} \nabla_1 u_{13}| q_2, q_1, j\rangle$, $\sigma_1$ can be either parallel or anti-parallel to $\sigma_2 = \sigma$. In this case, while $u_{13}$ should be the spin-parallel Jastrow function owing to $\sigma_2=\sigma$, $u_{12}$ is the spin-parallel and spin-anti-parallel Jastrow functions for $\sigma_1=\sigma_2$ and $\sigma_1\neq \sigma_2$, respectively. In the SCF calculation, we take a summation over occupied orbitals $q_1, q_2$:
\begin{align}
&\sum_{q_1, q_2}^{\mathrm{occupied}} \langle *, q_1, q_2 | \nabla_1 u_{12} \nabla_1 u_{13}| q_1, q_2, j\rangle \equiv \notag \\
&\sum_{q_1, q_2} \int \mathrm{d}{\bm r}_2 \mathrm{d}{\bm r}_3\ \chi^*_{q_1} ({\bm r}_2) \chi^*_{q_2} ({\bm r}_3)
\nabla_1 u_{\sigma_1, \sigma_2}(|{\bm r}_1 - {\bm r}_2|) \nabla_1 u_{\sigma_1, \sigma}(|{\bm r}_1 - {\bm r}_3|) \notag \\
 &\ \ \ \ \ \ \ \ \phi_{q_1} ({\bm r}_1) \phi_{q_2} ({\bm r}_2) \phi_j ({\bm r}_3)f_{q_1}f_{q_2},
\end{align}
where $f_q$ is the occupation number of the state $q$, satisfying $0\leq f_q \leq 1$.

\subsubsection{Computational treatment of one-electron orbitals\label{sec:comput_one_orb}}

For representing a cell-periodic function $f({\bm r})$, such as a cell-periodic part of the one-electron orbital $\phi_{\mathrm{periodic}}$, as a discrete numerical array $f_i$ in our computational code, we used a dimensionless array defined as
\begin{equation}
f_i = \sqrt{\Omega} f({\bm r}_i)
\end{equation}
so that the normalization condition satisfies
\begin{equation}
\frac{1}{N_{\mathrm{pw}}}\sum_i |f_i|^2 = 1 = \int_{\Omega} \mathrm{d}{\bm r}\ |f({\bm r})|^2,
\end{equation}
where $N_{\mathrm{pw}}$ is the number of plane waves for Fourier transform, which equals the number of discrete points in the real-space mesh.
This equality can be understood by 
\begin{equation}
\int_{\Omega} \mathrm{d}{\bm r} \simeq (\Delta x\Delta y\Delta z) \sum_i = \frac{\Omega}{N_{\mathrm{pw}}} \sum_i.
\end{equation}
To represent $f({\bm r})$ by $f_i$, we often divide calculated quantities by $\Omega$ and/or $N_{\mathrm{pw}}$ in our code,
but we did not use $f_i$ in this paper and so did not show such factors in equations shown in this paper.

We also mention how the crystal symmetry is applied to one-electron orbitals.
Suppose that a system has a symmetry operation,
\begin{equation}
{\bm r} \to {\bm r}' = S({\bm r} + {\bm t}),\ (\mathrm{i.e.},\  {\bm r} = S^{-1}{\bm r}' - {\bm t}) \label{eq:symmetry_operation}
\end{equation}
where $S$ and ${\bm t}$ are a symmorphic symmetry operator and a translation vector, respectively.
Then, by applying this symmetry operation to the one-electron orbital for the state $j_0=(\sigma, {\bm k}_0, \mu_j)$,
\begin{equation}
\phi_{j_0}({\bm r}) = \frac{\mathrm{e}^{i {\bm k}_0\cdot {\bm r}}}{\sqrt{N_{\bm k}}} \phi_{\mathrm{periodic}, j_0}({\bm r})
= \frac{\mathrm{e}^{i {\bm k}_0\cdot {\bm r}}}{\sqrt{N_{\bm k}}} \sum_{{\bm G}_0} \tilde{\phi}_{\mathrm{periodic},j_0}({\bm G}_0) \mathrm{e}^{i {\bm G}_0\cdot {\bm r}},
\end{equation}
we can suppose that $\phi_{j_0}({\bm r}')$ is also the eigenstate of ${\hat h}$~\cite{note_symmetry}.
Note that this is not always true, e.g., when a symmetry breaking for the electronic state takes place. Thus, we can choose whether the symmetry operations are used in an input file of calculation.
By using this symmetry operation, we can get
\begin{align}
\phi_{j_0}({\bm r}') &=  \frac{\mathrm{e}^{i {\bm k}_0\cdot S({\bm r}+{\bm t})}}{\sqrt{N_{\bm k}}} \sum_{{\bm G}_0} \tilde{\phi}_{\mathrm{periodic},j_0}({\bm G}_0) \mathrm{e}^{i {\bm G}_0\cdot S({\bm r}+{\bm t})} \\
&\propto \frac{\mathrm{e}^{i (S^{\dag}{\bm k}_0)\cdot {\bm r}}}{\sqrt{N_{\bm k}}}  \sum_{{\bm G}_0} \tilde{\phi}_{\mathrm{periodic},j_0}({\bm G}_0) \mathrm{e}^{i (S^{\dag} {\bm G}_0)\cdot {\bm t}}  \mathrm{e}^{i (S^{\dag}{\bm G}_0)\cdot {\bm r}}, \label{eq:symmetry_orbital1}
\end{align}
where we discard a constant phase of $\mathrm{exp}[i{\bm k}_0\cdot S{\bm t}]$ in the second line. By defining
\begin{equation}
{\bm k} = S^{\dag}{\bm k}_0,\ \ {\bm G} = S^{\dag}{\bm G}_0,
\end{equation}
we can rewrite Eq.~(\ref{eq:symmetry_orbital1}) as follows:
\begin{equation}
\phi_{j_0}({\bm r}') \propto \frac{\mathrm{e}^{i {\bm k}\cdot {\bm r}}}{\sqrt{N_{\bm k}}}  \sum_{{\bm G}_0} \tilde{\phi}_{\mathrm{periodic},j_0}({\bm G}_0) \mathrm{e}^{i {\bm G}\cdot {\bm t}}  \mathrm{e}^{i {\bm G}\cdot {\bm r}}, 
\end{equation}
which can be regarded as a one-electron orbital for the state $j=(\sigma, {\bm k}, \mu_j)$:
\begin{equation}
\phi_{j}({\bm r}) = \frac{\mathrm{e}^{i {\bm k}\cdot {\bm r}}}{\sqrt{N_{\bm k}}}  \sum_{{\bm G}} \tilde{\phi}_{\mathrm{periodic},j}({\bm G})  \mathrm{e}^{i {\bm G}\cdot {\bm r}},
\end{equation}
where
\begin{equation}
\tilde{\phi}_{\mathrm{periodic},j}({\bm G}) = \tilde{\phi}_{\mathrm{periodic},j_0}({\bm G}_0) \mathrm{e}^{i {\bm G}\cdot {\bm t}}.
\end{equation}

One often considers that $T[\phi_{j}({\bm r}')]$ is the eigenstate of ${\hat h}$, where $T$ is a time-reversal operation.
In this case, the following equalities instead hold:
\begin{align}
j = (-\sigma, {\bm k}, \mu_j),\ \ {\bm k} = -S^{\dag}{\bm k}_0,\ \ {\bm G} = -S^{\dag}{\bm G}_0,\\
\phi_{j}({\bm r}) = \frac{\mathrm{e}^{i {\bm k}\cdot {\bm r}}}{\sqrt{N_{\bm k}}}  \sum_{{\bm G}} \tilde{\phi}_{\mathrm{periodic},j}({\bm G})  \mathrm{e}^{i {\bm G}\cdot {\bm r}},\\
\tilde{\phi}_{\mathrm{periodic},j}({\bm G}) = (\tilde{\phi}_{\mathrm{periodic},j_0}({\bm G}_0))^* \mathrm{e}^{-i {\bm G}\cdot {\bm t}}.
\end{align}

\subsubsection{One-body terms}

One-body terms included in TC Hamiltonian are the kinetic-energy term and the pseudopotential term, which are evaluated in the same way as that adopted in many calculation codes.
The kinetic-energy operator can be easily applied to the one-electron orbital of the state $j = (\sigma, {\bm k}, \mu)$,
\begin{equation}
\phi_{j}({\bm r}) = \frac{\mathrm{e}^{i {\bm k}\cdot {\bm r}}}{\sqrt{N_{\bm k}}}  \phi_{\mathrm{periodic},j}({\bm r})
=  \frac{\mathrm{e}^{i {\bm k}\cdot {\bm r}}}{\sqrt{N_{\bm k}}}  \sum_{{\bm G}} \tilde{\phi}_{\mathrm{periodic},j}({\bm G}) \mathrm{e}^{i {\bm G}\cdot {\bm r}},
\end{equation}
and we get
\begin{equation}
-\frac{\nabla^2}{2} \phi_{j}({\bm r}) 
=  \frac{\mathrm{e}^{i {\bm k}\cdot {\bm r}}}{\sqrt{N_{\bm k}}} \sum_{{\bm G}} \frac{({\bm k} + {\bm G})^2}{2} \tilde{\phi}_{\mathrm{periodic},j}({\bm G}) \mathrm{e}^{i {\bm G}\cdot {\bm r}}.
\end{equation}
TC\verb!++! at present only accepts a norm-conserving pseudopotential without the partial core correction. For such a pseudopotential, a pseudopotential operator consists of the local and non-local terms for each atom $\tau$:
\begin{equation}
V^{\tau}_{\mathrm{pp}} = V^{\tau}_{\mathrm{loc}}+ V^{\tau}_{\mathrm{nloc}}.
\end{equation}
The local potential $V^{\tau}_{\mathrm{loc}} =  V^{\tau}_{\mathrm{loc}}({\bm r})$ asymptotically behaves as $-Z^{\tau}/|{\bm r}-{\bm r_{\tau}}|$ for a large $|{\bm r}-{\bm r_{\tau}}|$, where ${\bm r_{\tau}}$ and $Z^{\tau}$ are the position and the number of valence electrons for the atom $\tau$.
We assume that the local potential is spherically symmetric, and thus given as $V^{\tau}_{\mathrm{loc}}(|{\bm r}-{\bm r_{\tau}}| )$.
We evaluate the local potential in the following way. First, the following formula is well known in the context of the Ewald summation:
\begin{equation}
\sum_{{\bm R}} \frac{\mathrm{erf}(a|{\bm r} - {\bm r_{\tau}} - {\bm R}|)}{|{\bm r} - {\bm r_{\tau}} - {\bm R}|}
= \frac{4\pi}{\Omega} \sum_{{\bm G}} \frac{\mathrm{exp}(-{\bm G}^2/(4a^2))}{{\bm G}^2} \mathrm{e}^{i{\bm G}\cdot({\bm r} - {\bm r_{\tau}})},
\end {equation}
where ${\bm R}$ and $a$ are the lattice vector and the Ewald parameter, respectively.
Next, the local potential is decomposed into long-ranged and short-ranged functions using the above formula:
\begin{align}
&\sum_{{\bm R}} V^{\tau}_{\mathrm{loc}}(|{\bm r}-{\bm r_{\tau}} - {\bm R}|) =  V^{\tau}_{\mathrm{loc},1}({\bm r}-{\bm r_{\tau}}) +  V^{\tau}_{\mathrm{loc}, 2}({\bm r}-{\bm r_{\tau}}),\\ 
&\ \ \ \ \ \ V^{\tau}_{\mathrm{loc},1}({\bm r}-{\bm r_{\tau}}) = \sum_{{\bm R}} \bigg[ V^{\tau}_{\mathrm{loc}}(|{\bm r}-{\bm r_{\tau}} - {\bm R}|) + \frac{Z^{\tau}\mathrm{erf}(a|{\bm r} - {\bm r_{\tau}} - {\bm R}|)}{|{\bm r} - {\bm r_{\tau}} - {\bm R}|} \bigg],\\
&\ \ \ \ \ \ V^{\tau}_{\mathrm{loc},2}({\bm r}-{\bm r_{\tau}}) = -\frac{4\pi Z^{\tau}}{\Omega} \sum_{{\bm G}} \frac{\mathrm{exp}(-{\bm G}^2/(4a^2))}{{\bm G}^2} \mathrm{e}^{i{\bm G}\cdot({\bm r} - {\bm r_{\tau}})}.
\end{align}
Because $V^{\tau}_{\mathrm{loc},1}$ is a short-ranged function with the lattice periodicity (i.e., $V^{\tau}_{\mathrm{loc},1}({\bm r}-{\bm r_{\tau}}+{\bm R}) = V^{\tau}_{\mathrm{loc},1}({\bm r}-{\bm r_{\tau}})$ for an arbitrary lattice vector ${\bm R}$), we can safely perform Fourier transformation of $V^{\tau}_{\mathrm{loc},1}$:
\begin{equation}
V^{\tau}_{\mathrm{loc},1}({\bm r}-{\bm r_{\tau}}) = \sum_{{\bm G}} {\tilde{V}}^{\tau}_{\mathrm{loc},1} ({\bm G}) \mathrm{e}^{i{\bm G}\cdot{(\bm r}-{\bm r_{\tau}})},
\end{equation}
where
\begin{align}
{\tilde{V}}^{\tau}_{\mathrm{loc},1} ({\bm G})
&= \frac{1}{\Omega N_{\bm R}} \int \mathrm{d}{\bm r}\  V^{\tau}_{\mathrm{loc},1}({\bm r}-{\bm r_{\tau}})  \mathrm{e}^{-i{\bm G}\cdot({\bm r}-{\bm r_{\tau}})} \label{eq:1st_line_Vloc1G}\\
&=  \frac{4\pi}{\Omega} \int_0^{\infty} \mathrm{d}\tilde{r}\  
\left( V^{\tau}_{\mathrm{loc}}(\tilde{r}) + \frac{Z^{\tau}\mathrm{erf}(a\tilde{r})}{\tilde{r}}  \right)
\frac{\sin(|{\bm G}|\tilde{r})}{|{\bm G}|\tilde{r}} \tilde{r}^2.
\end{align}
The integration in Eq.~(\ref{eq:1st_line_Vloc1G}) is defined in the supercell with a volume $\Omega N_{\bm R}$, to consider $\int_0^{\infty}\mathrm{d}\tilde{r}$ integration. $N_{\bm R}$ is the number of ${\bm R}$ vectors, and we consider the $N_{\bm R} \to \infty$ limit.
In $V^{\tau}_{\mathrm{loc}, 2}$, we can exclude a $1/{\bm G}^2$ divergence in the ${\bm G}={\bm 0}$ component because it should be canceled with ${\bm G}={\bm 0}$ components of the ion-ion and electron-electron (Hartree) Coulomb potential terms under the charge neutrality.
This is considered in the usual Ewald summation~\cite{memo_ewald}. Therefore, $V^{\tau}_{\mathrm{loc},1}({\bm r}) +  V^{\tau}_{\mathrm{loc}, 2}({\bm r})$ can be calculated as
\begin{align}
 &\sum_{{\bm G}\ne {\bm 0}} \bigg[ \left( {\tilde{V}}^{\tau}_{\mathrm{loc},1} ({\bm G}) 
 -\frac{4\pi Z^{\tau}}{\Omega} \frac{\mathrm{exp}(-{\bm G}^2/(4a^2))}{{\bm G}^2} \right) \mathrm{e}^{i{\bm G}\cdot{(\bm r}-{\bm r_{\tau}})}\bigg] \\
&+ \lim_{{\bm G} \to {\bm 0}} \left( {\tilde{V}}^{\tau}_{\mathrm{loc},1} ({\bm G}) 
 -\frac{4\pi Z^{\tau}}{\Omega} \frac{\mathrm{exp}(-{\bm G}^2/(4a^2))-1}{{\bm G}^2} \right)\\
 &= \sum_{{\bm G}\ne {\bm 0}} \bigg[ \left( {\tilde{V}}^{\tau}_{\mathrm{loc},1} ({\bm G}) 
 -\frac{4\pi Z^{\tau}}{\Omega} \frac{\mathrm{exp}(-{\bm G}^2/(4a^2))}{{\bm G}^2} \right) \mathrm{e}^{i{\bm G}\cdot{(\bm r}-{\bm r_{\tau}})}\bigg] \\
&\ \ \ \ +  {\tilde{V}}^{\tau}_{\mathrm{loc},1} ({\bm G}={\bm 0}) + \frac{\pi Z^{\tau}}{a^2\Omega}.
\end{align}
Finally, the summation over the atom index $\tau$ is performed, and then we get the local part of the pseudopotential.

The non-local part of the pseudopotential with the Kleinman-Bylander form~\cite{KBpp} for an atom at ${\bm r_{\tau}}$ is given as follows:
\begin{equation}
\sum_{l,m,i_1,i_2} |\beta_{l,m,i_1}(|{\bm r}-{\bm r_{\tau}}|)Y_{lm}(\hat{{\bm e}}_{{\bm r} - {\bm r_{\tau}}})\rangle D_{i_1, i_2} \langle \beta_{l,m,i_2}(|{\bm r}-{\bm r_{\tau}}|)Y_{lm}(\hat{{\bm e}}_{{\bm r} - {\bm r_{\tau}}})|,
\end{equation}
where $\beta$, $Y_{lm}$, $D$, and $\hat{{\bm e}}_{{\bm v}}$ are the short-ranged (i.e., zero outside the cutoff radius) radial projector function, the spherical harmonics, the coefficient, and the unit vector along the vector ${\bm v}$, respectively.
To evaluate the non-local terms, using the Rayleigh expansion,
\begin{equation}
 \mathrm{e}^{i({\bm k}+{\bm G})\cdot ({\bm r}-{\bm r_{\tau}})}
= \sum_{l,m} 4\pi i^l j_l(|{\bm k}+{\bm G}| |{\bm r} - {\bm r_{\tau}}|) Y_{lm}^*(\hat{{\bm e}}_{{\bm k}+{\bm G}}) Y_{lm}(\hat{{\bm e}}_{{\bm r}-{\bm r_{\tau}}}),
\end{equation}
where $j_l$ is the spherical Bessel function, we get the formula
\begin{align}
&\langle \beta(|{\bm r}-{\bm r_{\tau}}|)Y_{lm}(\hat{{\bm e}}_{{\bm r} - {\bm r_{\tau}}}) |  {\bm k} + {\bm G} \rangle \\
&= \int \mathrm{d}{\bm r}\ \beta^*(|{\bm r}-{\bm r_{\tau}}|) Y^*_{lm}(\hat{{\bm e}}_{{\bm r} - {\bm r_{\tau}}})  \mathrm{e}^{i({\bm k}+{\bm G})\cdot {\bm r}}\\
&=  \mathrm{e}^{i({\bm k}+{\bm G})\cdot {\bm r_{\tau}}} Y_{lm}^*(\hat{{\bm e}}_{{\bm k}+{\bm G}}) 4\pi i^l \int \mathrm{d}\tilde{r}\ \tilde{r}^2\beta^*(\tilde{r}) j_l(|{\bm k}+{\bm G}|\tilde{r}).
\end{align}
Using this formula, we apply the non-local operator to the one-electron orbitals expanded with the plane-wave basis set.

\subsubsection{Two-body terms}

We classify the two-body Hartree (h) terms in TC Hamiltonian as follows:
\begin{description}
\item[2ah] $\sum_{q}^{\mathrm{occupied}} \langle *, q |V_{\mathrm{2a}} | j, q\rangle$
\item[2bh1] $\sum_{q}^{\mathrm{occupied}} \langle *, q | \nabla_1 u_{12} \cdot \nabla_1 | j, q\rangle$
\item[2bh2] $\sum_{q}^{\mathrm{occupied}} \langle *, q | \nabla_2 u_{12} \cdot \nabla_2 | j, q\rangle$,
\end{description}
where
\begin{align}
V_{\mathrm{2a}}(x_1, x_2) &= V_{\mathrm{2a}}^{\sigma_1, \sigma_2} (|{\bm r}_1 - {\bm r}_2|) \\
&= \frac{1}{|{\bm r}_1 - {\bm r}_2|} + \frac{1}{2}\bigg[
\nabla_1^2 u_{\sigma_1, \sigma_2}(|{\bm r}_1 - {\bm r}_2|)
+ \nabla_2^2 u_{\sigma_1, \sigma_2}(|{\bm r}_1 - {\bm r}_2|)  \notag\\
&\ \ \ - (\nabla_1 u_{\sigma_1, \sigma_2}(|{\bm r}_1 - {\bm r}_2|))^2
- (\nabla_2 u_{\sigma_1, \sigma_2}(|{\bm r}_1 - {\bm r}_2|))^2 \bigg].
\end{align}
In the same way, the two-body exchange (x) terms are defined as follows:
\begin{description}
\item[2ax] $-\sum_{q}^{\mathrm{occupied}} \langle *, q | V_{\mathrm{2a}} | q, j\rangle$
\item[2bx1] $-\sum_{q}^{\mathrm{occupied}} \langle *, q | \nabla_1 u_{12} \cdot \nabla_1 | q, j \rangle$
\item[2bx2] $-\sum_{q}^{\mathrm{occupied}} \langle *, q | \nabla_2 u_{12} \cdot \nabla_2 | q, j \rangle$.
\end{description}
Here we present how to calculate each term.
The {\bf 2ah} term is calculated as follows: (1) calculate the spin density,
\begin{equation}
n_{\sigma'}({\bm r}) = \sum_{{\bm q},\mu} \chi_q^*({\bm r}) \phi_q ({\bm r}) f_{{\bm q},\mu},\ \ \ (q =  (\sigma', {\bm q},\mu))
\end{equation}
(2) use the convolution formula, Eq.~(\ref{eq:convolution}), as
\begin{align}
\sum_{q}^{\mathrm{occupied}} \langle *, q |V_{\mathrm{2a}} | *, q\rangle
&= \sum_{\sigma'} \int \mathrm{d}{\bm r}_2\  V_{\mathrm{2a}}^{\sigma, \sigma'}(|{\bm r}_1 - {\bm r}_2|) n_{\sigma'}({\bm r}_2) \\
&= \sum_{\sigma'} \sum_{{\bm G}\ne {\bm 0}} \tilde{V}_{\mathrm{2a}}^{\sigma, \sigma'}({\bm G}) \tilde{n}_{\sigma'}({\bm G})
\mathrm{e}^{i{\bm G}\cdot {\bm x}_1}, \label{eq:2ah}
\end{align}
and (3) multiply $\phi_j({\bm x}_1)$ with Eq.~(\ref{eq:2ah}). In our implementation, the spin density is calculated and saved before evaluation of several terms in the TC Hamiltonian.
A pseudocode for calculating the {\bf 2ah} term is shown in Algorithm~\ref{algo:2ah}.
\begin{figure}[h]
\begin{algorithm}[H]
  \caption{Calculate {\bf 2ah}: $g_j({\bm r}_1) = \sum_{q}^{\mathrm{occupied}} \langle *, q | V_{\mathrm{2a}} | j, q\rangle \times \mathrm{exp}(-i{\bm k}\cdot {\bm r}_1)$}
  \label{algo:2ah}
   \begin{algorithmic}[1]
    	\For {$j = 1$ to $N$ (MPI parallelized)}
		\State $\tilde{h}^{\sigma}_1({\bm G}) \gets \sum_{\sigma'} \tilde{V}_{\mathrm{2a}}^{\sigma, \sigma'}({\bm G})\tilde{n}_{\sigma'}({\bm G})$
		\State $h^{\sigma}_1({\bm r}_1) \gets$ {\sc FT}$^{-1}$[$\tilde{h}^{\sigma}_1({\bm G})$]
		\State $g_j({\bm r}_1) \gets h^{\sigma}_1({\bm r}_1)\phi_{\mathrm{periodic},j}({\bm r}_1)$
	\EndFor
   \end{algorithmic}
\end{algorithm}
\end{figure} \noindent
Here we omit the ${\bm G} = {\bm 0}$ component in Eq.~(\ref{eq:2ah}) because that for the Coulomb potential is already considered in the Ewald summation and we assume $u({\bm G} = {\bm 0})=0$. For the Hartree terms, this treatment is valid because a constant shift of the Jastrow function $u$ (i.e., a constant multiplication with the Jastrow factor $F$) does not change TC Hamiltonian $\mathcal{H}_{\mathrm{TC}} = F^{-1}\mathcal{H}F$. A bit different situation for the exchange terms shall be described later.
The {\bf 2bh1} and {\bf 2bh2} terms are calculated in the same way. 
Note that a derivative of the one-electron orbital is easily calculated in reciprocal space.
The remaining problem is to calculate the Fourier transform of the effective interactions.
For the Jastrow factor shown in Eq.~(\ref{eq:Jastrow}), its Fourier transform is calculated as follows:
\begin{equation}
\tilde{u}(G) = 4\pi A \left( \frac{1}{G^2} - \frac{1}{G^2+1/C^2} \right) = 4\pi A \frac{1/C^2}{G^2(G^2+1/C^2)},
\end{equation}
which immediately yields
\begin{equation}
\widetilde{\nabla^2 u} (G) = -G^2 \tilde{u}(G)  = -4\pi A \frac{1/C^2}{G^2+1/C^2}.
\end{equation}
The Fourier transform of $(\nabla u)^2$ is a bit complicated. It is given as
\begin{equation}
\widetilde{(\nabla u)^2} (G)= \frac{4\pi A^2}{Cg} \bigg[ -\frac{\pi}{4}g^2 - \left( 1+\frac{g^2}{2}\right) \arctan{\frac{g}{2}} + (1+g^2)\arctan{g}\bigg]\ \ \ (g=CG) \label{eq:fourier_transform_nabla2u},
\end{equation}
the derivation of which is shown in Appendix B.

The exchange terms are calculated in the following way.
For calculating the {\bf 2ax} term, we use the convolution formula, Eq.~(\ref{eq:convolution}), as follows:
\begin{align}
&\int \mathrm{d}{\bm r}_2\  V_{\mathrm{2a}}^{\sigma, \sigma}(|{\bm r}_1 - {\bm r}_2|)  \chi^*_q({\bm r}_2)\phi_j({\bm r}_2)\\
&=\frac{1}{N_{\bm k}} \int \mathrm{d}{\bm r}_2\  V_{\mathrm{2a}}^{\sigma, \sigma}(|{\bm r}_1 - {\bm r}_2|) \chi^*_{\mathrm{periodic},q}({\bm r}_2) \phi_{\mathrm{periodic},j}({\bm r}_2) \mathrm{e}^{i({\bm k}- {\bm q})\cdot {\bm r}_2}\\
&= \frac{1}{N_{\bm k}} \sum_{{\bm G}} \tilde{V}_{\mathrm{2a}}^{\sigma, \sigma}({\bm k} - {\bm q} + {\bm G}) \mathrm{FT}[\chi_{\mathrm{periodic},q}^* \phi_{\mathrm{periodic},j}]({\bm G}) \mathrm{e}^{i({\bm k} - {\bm q} + {\bm G})\cdot{\bm r}_1},
\end{align}
for the states $q =  (\sigma'=\sigma, {\bm q},\mu)$ and $j =  (\sigma, {\bm k},\mu_j)$. Note that the spin indices for $q$ and $j$ should be the same in the exchange terms.
A pseudocode for calculating the {\bf 2ax} term is shown in Algorithm~\ref{algo:2ax}.
\begin{figure}[h]
\begin{algorithm}[H]
  \caption{Calculate {\bf 2ax}: $g_j({\bm r}_1) = -\sum_{q}^{\mathrm{occupied}} \langle *, q | V_{\mathrm{2a}} | q, j\rangle  \times \mathrm{exp}(-i{\bm k}\cdot {\bm r}_1)$ (except the divergence correction terms)}
  \label{algo:2ax}
   \begin{algorithmic}[1]
    	\For {$j = 1$ to $N$ (MPI parallelized)}
		\State $g_j({\bm r}_1)  \gets 0$
		\For {$q = 1$ to $N$}	
			\If{$\sigma \neq \sigma'$}
				\State continue
			\EndIf
			\State $h^{\sigma}_1({\bm r}_2) \gets  \chi^*_{\mathrm{periodic},q}({\bm r}_2) \phi_{\mathrm{periodic},j}({\bm r}_2)$
			\State $\tilde{h}^{\sigma}_1({\bm G}) \gets$ {\sc FT}[$h^{\sigma}_1({\bm r}_2)$]
			\State $\tilde{h}^{\sigma}_2({\bm G}) \gets \tilde{V}_{\mathrm{2a}}^{\sigma, \sigma}({\bm k} - {\bm q} + {\bm G})\tilde{h}^{\sigma}_1({\bm G})$
			\State $h^{\sigma}_2({\bm r}_1) \gets$  {\sc FT}$^{-1}$[$\tilde{h}^{\sigma}_2({\bm G})$]
			\State $g_j({\bm r}_1) \gets g_j({\bm r}_1) - h^{\sigma}_2({\bm r}_1)\phi_{\mathrm{periodic},q}({\bm r}_1) f_q/N_{\bm k}$
		\EndFor
	\EndFor
   \end{algorithmic}
\end{algorithm}
\end{figure} \noindent
Here, the function $h^{\sigma}_2({\bm r}_1)$ in the pseudocode is defined as
\begin{equation}
h^{\sigma}_2({\bm r}_1) = \sum_{{\bm G}} \tilde{V}_{\mathrm{2a}}^{\sigma, \sigma}({\bm k} - {\bm q} + {\bm G})\mathrm{FT}[\chi_{\mathrm{periodic},q}^* \phi_{\mathrm{periodic},j}]({\bm G})  \mathrm{e}^{i{\bm G}\cdot{\bm r}_1},
\end{equation}
and $g_j({\bm r}_1)$ in Algorithm~\ref{algo:2ax} is $-\sum_{q}^{\mathrm{occupied}} \langle *, q | V_{\mathrm{2a}} | q, j\rangle$ multiplied by $\mathrm{e}^{-i{\bm k}\cdot{\bm r}_1}$.
This algorithm is the same as that for calculating the exchange term in the HF method using the plane-wave basis set (see, e.g., Ref.~\cite{div_corr2}).
As shown in the pseudocode, MPI parallelization is performed for the index $j$.

We note that the divergence correction is required for the term {\bf 2ax}. 
For the SCF calculation, the correction terms can be calculated using Eq.~(\ref{eq:correction2}) as follows:
\begin{align}
- \bigg[ \frac{\Omega}{\sqrt{\pi \alpha}} -
\frac{4\pi}{N_{\bm k}} \bigg( -\alpha + \sum_{\substack{{\bm q}, {\bm G}\\ ({\bm k} - {\bm q} + {\bm G}\ne {\bm 0})}}  A_{\mathrm{aux}}({\bm k} - {\bm q}+{\bm G}) \bigg)
 \bigg] \notag \\
 \times \sum_{\mu} \bigg[ \int \mathrm{d}{\bm r}_2\ \chi^*_{\mathrm{periodic},q_k}({\bm r}_2) \phi_{\mathrm{periodic},j}({\bm r}_2) \bigg]
 \phi_{\mathrm{periodic},q_k}({\bm r}_1) f_{q_k}, \label{eq:divcorr_2ax_scf}
 \end{align}
where $q_k =  (\sigma, {\bm k},\mu)$ belongs to the same ${\bm k}$-point with that for $j = (\sigma, {\bm k},\mu_j)$. This is because the coefficient $f({\bm 0})$ in Eq.~(\ref{eq:correction2}) should be calculated at the diverging point ${\bm q}={\bm k}$ for the ${\bm q}$-integration.
A factor of $4\pi$ is multiplied with the whole terms because the Coulomb potential exhibits a $4\pi/q^2$ divergence in reciprocal space, while Eq.~(\ref{eq:correction2}) represents the correction term for the $1/q^2$ divergence.
Here, we consider the divergence correction only for the Coulomb potential because both $\nabla^2 u$ and $(\nabla u)^2$ does not exhibit divergence at the origin in reciprocal space. 

The divergence correction for the band-structure calculation (or for zero-weight ${\bm k}$-points even in SCF calculation) is a bit different. In the band-structure calculation, while ${\bm q}$ belongs to the SCF ${\bm k}$-mesh, ${\bm k}$ is defined along the band ${\bm k}$-path and so is not necessarily included in the SCF ${\bm k}$-mesh.
In this case, the summation over ${\bm q}, {\bm G}$ in the divergence correction does not necessarily include the diverging point ${\bm k}-{\bm q}+{\bm G}={\bm 0}$.
For band ${\bm k}$-points ${\bm k}$ included in the SCF ${\bm k}$-mesh, the divergence correction term is the same as that for the SCF calculation, Eq.~(\ref{eq:divcorr_2ax_scf}). For band ${\bm k}$-points ${\bm k}$ not included in the SCF ${\bm k}$-mesh, we should instead use Eq.~(\ref{eq:correction}), and then consider the following correction terms,
\begin{align}
- \bigg[ \frac{\Omega}{\sqrt{\pi \alpha}} -
\frac{4\pi}{N_{\bm k}} \sum_{{\bm q}, {\bm G}}  A_{\mathrm{aux}}({\bm k} - {\bm q}+{\bm G}) \bigg] \notag \\
\times \sum_{\mu} \bigg[ \int \mathrm{d}{\bm r}_2\ \chi^*_{\mathrm{periodic},q_k}({\bm r}_2) \phi_{\mathrm{periodic},j}({\bm r}_2) \bigg]
 \phi_{\mathrm{periodic},q_k}({\bm r}_1) f_{q_k}. \label{eq:divcorr_2ax_band}
 \end{align}
 
The {\bf 2bx1} and {\bf 2bx2} terms are calculated in the same way as that for the {\bf 2ax} term, except the divergence correction. 
For the {\bf 2bx1} term, $-\sum_{q}^{\mathrm{occupied}} \langle *, q | \nabla_1 u_{12} \cdot \nabla_1 | q, j \rangle$, the correction terms are calculated using Eq.~(\ref{eq:correction_nabla}) as follows:
\begin{align}
\bigg[ - \frac{4\pi A_{\sigma,\sigma}}{N_{\bm k}}\sum_{\substack{{\bm q}, {\bm G}\\ ({\bm q} + {\bm G}\ne {\bm 0})}}  ({\bm q}+{\bm G})A_{\mathrm{aux}}({\bm q}+{\bm G}) 
 \bigg]  \notag \\
\cdot \sum_{\mu} \bigg[ \sum_{\bm G'} ({\bm k}+{\bm G'}) \mathrm{FT} [ \phi_{\mathrm{periodic},q_k} ] ({\bm G'}) \mathrm{e}^{i{\bm G'}\cdot{\bm r}_1} \bigg] \notag \\
 \times  \bigg[ \int \mathrm{d}{\bm r}_2\ \chi^*_{\mathrm{periodic},q_k}({\bm r}_2) \phi_{\mathrm{periodic},j}({\bm r}_2) \bigg]  f_{q_k}. \label{eq:divcorr_2bx1}
\end{align}
A factor of $4\pi A_{\sigma,\sigma}$ is multiplied with the whole terms because the Jastrow function $u$ exhibits a $4\pi A_{\sigma,\sigma}/q^2$ divergence in reciprocal space, while Eq.~(\ref{eq:correction_nabla}) represents the correction term for the $1/q^2$ divergence.
In addition, a factor of $-1$ originating from the sign of the exchange term ($-\sum_{q}^{\mathrm{occupied}} \dots$) and another $-1$ originating from Fourier transform of two $\nabla$ in $\nabla u\cdot \nabla$: $i^2 = -1$, are multiplied.
The correction terms for the {\bf 2bx2} term, $-\sum_{q}^{\mathrm{occupied}} \langle *, q | \nabla_2 u_{12} \cdot \nabla_2 | q, j \rangle$ are similarly calculated as follows:
\begin{align}
-\bigg[ - \frac{4\pi A_{\sigma,\sigma}}{N_{\bm k}}\sum_{\substack{{\bm q}, {\bm G}\\ ({\bm q} + {\bm G}\ne {\bm 0})}}  ({\bm q}+{\bm G})A_{\mathrm{aux}}({\bm q}+{\bm G}) 
 \bigg] \notag \\
 \cdot \sum_{\mu}\bigg[ \int \mathrm{d}{\bm r}_2\ \chi^*_{\mathrm{periodic},q_k}({\bm r}_2) \sum_{\bm G'} ({\bm k}+{\bm G'}) {\mathrm FT} [ \phi_{\mathrm{periodic},j} ] ({\bm G'}) \mathrm{e}^{i{\bm G'}\cdot{\bm r}_1} \bigg]\notag \\
 \times  \phi_{\mathrm{periodic},q_k}({\bm r}_1) f_{q_k}, \label{eq:divcorr_2bx2}
\end{align}
where an additional factor of $-1$ is multiplied because of $\nabla_2 u_{21} = -\nabla_1 u_{12}$.

We note two things here. One is that these divergence corrections for $\nabla u \cdot \nabla$ are not required when ${\bm k}$ is included in the SCF ${\bm k}$-mesh, because the correction term shown in Eq.~(\ref{eq:correction_nabla}) becomes zero due to the symmetry of the auxiliary function $A_{\mathrm{aux}}({\bm G})$.
We here assume that the SCF ${\bm k}$-mesh is uniform.
The other thing is that the divergence correction is considered only for the exchange terms that include ${\bm q}, {\bm G}$-summation, but not so for the Hartree terms because they only have a discrete ${\bm G}$-summation. 
This is because the former summation gets closer to the integration over a continuous variable in the $N_{\bm k}\to \infty$ limit, while the latter summation does not.
This is also true for three-body terms as we shall see next.

\subsubsection{Three-body terms}

We classify the three-body terms including $\nabla_1 u_{12} \cdot \nabla_1 u_{13}$, called {\bf 3a*} terms in this paper, as follows:
\begin{description}
\item[3a1] $-\dfrac{1}{2}\sum_{q_1, q_2}^{\mathrm{occupied}} \langle *, q_1, q_2 | \nabla_1 u_{12} \cdot \nabla_1 u_{13} | j, q_1, q_2 \rangle$
\item[3a2] $\dfrac{1}{2}\sum_{q_1, q_2}^{\mathrm{occupied}} \langle *, q_1, q_2 | \nabla_1 u_{12} \cdot \nabla_1 u_{13} | j, q_2, q_1 \rangle$
\item[3a3] $\dfrac{1}{2}\sum_{q_1, q_2}^{\mathrm{occupied}} \langle *, q_1, q_2 | \nabla_1 u_{12} \cdot \nabla_1 u_{13} | q_1, j, q_2 \rangle$
\item[3a4] $-\dfrac{1}{2}\sum_{q_1, q_2}^{\mathrm{occupied}} \langle *, q_1, q_2 | \nabla_1 u_{12} \cdot \nabla_1 u_{13} | q_1, q_2, j \rangle$
\item[3a5] $-\dfrac{1}{2}\sum_{q_1, q_2}^{\mathrm{occupied}} \langle *, q_1, q_2 | \nabla_1 u_{12} \cdot \nabla_1 u_{13} | q_2, j, q_1 \rangle$
\item[3a6] $\dfrac{1}{2}\sum_{q_1, q_2}^{\mathrm{occupied}} \langle *, q_1, q_2 | \nabla_1 u_{12} \cdot \nabla_1 u_{13} | q_2, q_1, j\rangle$.
\end{description}
Here, {\bf 3a3} and {\bf 3a6} are equivalent by the simultaneous permutations of $q_1 \leftrightarrow q_2$ and ${\bm x}_2 \leftrightarrow {\bm x}_3$.
Also, {\bf 3a4} and {\bf 3a5} are equivalent by the same operation.
The three-body terms including $\nabla_2 u_{21} \cdot \nabla_2 u_{23}$, called {\bf 3b*} terms, are classified as
\begin{description}
\item[3b1] $-\dfrac{1}{2}\sum_{q_1, q_2}^{\mathrm{occupied}} \langle *, q_1, q_2 | \nabla_2 u_{21} \cdot \nabla_2 u_{23} | j, q_1, q_2 \rangle$
\item[3b2] $\dfrac{1}{2}\sum_{q_1, q_2}^{\mathrm{occupied}} \langle *, q_1, q_2 | \nabla_2 u_{21} \cdot \nabla_2 u_{23} | j, q_2, q_1 \rangle$
\item[3b3] $\dfrac{1}{2}\sum_{q_1, q_2}^{\mathrm{occupied}} \langle *, q_1, q_2 | \nabla_2 u_{21} \cdot \nabla_2 u_{23} | q_1, j, q_2 \rangle$
\item[3b4] $-\dfrac{1}{2}\sum_{q_1, q_2}^{\mathrm{occupied}} \langle *, q_1, q_2 | \nabla_2 u_{21} \cdot \nabla_2 u_{23}  | q_1, q_2, j \rangle$
\item[3b5] $-\dfrac{1}{2}\sum_{q_1, q_2}^{\mathrm{occupied}} \langle *, q_1, q_2 |\nabla_2 u_{21} \cdot \nabla_2 u_{23} | q_2, j, q_1 \rangle$
\item[3b6] $\dfrac{1}{2}\sum_{q_1, q_2}^{\mathrm{occupied}} \langle *, q_1, q_2 | \nabla_2 u_{21} \cdot \nabla_2 u_{23} | q_2, q_1, j\rangle$.
\end{description}
The remaining three-body terms including $\nabla_3 u_{31} \cdot \nabla_3 u_{32}$ are equivalent to these {\bf 3b*} terms, which is shown by the permutation of ${\bm x}_2 \leftrightarrow {\bm x}_3$.
Therefore, we should consider ten kinds of three-body terms in total: {\bf 3a[1-4]} and {\bf 3b[1-6]}.
We shall present how to calculate each term. For efficient computation, we used the algorithm we developed for solid-state calculation~\cite{TCaccel}, by which the computational time of the (BI)TC method involving the three-body terms is the same order as that for the HF method involving up-to the two-body terms.

A pseudocode for calculating the {\bf 3a1} term is shown in Algorithm~\ref{algo:3a1}.
\begin{figure}[h]
\begin{algorithm}[H]
  \caption{Calculate {\bf 3a1}: $g_j({\bm r}_1) = -\dfrac{1}{2}\sum_{q_1, q_2}^{\mathrm{occupied}} \langle *, q_1, q_2 | \nabla_1 u_{12} \cdot \nabla_1 u_{13} | j, q_1, q_2 \rangle \times \mathrm{exp}(-i{\bm k}\cdot {\bm r}_1) $}
  \label{algo:3a1}
   \begin{algorithmic}[1]
    	\For {$j = 1$ to $N$ (MPI parallelized)}
		\State $\tilde{{\bm h}}^{\sigma}_1({\bm G}) \gets \sum_{\sigma'} i{\bm G}\tilde{u}^{\sigma, \sigma'}({\bm G}) \tilde{n}_{\sigma'}({\bm G})$
		\State ${\bm h}^{\sigma}_1({\bm r}_1) \gets$ {\sc FT}$^{-1}$[$\tilde{\bm h}^{\sigma}_1({\bm G})$]
		\State $g_j({\bm r}_1) \gets -|{\bm h}^{\sigma}_1({\bm r}_1)|^2 \phi_{\mathrm{periodic},j}({\bm r}_1)/2$
	\EndFor
   \end{algorithmic}
\end{algorithm}
\end{figure} \noindent
This term does not need the divergence correction by the same reason as the two-body Hartree terms.

A pseudocode for calculating the {\bf 3a2} term is shown in Algorithm~\ref{algo:3a2}.
\begin{figure}[h]
\begin{algorithm}[H]
  \caption{Calculate {\bf 3a2}: $g_j({\bm r}_1) = \dfrac{1}{2}\sum_{q_1, q_2}^{\mathrm{occupied}} \langle *, q_1, q_2 | \nabla_1 u_{12} \cdot \nabla_1 u_{13} | j, q_2, q_1 \rangle  \times \mathrm{exp}(-i{\bm k}\cdot {\bm r}_1)$ (except the divergence correction terms)}
  \label{algo:3a2}
   \begin{algorithmic}[1]
	\State $h^{\sigma}_6({\bm r}_1)  \gets 0$
	\State $h^{\sigma, \sigma_2}_5({\bm r}_1; q_2)$ ($q_2 = 1$ to $N$) $\gets 0$
	\For {$q_{2;0} = 1$ to $N_{\mathrm{irred}}$}	
		\For {$q_1 = 1$ to $N$ (MPI parallelized)}
			\If{$\sigma_1\neq \sigma_2$}
				\State continue
			\EndIf
			\State $h_1^{\sigma_2}({\bm r}_3) \gets  \chi^*_{\mathrm{periodic},q_{2;0}}({\bm r}_3) \phi_{\mathrm{periodic},q_1}({\bm r}_3)$
			\State $\tilde{h}^{\sigma_2}_1({\bm G}) \gets$ {\sc FT}[$h^{\sigma_2}_1({\bm r}_3)$]
			\State $\tilde{\bm h}^{\sigma,\sigma_2}_2({\bm G}) \gets i({\bm q}_1 - {\bm q}_{2;0} + {\bm G}) \tilde{u}^{\sigma, \sigma_2}({\bm q}_1 - {\bm q}_{2;0} + {\bm G})\tilde{h}^{\sigma_2}_1({\bm G})$
			\State ${\bm h}^{\sigma,\sigma_2}_2({\bm r}_1) \gets$  {\sc FT}$^{-1}$[$\tilde{\bm h}^{\sigma,\sigma_2}_2({\bm G})$]
			\State $h_3^{\sigma_2}({\bm r}_2) \gets  \chi^*_{\mathrm{periodic},q_1}({\bm r}_2) \phi_{\mathrm{periodic},q_{2;0}}({\bm r}_2)$
			\State $\tilde{h}^{\sigma_2}_3({\bm G}) \gets$ {\sc FT}[$h^{\sigma_2}_3({\bm r}_2)$]
			\State $\tilde{\bm h}^{\sigma,\sigma_2}_4({\bm G}) \gets i({\bm q}_{2;0} - {\bm q}_1 + {\bm G}) \tilde{u}^{\sigma, \sigma_2}({\bm q}_{2;0} - {\bm q}_1 + {\bm G})\tilde{h}^{\sigma_2}_3({\bm G})$
			\State ${\bm h}^{\sigma,\sigma_2}_4({\bm r}_1) \gets$  {\sc FT}$^{-1}$[$\tilde{\bm h}^{\sigma,\sigma_2}_4({\bm G})$]
			\State $h^{\sigma, \sigma_2}_5({\bm r}_1; q_{2;0})  \gets h^{\sigma, \sigma_2}_5({\bm r}_1; q_{2;0}) + {\bm h}^{\sigma,\sigma_2}_2({\bm r}_1)\cdot{\bm h}^{\sigma,\sigma_2}_4({\bm r}_1) f_{q_1} /N_{\bm k}$
		\EndFor
		\State MPI Allreduce for $q_1$-parallelization
		\For {symmetry operation ($q_{2;0}\to q_2$)}
			\State make $h^{\sigma, \sigma_2}_5({\bm r}_1; q_2)$ from $h^{\sigma, \sigma_2}_5({\bm r}_1; q_{2;0})$ by symmetry operation
			\State $h^{\sigma}_6({\bm r}_1)  \gets h^{\sigma}_6({\bm r}_1) + \sum_{\sigma_2} h^{\sigma, \sigma_2}_5({\bm r}_1;q_2) f_{q_2} /N_{\bm k}$
		\EndFor
	\EndFor
    	\For {$j = 1$ to $N$ (MPI parallelized)}
		\State $g_j({\bm r}_1) \gets h^{\sigma}_6({\bm r}_1)  \phi_{\mathrm{periodic},j}({\bm r}_1)/2$
	\EndFor
   \end{algorithmic}
\end{algorithm}
\end{figure} \noindent
There is at most a doubly nested loop of ($q_1, q_{2;0}$) even though we handle three orbital indices $(j, q_1, q_2)$ for the three-body terms, which is an important advantage of the (BI)TC method~\cite{TCaccel}.
To reduce computational cost, we consider $q_{2;0}=(\sigma_2, {\bm q}_{2;0}, \mu_2)$, where ${\bm q}_{2;0}$ is the irreducible ${\bm k}$-point corresponding to ${\bm q}_2$: there exists a symmorphic symmetry operator $S$ satisfying ${\bm q}_2 = S^{\dag}{\bm q}_{2;0}$ (see Sec.~\ref{sec:comput_one_orb}).
The number of the states $q_{2;0}$ with irreducible ${\bm k}$-points is $N_{\mathrm{irred}}$, which is smaller than $N$.
Considering the symmetry, Eq.~(\ref{eq:symmetry_operation}),
\begin{equation}
h^{\sigma, \sigma_2}_5({\bm r}_1; q_2) = N_{\bm k}\sum_{q_1}^{\mathrm{occupied}} \langle *, q_1, q_2 | \nabla_1 u_{12} \cdot \nabla_1 u_{13} | *, q_2, q_1 \rangle
\end{equation}
satisfies $h^{\sigma, \sigma_2}_5({\bm r}_1; q_2) = h^{\sigma, \sigma_2}_5(S({\bm r}+{\bm t}); q_{2;0})$, similarly to $\phi_{q_2}({\bm r}) \propto \phi_{q_{2;0}}(S({\bm r}+{\bm t}))$.
Thus,
\begin{align}
\tilde{h}^{\sigma, \sigma_2}_5(S^{\dag} {\bm G}; q_{2})&= \frac{1}{\Omega}\int_{\Omega} \mathrm{d}{\bm r}\ h^{\sigma, \sigma_2}_5({\bm r}; q_{2})\mathrm{e}^{-i(S^{\dag}{\bm G})\cdot {\bm r}} \\
&= \frac{1}{\Omega}\int_{\Omega} \mathrm{d}{\bm r}\ h^{\sigma, \sigma_2}_5(S({\bm r}+{\bm t}); q_{2;0})\mathrm{e}^{-i{\bm G}\cdot S({\bm r}+{\bm t})} \mathrm{e}^{i(S^{\dag}{\bm G})\cdot {\bm t}}  \\
&= \tilde{h}^{\sigma, \sigma_2}_5({\bm G}; q_{2;0}) \mathrm{e}^{i(S^{\dag} {\bm G})\cdot {\bm t}}
\end{align}
holds, which is used to obtain $\tilde{h}^{\sigma, \sigma_2}_5(S^{\dag} {\bm G}; q_{2})$ from $\tilde{h}^{\sigma, \sigma_2}_5({\bm G}; q_{2;0})$.
When the time-reversal symmetry is applied to the state,
\begin{equation}
\tilde{h}^{\sigma, \sigma_2}_5(-S^{\dag} {\bm G}; q_{2})
= (\tilde{h}^{\sigma, \sigma_2}_5({\bm G}; q_{2;0}))^* \mathrm{e}^{-i(S^{\dag} {\bm G})\cdot {\bm t}}
\end{equation}
is used instead. 
Here we do not directly use $h^{\sigma, \sigma_2}_5({\bm r}_1; q_2) = h^{\sigma, \sigma_2}_5(S({\bm r}+{\bm t}); q_{2;0})$ to get $h^{\sigma, \sigma_2}_5({\bm r}_1; q_2)$ in our computational code, because $S({\bm r}+{\bm t})$ is not necessarily included in the real-space grid.

The divergence correction for $h^{\sigma, \sigma_2}_5({\bm r}_1; q_{2;0})$ in {\bf 3a2} is considered as follows.
$h^{\sigma, \sigma_2}_5({\bm r}_1; q_{2;0})$ can be written as
\begin{align}
&h^{\sigma, \sigma_2}_5({\bm r}_1; q_{2;0}) = \notag \\
&-\frac{1}{N_{\bm k}}\sum_{q_1, {\bm G}, {\bm G}'} ({\bm q}_1 - {\bm q}_{2;0} + {\bm G}) \tilde{u}^{\sigma, \sigma_2}({\bm q}_1 - {\bm q}_{2;0} + {\bm G})
\mathrm{FT} [ \chi^*_{\mathrm{periodic},q_{2;0}} \phi_{\mathrm{periodic},q_1} ] ({\bm G}) \notag \\
&\cdot ({\bm q}_{2;0} - {\bm q}_{1} + {\bm G}') \tilde{u}^{\sigma, \sigma_2}({\bm q}_{2;0} - {\bm q}_{1} + {\bm G}')
\mathrm{FT} [ \chi^*_{\mathrm{periodic},q_1} \phi_{\mathrm{periodic},q_{2;0}} ] ({\bm G}') \notag \\
&\times \mathrm{e}^{i({\bm G}+{\bm G}')\cdot {\bm r}_1} f_{q_1} \delta_{\sigma_1, \sigma_2}, \label{eq:3a2_h5_div} 
\end{align}
where the negative sign comes from the square of the Fourier transform of $\nabla$.
Here we should consider the following two types of the correction terms.
One is that for ${\bm G}={\bm G}'={\bm 0}$ contribution, where two Jastrow functions simultaneously diverge.
By considering the divergence at ${\bm q}_1 = {\bm q}_{2;0}$, we obtain the following correction terms,
\begin{align}
4\pi A_{\sigma,\sigma_2} \bigg[ \frac{A_{\sigma,\sigma_2}\Omega}{\sqrt{\pi \alpha}} -
\frac{4\pi A_{\sigma,\sigma_2}}{N_{\bm k}} \bigg( -\alpha + \sum_{\substack{{\bm q}, {\bm G}\\ ({\bm k} - {\bm q} + {\bm G}\ne {\bm 0})}}  A_{\mathrm{aux}}({\bm k} - {\bm q}+{\bm G}) \bigg)
+ \frac{\tilde{u}^{\sigma,\sigma_2}_{\mathrm{short}}({\bm 0})}{N_{\bm k}}
 \bigg] f_{q_{2;0}},   \label{eq:divcor_3a2_first}
 \end{align}
where $\tilde{u}^{\sigma,\sigma_2}_{\mathrm{short}}$ is a short-range component of $\tilde{u}$ defined as
\begin{equation}
\tilde{u}^{\sigma,\sigma_2}_{\mathrm{short}}({\bm 0})=\lim_{{\bm q}\to {\bm 0}} \bigg[ \tilde{u}^{\sigma,\sigma_2}({\bm q}) - \frac{4\pi A_{\sigma,\sigma_2}}{q^2} \bigg] ,\label{eq:u_short}
\end{equation}
and we use
\begin{equation}
\lim_{{\bm q}_1- {\bm q}_{2;0},{\bm G},{\bm G}' \to {\bm 0}}  ({\bm q}_1 - {\bm q}_{2;0} + {\bm G})\cdot ({\bm q}_{2;0} - {\bm q}_{1} + {\bm G}')  \tilde{u}^{\sigma, \sigma_2}({\bm q}_1 - {\bm q}_{2;0} + {\bm G}) = -4\pi A_{\sigma,\sigma_2},
\end{equation}
\begin{equation}
\lim_{{\bm q}_1 \to {\bm q}_{2;0}} \mathrm{FT} [ \chi^*_{\mathrm{periodic},q_{2;0}} \phi_{\mathrm{periodic},q_1} ] ({\bm 0}) = \delta_{\mu_1, \mu_2}\ \ (\mathrm{for}\ \sigma_1 = \sigma_2),\label{eq:3a2orthonorm}
\end{equation}
and Eq.~(\ref{eq:correction2}).
The other correction terms come from ${\bm G}\neq {\bm G}'$ contribution, where one of the two Jastrow functions in Eq.~(\ref{eq:3a2_h5_div}) diverges. By considering the divergence at ${\bm q}_1 = {\bm q}_{2;0}$, we obtain the following correction terms, 
\begin{equation}
2\times \frac{4\pi A_{\sigma,\sigma_2}}{N_{\bm k}} \sum_{{\bm G}\neq {\bm 0}} \tilde{u}^{\sigma, \sigma_2} ({\bm G}) \mathrm{FT} [ \chi^*_{\mathrm{periodic},q_{2;0}} \phi_{\mathrm{periodic},q_{2;0}} ] ({\bm G})
 \mathrm{e}^{i {\bm G}\cdot {\bm r}_1} f_{q_{2;0}},  \label{eq:divcor_3a2_second}
\end{equation}
where the factor of two comes from two contributions for the ${\bm G}\neq {\bm G}'$ divergence correction: ${\bm G}=0$ while ${\bm G}'\neq 0$ and vice versa.
For deriving this correction term, we decompose
\begin{equation}
({\bm q}_1 - {\bm q}_{2;0} + {\bm G}) \cdot ({\bm q}_{2;0} - {\bm q}_{1} + {\bm 0}) = -({\bm q}_1 - {\bm q}_{2;0})^2 + {\bm G}\cdot ({\bm q}_{2;0} - {\bm q}_{1}) \label{eq:divcor_3a2_decompose}
\end{equation}
for the ${\bm G}'={\bm 0}$ contribution, and consider the divergence correction for the first term in the right-hand side of Eq.~(\ref{eq:divcor_3a2_decompose}): $-({\bm q}_1 - {\bm q}_{2;0})^2 \tilde{u}^{\sigma, \sigma_2}({\bm q}_1 - {\bm q}_{2;0}) \to -4\pi A_{\sigma,\sigma_2}$.
Note that the second term in the right-hand side of Eq.~(\ref{eq:divcor_3a2_decompose}) does not require the divergence correction.
The reason for it is the same as the treatment of $\nabla u$ in the {\bf 2bx1} and {\bf 2bx2} terms. Namely, for the divergence correction of $({\bm q}_1 - {\bm q}_{2;0}) \tilde{u}^{\sigma, \sigma_2}({\bm q}_1 - {\bm q}_{2;0})$, Eq.~(\ref{eq:correction_nabla}) becomes zero due to the symmetry of the auxiliary function $A_{\mathrm{aux}}({\bm G})$.

We note that Eq.~(\ref{eq:3a2orthonorm}) breaks when one uses the density-matrix mixing (see, Sec.~\ref{sec:diagonalization}), because the electron orbitals in two different SCF loops (i.e., `new' and `old' orbitals) are not orthogonalized. In that case, Eqs.~(\ref{eq:divcor_3a2_first}) and (\ref{eq:divcor_3a2_second}) should be replaced with
\begin{align}
&4\pi A_{\sigma,\sigma_2} \bigg[ \frac{A_{\sigma,\sigma_2}\Omega}{\sqrt{\pi \alpha}} -
\frac{4\pi A_{\sigma,\sigma_2}}{N_{\bm k}} \bigg( -\alpha + \sum_{\substack{{\bm q}, {\bm G}\\ ({\bm k} - {\bm q} + {\bm G}\ne {\bm 0})}}  A_{\mathrm{aux}}({\bm k} - {\bm q}+{\bm G}) \bigg)
+ \frac{\tilde{u}^{\sigma,\sigma_2}_{\mathrm{short}}({\bm 0})}{N_{\bm k}}
 \bigg]\notag\\
 &\times  \sum_{\mu_1}
 \mathrm{FT} [ \chi^*_{\mathrm{periodic},q_{2;0}} \phi_{\mathrm{periodic},{\tilde q}_1} ] ({\bm 0})
\mathrm{FT} [ \chi^*_{\mathrm{periodic},{\tilde q}_1} \phi_{\mathrm{periodic},q_{2;0}} ] ({\bm 0})  
 f_{{\tilde q}_1}, \label{eq:divcor_3a2_first_density_matrix_mixing}
 \end{align}
where ${\tilde q}_1 = \{ {\bm q}_{2;0}, \mu_1, \sigma_2\}$, and
\begin{align}
&\frac{4\pi A_{\sigma,\sigma_2}}{N_{\bm k}} \sum_{{\bm G}\neq {\bm 0}} \tilde{u}^{\sigma, \sigma_2} ({\bm G}) \notag\\
&\sum_{\mu_1} \bigg( \mathrm{FT} [ \chi^*_{\mathrm{periodic},q_{2;0}} \phi_{\mathrm{periodic},{\tilde q}_1} ] ({\bm G})
\mathrm{FT} [ \chi^*_{\mathrm{periodic},{\tilde q}_1} \phi_{\mathrm{periodic},q_{2;0}} ] ({\bm 0})  +\notag\\
& \mathrm{FT} [ \chi^*_{\mathrm{periodic},q_{2;0}} \phi_{\mathrm{periodic},{\tilde q}_1} ] ({\bm 0})
\mathrm{FT} [ \chi^*_{\mathrm{periodic},{\tilde q}_1} \phi_{\mathrm{periodic},q_{2;0}} ] ({\bm G})  \bigg)
\mathrm{e}^{i {\bm G}\cdot {\bm r}_1} f_{\tilde{q}_1}, \label{eq:divcor_3a2_second_density_matrix_mixing}
 \end{align}
respectively. Since our implementation of the density-matrix mixing is applied only to the orbitals included in the SCF ${\bm k}$-mesh, such a replacement is not necessary for other correction terms that are non-zero only for the orbitals not included in the SCF ${\bm k}$-mesh.

A pseudocode for calculating the {\bf 3a3} term is shown in Algorithm~\ref{algo:3a3}.
\begin{figure}[h]
\begin{algorithm}[H]
  \caption{Calculate {\bf 3a3}: $g_j({\bm r}_1) = \dfrac{1}{2}\sum_{q_1, q_2}^{\mathrm{occupied}} \langle *, q_1, q_2 | \nabla_1 u_{12} \cdot \nabla_1 u_{13} |  q_1, j, q_2 \rangle  \times \mathrm{exp}(-i{\bm k}\cdot {\bm r}_1)$ (except the divergence correction terms)}
  \label{algo:3a3}
   \begin{algorithmic}[1]
	\State $\tilde{{\bm h}}^{\sigma}_1({\bm G}) \gets \sum_{\sigma_2} i{\bm G}\tilde{u}^{\sigma, \sigma_2}({\bm G}) \tilde{n}_{\sigma_2}({\bm G})$
	\State ${\bm h}^{\sigma}_1({\bm r}_1) \gets$ {\sc FT}$^{-1}$[$\tilde{\bm h}^{\sigma}_1({\bm G})$]
    	\For {$j = 1$ to $N$ (MPI parallelized)}
		\State ${\bm h}^{\sigma}_4({\bm r}_1) \gets 0$
		\For {$q_1 = 1$ to $N$}
			\If{$\sigma_1\neq \sigma$}
				\State continue
			\EndIf
			\State $h_2^{\sigma}({\bm r}_2) \gets  \chi^*_{\mathrm{periodic},q_1}({\bm r}_2) \phi_{\mathrm{periodic},j}({\bm r}_2)$
			\State $\tilde{h}^{\sigma}_2({\bm G}) \gets$ {\sc FT}[$h^{\sigma}_2({\bm r}_2)$]
			\State $\tilde{\bm h}^{\sigma}_3({\bm G}) \gets i({\bm k} - {\bm q}_1 + {\bm G}) \tilde{u}^{\sigma, \sigma}({\bm k} - {\bm q}_1 + {\bm G})\tilde{h}^{\sigma}_2({\bm G})$
			\State ${\bm h}^{\sigma}_3({\bm r}_1) \gets$ {\sc FT}$^{-1}$[$\tilde{\bm h}^{\sigma}_3({\bm G})$]
			\State ${\bm h}^{\sigma}_4({\bm r}_1) \gets {\bm h}^{\sigma}_4({\bm r}_1) + {\bm h}^{\sigma}_3({\bm r}_1)\phi_{\mathrm{periodic},q_1}({\bm r}_1)f_{q_1}/N_{\bm k}$
		\EndFor
		\State $g_j({\bm r}_1) \gets {\bm h}^{\sigma}_1({\bm r}_1)\cdot {\bm h}^{\sigma}_4({\bm r}_1)/2$
	\EndFor
   \end{algorithmic}
\end{algorithm}
\end{figure} \noindent
Because ${\bm h}^{\sigma}_4({\bm r}_1) = \sum_{q_1}^{\mathrm{occupied}} \langle *, q_1 | \nabla_1 u_{12}  |  q_1, j \rangle\ (\times \mathrm{exp}[-i{\bm k}\cdot{\bm r}_1])$ in Algorithm~\ref{algo:3a3} is similar to the {\bf 2bx1} term, $-\sum_{q}^{\mathrm{occupied}} \langle *, q | \nabla_1 u_{12} \cdot \nabla_1 | q, j \rangle$, the divergence correction for {\bf 3a3} is calculated in a similar way.
Namely, the correction term for ${\bm h}^{\sigma}_4({\bm r}_1)$ in {\bf 3a3} is
\begin{align}
\bigg[ -i \frac{4\pi A_{\sigma,\sigma}}{N_{\bm k}}\sum_{\substack{{\bm q}, {\bm G}\\ ({\bm q} + {\bm G}\ne {\bm 0})}}  ({\bm q}+{\bm G})A_{\mathrm{aux}}({\bm q}+{\bm G}) 
 \bigg]  \notag \\
 \times  \sum_{\mu_1} \bigg[ \int \mathrm{d}{\bm r}_2\ \chi^*_{\mathrm{periodic},q_{1k}}({\bm r}_2) \phi_{\mathrm{periodic},j}({\bm r}_2) \bigg]   \phi_{\mathrm{periodic},q_{1k}}({\bm r}_1) f_{q_{1k}}, \label{eq:divcorr_3a3}
\end{align}
where $q_{1k} =  (\sigma, {\bm k},\mu_1)$ belongs to the same ${\bm k}$-point with that for $j = (\sigma, {\bm k},\mu_j)$.
This correction term becomes zero when ${\bm k}$ is included in the SCF ${\bm k}$-mesh.

A pseudocode for calculating the {\bf 3a4} term is shown in Algorithm~\ref{algo:3a4}.
\begin{figure}[h]
\begin{algorithm}[H]
  \caption{Calculate {\bf 3a4}: $g_j({\bm r}_1) = -\dfrac{1}{2}\sum_{q_1, q_2}^{\mathrm{occupied}} \langle *, q_1, q_2 | \nabla_1 u_{12} \cdot \nabla_1 u_{13} |  q_1, q_2, j \rangle  \times \mathrm{exp}(-i{\bm k}\cdot {\bm r}_1)$ (except the divergence correction terms)}
  \label{algo:3a4}
   \begin{algorithmic}[1]
	\State $g_j({\bm r}_1) \gets 0$
	\State ${\bm h}^{\sigma_2}_3({\bm r}_1; q_2)$ ($q_2 = 1$ to $N$) $\gets 0$
	\For {$q_{2;0} = 1$ to $N_{\mathrm{irred}}$}
		\For {$q_1 = 1$ to $N$ (MPI parallelized)}
			\If{$\sigma_1\neq \sigma_2$}
				\State continue
			\EndIf
			\State $h_1^{\sigma_2}({\bm r}_2) \gets  \chi^*_{\mathrm{periodic},q_1}({\bm r}_2) \phi_{\mathrm{periodic},q_{2;0}}({\bm r}_2)$
			\State $\tilde{h}^{\sigma_2}_1({\bm G}) \gets$ {\sc FT}[$h^{\sigma_2}_1({\bm r}_2)$]
			\State $\tilde{\bm h}^{\sigma_2}_2({\bm G}) \gets i({\bm q}_{2;0} - {\bm q}_1 + {\bm G}) \tilde{u}^{\sigma_2, \sigma_2}({\bm q}_{2;0} - {\bm q}_1 + {\bm G})\tilde{h}^{\sigma_2}_1({\bm G})$
			\State ${\bm h}^{\sigma_2}_2({\bm r}_1) \gets$  {\sc FT}$^{-1}$[$\tilde{\bm h}^{\sigma_2}_2({\bm G})$]
			\State ${\bm h}^{\sigma_2}_3({\bm r}_1; q_{2;0})  \gets {\bm h}^{\sigma_2}_3({\bm r}_1; q_{2;0})  + {\bm h}^{\sigma_2}_2({\bm r}_1)\phi_{\mathrm{periodic},q_1}({\bm r}_1) f_{q_1} /N_{\bm k}$
		\EndFor
		\State MPI Allreduce for $q_1$-parallelization
		\For {symmetry operation ($q_{2;0}\to q_2$)}
			\State make ${\bm h}^{\sigma_2}_3({\bm r}_1; q_2)$ from ${\bm h}^{\sigma_2}_3({\bm r}_1; q_{2;0})$ by symmetry operation
    			\For {$j = 1$ to $N$ (MPI parallelized)}
				\If{$\sigma \neq \sigma_2$}
					\State continue
				\EndIf
				\State $h^{\sigma_2}_4({\bm r}_3; q_2) \gets  \chi^*_{\mathrm{periodic},q_2}({\bm r}_3) \phi_{\mathrm{periodic},j}({\bm r}_3)$
				\State $\tilde{h}^{\sigma_2}_4({\bm G}) \gets$ {\sc FT}[$h^{\sigma_2}_4({\bm r}_3)$]
				\State $\tilde{\bm h}^{\sigma_2}_5({\bm G}) \gets i({\bm k} - {\bm q}_2 + {\bm G}) \tilde{u}^{\sigma_2, \sigma_2}({\bm k} - {\bm q}_2 + {\bm G})\tilde{h}^{\sigma_2}_4({\bm G})$
				\State ${\bm h}^{\sigma_2}_5({\bm r}_1) \gets$  {\sc FT}$^{-1}$[$\tilde{\bm h}^{\sigma_2}_5({\bm G})$]
				\State $g_j({\bm r}_1) \gets g_j({\bm r}_1) -(1/2){\bm h}^{\sigma_2}_3({\bm r}_1; q_2) \cdot {\bm h}^{\sigma_2}_5({\bm r}_1) f_{q_2} /N_{\bm k}$
			\EndFor
		\EndFor
	\EndFor
   \end{algorithmic}
\end{algorithm}
\end{figure} \noindent
To obtain ${\bm h}^{\sigma_2}_3({\bm r}_1; q_2)$ from ${\bm h}^{\sigma_2}_3({\bm r}_1; q_{2;0})$ in Algorithm~\ref{algo:3a4}, we used the following relation in the same way as {\bf 3a2}:
\begin{equation}
\tilde{\bm h}^{\sigma_2}_3(S^{\dag} {\bm G}; q_{2}) = S^{\dag} \tilde{\bm h}^{\sigma_2}_3({\bm G}; q_{2;0}) \mathrm{e}^{i(S^{\dag} {\bm G})\cdot {\bm t}}, \label{eq:symope_3a2_1}
\end{equation}
for the case when the time-reversal symmetry is not used, and 
\begin{equation}
\tilde{\bm h}^{\sigma_2}_3(-S^{\dag} {\bm G}; q_{2}) = -S^{\dag} (\tilde{\bm h}^{\sigma_2}_3({\bm G}; q_{2;0}))^* \mathrm{e}^{-i(S^{\dag} {\bm G})\cdot {\bm t}} \label{eq:symope_3a2_2}
\end{equation}
for the case when the time-reversal symmetry is used.
We note that $(-)S^{\dag}$ in the right-hand side comes from the fact that $\tilde{\bm h}^{\sigma_2}_3$ is a vector quantity.
More concretely, calculation of $\tilde{\bm h}^{\sigma_2}_3$ involves the Fourier transform of $\nabla u$, which is proportional to ${\bm q}_{2;0} - {\bm q}_1 + {\bm G}$ where $(-)S^{\dag}$ should be operated.

The divergence correction for {\bf 3a4} is calculated in the following way.
$g_j({\bm r}_1)$ in Algorithm~\ref{algo:3a4} can be written as
\begin{align}
&g_j({\bm r}_1) = \notag \\
&-\frac{1}{2N^2_{\bm k}}\sum_{q_2, q_1,{\bm G}, {\bm G}'} ({\bm q}_2 - {\bm q}_1 + {\bm G}) \tilde{u}^{\sigma, \sigma}({\bm q}_2 - {\bm q}_1 + {\bm G})
\mathrm{FT} [ \chi^*_{\mathrm{periodic},q_1} \phi_{\mathrm{periodic},q_2} ] ({\bm G}) \notag \\
&\cdot ({\bm k} - {\bm q}_{2} + {\bm G}') \tilde{u}^{\sigma, \sigma}({\bm k}- {\bm q}_{2} + {\bm G}')
\mathrm{FT} [ \chi^*_{\mathrm{periodic},q_2} \phi_{\mathrm{periodic},j} ] ({\bm G}') \notag \\
&\times  \phi_{\mathrm{periodic},q_1}({\bm r}_1) \mathrm{e}^{i({\bm G}+{\bm G}')\cdot {\bm r}_1} f_{q_1} f_{q_2} \delta_{\sigma, \sigma_1} \delta_{\sigma, \sigma_2}.\label{eq:div_corr_3a4}
\end{align}
The $q_1$-summation of the former Jastrow function in Eq.~(\ref{eq:div_corr_3a4}) requires no correction since Eq.~(\ref{eq:correction_nabla}) becomes zero for this case.
Here, note that ${\bm q}_1, {\bm G}$-summation includes the diverging point ${\bm q}_2 - {\bm q}_1 + {\bm G} = {\bm 0}$ because both ${\bm q}_1$ and ${\bm q}_2$ are on the SCF ${\bm k}$-grid.
Thus, we only consider the divergence correction at ${\bm k} = {\bm q}_2$ with ${\bm G}'=0$. By substituting them into Eq.~(\ref{eq:div_corr_3a4}), we get the correction term for Eq.~(\ref{eq:div_corr_3a4}):
\begin{align}
&-\frac{1}{2N^2_{\bm k}}\sum_{\mu_2} \sum_{q_1,{\bm G}} ({\bm k} - {\bm q}_1 + {\bm G}) \tilde{u}^{\sigma, \sigma}({\bm k} - {\bm q}_1 + {\bm G})
\mathrm{FT} [ \chi^*_{\mathrm{periodic},q_1} \phi_{\mathrm{periodic},q_{2k}} ] ({\bm G}) \notag \\
&\cdot \bigg[ - 4\pi A_{\sigma,\sigma} \sum_{\substack{{\bm q}, {\bm G}'\\ ({\bm q} + {\bm G}'\ne {\bm 0})}}  ({\bm q}+{\bm G}')A_{\mathrm{aux}}({\bm q}+{\bm G}')   \bigg]
 \bigg[ \int \mathrm{d}{\bm r}_3\ \chi^*_{\mathrm{periodic},q_{2k}}({\bm r}_3) \phi_{\mathrm{periodic},j}({\bm r}_3) \bigg]  \notag \\
&\times \phi_{\mathrm{periodic},q_1}({\bm r}_1) \mathrm{e}^{i{\bm G}\cdot {\bm r}_1} f_{q_1} f_{q_{2k}} \delta_{\sigma, \sigma_1},\label{eq:div_corr_3a4_2}
\end{align}
where $q_{2k} =  (\sigma, {\bm k},\mu_2)$ belongs to the same ${\bm k}$-point as that for $j = (\sigma, {\bm k},\mu_j)$, and
\begin{equation}
\sum_{q_2, {\bm G}'} ({\bm k} - {\bm q}_{2} + {\bm G}') \tilde{u}^{\sigma, \sigma}({\bm k}- {\bm q}_{2} + {\bm G}')
\end{equation}
 in Eq.~(\ref{eq:div_corr_3a4}) is replaced with
\begin{equation}
- 4\pi A_{\sigma,\sigma} \sum_{\mu_2} \sum_{\substack{{\bm q}, {\bm G}'\\ ({\bm q} + {\bm G}'\ne {\bm 0})}}  ({\bm q}+{\bm G}')A_{\mathrm{aux}}({\bm q}+{\bm G}') 
\end{equation}
by taking a limit of ${\bm k} - {\bm q}_{2} + {\bm G}' \to {\bm 0}$ and considering the correction term as in Eq.~(\ref{eq:correction_nabla}).
We note that the correction term, Eq.~(\ref{eq:div_corr_3a4_2}), should also be corrected for considering the diverging behavior of the Jastrow function therein.
Thus, we should also consider the additional correction for Eq.~(\ref{eq:div_corr_3a4_2}),
\begin{align}
&-\frac{1}{2N^2_{\bm k}}\sum_{\mu_1, \mu_2} 
 \bigg[ - 4\pi A_{\sigma,\sigma} \sum_{\substack{{\bm q}, {\bm G}'\\ ({\bm q} + {\bm G}'\ne {\bm 0})}}  ({\bm q}+{\bm G}')A_{\mathrm{aux}}({\bm q}+{\bm G}')   \bigg]^2 \notag \\
& \times \bigg[ \int \mathrm{d}{\bm r}_2\ \chi^*_{\mathrm{periodic},q_{1k}}({\bm r}_2) \phi_{\mathrm{periodic},q_{2k}}({\bm r}_2) \bigg]
 \bigg[ \int \mathrm{d}{\bm r}_3\ \chi^*_{\mathrm{periodic},q_{2k}}({\bm r}_3) \phi_{\mathrm{periodic},j}({\bm r}_3) \bigg]  \notag \\
&\times \phi_{\mathrm{periodic},q_{1k}}({\bm r}_1)  f_{q_{1k}} f_{q_{2k}}, \\
&= -\frac{1}{2N^2_{\bm k}}\sum_{\mu_1} 
 \bigg[ - 4\pi A_{\sigma,\sigma} \sum_{\substack{{\bm q}, {\bm G}'\\ ({\bm q} + {\bm G}'\ne {\bm 0})}}  ({\bm q}+{\bm G}')A_{\mathrm{aux}}({\bm q}+{\bm G}')   \bigg]^2 \notag \\
&\times \bigg[ \int \mathrm{d}{\bm r}_3\ \chi^*_{\mathrm{periodic},q_{1k}}({\bm r}_3) \phi_{\mathrm{periodic},j}({\bm r}_3) \bigg]  
\phi_{\mathrm{periodic},q_{1k}}({\bm r}_1)  f_{q_{1k}}^2. \label{eq:div_corr_3a4_3}
\end{align}
For ${\bm k}$ included in the SCF ${\bm k}$-mesh, both of the divergence correction terms, Eqs.~(\ref{eq:div_corr_3a4_2}) and (\ref{eq:div_corr_3a4_3}), become zero.

A pseudocode for calculating the {\bf 3b1} term is shown in Algorithm~\ref{algo:3b1}.
\begin{figure}[h]
\begin{algorithm}[H]
  \caption{Calculate {\bf 3b1}: $g_j({\bm r}_1) = -\dfrac{1}{2}\sum_{q_1, q_2}^{\mathrm{occupied}} \langle *, q_1, q_2 | \nabla_2 u_{21} \cdot \nabla_2 u_{23} | j, q_1, q_2 \rangle  \times \mathrm{exp}(-i{\bm k}\cdot {\bm r}_1)$}
  \label{algo:3b1}
   \begin{algorithmic}[1]
    	\For {$j = 1$ to $N$ (MPI parallelized)}
		\State $\tilde{{\bm h}}^{\sigma_1}_1({\bm G}) \gets \sum_{\sigma_2} i{\bm G}\tilde{u}^{\sigma_1, \sigma_2}({\bm G}) \tilde{n}_{\sigma_2}({\bm G})$
		\State ${\bm h}^{\sigma_1}_1({\bm r}_2) \gets$ {\sc FT}$^{-1}$[$\tilde{\bm h}^{\sigma_1}_1({\bm G})$]
		\State ${\bm h}^{\sigma_1}_2 ({\bm r}_2) \gets {\bm h}^{\sigma_1}_1({\bm r}_2) n_{\sigma_1}({\bm r}_2)$
		\State $\tilde{{\bm h}}^{\sigma_1}_2({\bm G}) \gets$ {\sc FT}[$\bm h^{\sigma_1}_2({\bm r}_2)$]
		\State $\tilde{h}^{\sigma}_3({\bm G}) \gets \sum_{\sigma_1}- i{\bm G}\tilde{u}^{\sigma, \sigma_1}({\bm G})\cdot \tilde{{\bm h}}^{\sigma_1}_2({\bm G})$
		\State $h^{\sigma}_3({\bm r}_1) \gets$ {\sc FT}$^{-1}$[$\tilde{h}^{\sigma}_3({\bm G})$]
		\State $g_j({\bm r}_1) \gets -h^{\sigma}_3({\bm r}_1) \phi_{\mathrm{periodic},j}({\bm r}_1)/2$
	\EndFor
   \end{algorithmic}
\end{algorithm}
\end{figure} \noindent
This term does not need the divergence correction by the same reason as the two-body Hartree terms.

A pseudocode for calculating the {\bf 3b2} term is shown in Algorithm~\ref{algo:3b2}.
\begin{figure}[h]
\begin{algorithm}[H]
  \caption{Calculate {\bf 3b2}: $g_j({\bm r}_1) = \dfrac{1}{2}\sum_{q_1, q_2}^{\mathrm{occupied}} \langle *, q_1, q_2 | \nabla_2 u_{21} \cdot \nabla_2 u_{23} | j, q_2, q_1 \rangle \times \mathrm{exp}(-i{\bm k}\cdot {\bm r}_1)$}
  \label{algo:3b2}
   \begin{algorithmic}[1]
	\State ${\bm h}_5^{\sigma_2}({\bm r}_2) \gets 0$
	\State ${\bm h}^{\sigma_2}_3({\bm r}_2; q_2)$ ($q_2 = 1$ to $N$) $\gets 0$	
	\For {$q_{2;0} = 1$ to $N_{\mathrm{irred}}$}
		\For {$q_1 = 1$ to $N$ (MPI parallelized)}
			\If{$\sigma_1\neq \sigma_2$}
				\State continue
			\EndIf
			\State $h_1^{\sigma_2}({\bm r}_3) \gets  \chi^*_{\mathrm{periodic},q_{2;0}}({\bm r}_3) \phi_{\mathrm{periodic},q_1}({\bm r}_3)$
			\State $\tilde{h}^{\sigma_2}_1({\bm G}) \gets$ {\sc FT}[$h^{\sigma_2}_1({\bm r}_3)$]
			\State $\tilde{\bm h}^{\sigma_2}_2({\bm G}) \gets i({\bm q}_1 - {\bm q}_{2;0} + {\bm G}) \tilde{u}^{\sigma_2, \sigma_2}({\bm q}_1 - {\bm q}_{2;0} + {\bm G})\tilde{h}^{\sigma_2}_1({\bm G})$
			\State ${\bm h}^{\sigma_2}_2({\bm r}_2) \gets$  {\sc FT}$^{-1}$[$\tilde{\bm h}^{\sigma_2}_2({\bm G})$]
			\State ${\bm h}_3^{\sigma_2}({\bm r}_2; q_{2;0}) \gets {\bm h}_3^{\sigma_2}({\bm r}_2; q_{2;0}) + {\bm h}^{\sigma_2}_2({\bm r}_2) \chi^*_{\mathrm{periodic},q_1}({\bm r}_2) f_{q_1}/N_{\bm k}$
		\EndFor
		\State MPI Allreduce for $q_1$-parallelization
		\For {symmetry operation ($q_{2;0}\to q_2$)}
			\State make ${\bm h}^{\sigma_2}_3({\bm r}_2; q_2)$ from ${\bm h}^{\sigma_2}_3({\bm r}_2; q_{2;0})$ by symmetry operation
			\State ${\bm h}_4^{\sigma_2}({\bm r}_2) \gets {\bm h}^{\sigma_2}_3({\bm r}_2; q_2) \phi_{\mathrm{periodic},q_{2}}({\bm r}_2)$
			\State ${\bm h}_5^{\sigma_2}({\bm r}_2) \gets {\bm h}_5^{\sigma_2}({\bm r}_2)  + {\bm h}^{\sigma_2}_4({\bm r}_2) f_{q_2}/N_{\bm k}$
		\EndFor
	\EndFor
	\State $\tilde{\bm h}^{\sigma_2}_5({\bm G}) \gets$ {\sc FT}[${\bm h}_5^{\sigma_2}({\bm r}_2)$]
	\State $\tilde{h}^{\sigma}_6({\bm G}) \gets \sum_{\sigma_2} -i {\bm G} \tilde{u}^{\sigma, \sigma_2}({\bm G})\cdot \tilde{\bm h}^{\sigma_2}_5({\bm G})$
	\State $h^{\sigma}_6({\bm r}_1) \gets$  {\sc FT}$^{-1}$[$\tilde{h}^{\sigma}_6({\bm G})$]
    	\For {$j = 1$ to $N$ (MPI parallelized)}
		\State $g_j({\bm r}_1) \gets h^{\sigma}_6({\bm r}_1)  \phi_{\mathrm{periodic},j}({\bm r}_1)/2$
	\EndFor
   \end{algorithmic}
\end{algorithm}
\end{figure} \noindent
Symmetry operation for obtaining ${\bm h}^{\sigma_2}_3({\bm r}_1; q_2)$ from ${\bm h}^{\sigma_2}_3({\bm r}_1; q_{2;0})$ is exactly the same as Eqs.~(\ref{eq:symope_3a2_1}) and (\ref{eq:symope_3a2_2}).
The {\bf 3b2} term does not require the divergence correction because $\nabla_2 u_{21}$ only involves a ${\bm G}$-sum in reciprocal space (see $\tilde{h}^{\sigma}_6({\bm G})$ in Algorithm~\ref{algo:3b2}) and $\nabla_2 u_{23}$ yields a summation over a reciprocal-space grid including the zero vector, which makes no correction term: Eq.~(\ref{eq:correction_nabla}) becomes zero for this case.

A pseudocode for calculating the {\bf 3b3} term is shown in Algorithm~\ref{algo:3b3}.
\begin{figure}[h]
\begin{algorithm}[H]
  \caption{Calculate {\bf 3b3}: $g_j({\bm r}_1) = \dfrac{1}{2}\sum_{q_1, q_2}^{\mathrm{occupied}} \langle *, q_1, q_2 | \nabla_2 u_{21} \cdot \nabla_2 u_{23} | q_1, j, q_2 \rangle \times \mathrm{exp}(-i{\bm k}\cdot {\bm r}_1)$ (except the divergence correction terms)}
  \label{algo:3b3}
   \begin{algorithmic}[1]
	\State $\tilde{{\bm h}}^{\sigma}_1({\bm G}) \gets \sum_{\sigma_2} i{\bm G}\tilde{u}^{\sigma, \sigma_2}({\bm G}) \tilde{n}_{\sigma_2}({\bm G})$
	\State ${\bm h}^{\sigma}_1({\bm r}_2) \gets$ {\sc FT}$^{-1}$[$\tilde{\bm h}^{\sigma}_1({\bm G})$]
    	\For {$j = 1$ to $N$ (MPI parallelized)}
		\State $g_j({\bm r}_1) \gets 0$
		\For {$q_1 = 1$ to $N$}
			\If{$\sigma_1\neq \sigma$}
				\State continue
			\EndIf
			\State ${\bm h}_2^{\sigma}({\bm r}_2) \gets  {\bm h}^{\sigma}_1({\bm r}_2) \chi^*_{\mathrm{periodic},q_1}({\bm r}_2) \phi_{\mathrm{periodic},j}({\bm r}_2)$
			\State $\tilde{\bm h}^{\sigma}_2({\bm G}) \gets$ {\sc FT}[${\bm h}^{\sigma}_2({\bm r}_2)$]
			\State $\tilde{h}^{\sigma}_3({\bm G}) \gets -i({\bm k} - {\bm q}_1 + {\bm G}) \tilde{u}^{\sigma, \sigma}({\bm k} - {\bm q}_1 + {\bm G})\cdot \tilde{\bm h}^{\sigma}_2({\bm G})$
			\State ${h}^{\sigma}_3({\bm r}_1) \gets$ {\sc FT}$^{-1}$[$\tilde{h}^{\sigma}_3({\bm G})$]
			\State $g_j({\bm r}_1) \gets g_j({\bm r}_1) + (1/2){h}^{\sigma}_3({\bm r}_1)\phi_{\mathrm{periodic},q_1}({\bm r}_1)f_{q_1}/N_{\bm k}$
		\EndFor
	\EndFor
   \end{algorithmic}
\end{algorithm}
\end{figure} \noindent
The divergence correction for {\bf 3b3} is calculated as follows. $g_j({\bm r}_1)$ in Algorithm~\ref{algo:3b3} can be written as
\begin{align}
&g_j({\bm r}_1) = \frac{-i}{2N_{\bm k}}\sum_{q_1,{\bm G}} 
({\bm k} - {\bm q}_{1} + {\bm G}) \tilde{u}^{\sigma, \sigma}({\bm k}- {\bm q}_{1} + {\bm G}) \notag \\
&\cdot \mathrm{FT} [ \chi^*_{\mathrm{periodic},q_1} \phi_{\mathrm{periodic},j} {\bm h}_1^{\sigma} ] ({\bm G}) 
 \phi_{\mathrm{periodic},q_1}({\bm r}_1) \mathrm{e}^{i{\bm G}\cdot {\bm r}_1} f_{q_1} \delta_{\sigma, \sigma_1},\label{eq:div_corr_3b3}
\end{align}
the divergence correction for which is,
\begin{align}
 &\frac{-i}{2N_{\bm k}} \sum_{\mu_1}
  \bigg[ - 4\pi A_{\sigma,\sigma} \sum_{\substack{{\bm q}, {\bm G}'\\ ({\bm q} + {\bm G}'\ne {\bm 0})}}  ({\bm q}+{\bm G}')A_{\mathrm{aux}}({\bm q}+{\bm G}')   \bigg]
\notag \\
&\cdot  \bigg[ \int \mathrm{d}{\bm r}_2\ \chi^*_{\mathrm{periodic},q_{1k}}({\bm r}_2) \phi_{\mathrm{periodic},j} ({\bm r}_2) {\bm h}_1^{\sigma} ({\bm r}_2)\bigg] 
 \phi_{\mathrm{periodic},q_{1k}}({\bm r}_1)f_{q_{1k}},\label{eq:div_corr_3b3_2}
\end{align}
where $q_{1k} =  (\sigma, {\bm k},\mu_1)$ belongs to the same ${\bm k}$-point as that for $j = (\sigma, {\bm k},\mu_j)$, and
\begin{equation}
 \sum_{q_1,{\bm G}} 
({\bm k} - {\bm q}_{1} + {\bm G}) \tilde{u}^{\sigma, \sigma}({\bm k}- {\bm q}_{1} + {\bm G}) 
\end{equation}
in Eq.~(\ref{eq:div_corr_3b3}) is replaced with
\begin{equation}
-4\pi A_{\sigma,\sigma} \sum_{\mu_1} \sum_{\substack{{\bm q}, {\bm G}'\\ ({\bm q} + {\bm G}'\ne {\bm 0})}}  ({\bm q}+{\bm G}')A_{\mathrm{aux}}({\bm q}+{\bm G}')
\end{equation}
by taking a limit of ${\bm k} - {\bm q}_{1} + {\bm G}  \to {\bm 0}$ and considering the correction term as in Eq.~(\ref{eq:correction_nabla}).
For ${\bm k}$ included in the SCF ${\bm k}$-mesh, the divergence correction term, Eq.~(\ref{eq:div_corr_3b3_2}), becomes zero: Eq.~(\ref{eq:correction_nabla}) becomes zero due to the symmetry of the auxiliary function $A_{\mathrm{aux}}({\bm G})$.

A pseudocode for calculating the {\bf 3b4} term is shown in Algorithm~\ref{algo:3b4}.
\begin{figure}[h]
\begin{algorithm}[H]
  \caption{Calculate {\bf 3b4}: $g_j({\bm r}_1) = -\dfrac{1}{2}\sum_{q_1, q_2}^{\mathrm{occupied}} \langle *, q_1, q_2 | \nabla_2 u_{21} \cdot \nabla_2 u_{23} |  q_1, q_2, j \rangle \times \mathrm{exp}(-i{\bm k}\cdot {\bm r}_1)$ (except the divergence correction terms)}
 \label{algo:3b4}
   \begin{algorithmic}[1]
	\State $g_j({\bm r}_1) \gets {\bm 0}$
    	\For {$j = 1$ to $N$ (MPI parallelized)}
	\State ${\bm h}^{\sigma}_3({\bm r}_2) \gets {\bm 0}$
		\For {$q_2 = 1$ to $N$}
			\If{$\sigma_2\neq \sigma$}
				\State continue
			\EndIf
			\State $h_1^{\sigma}({\bm r}_3) \gets  \chi^*_{\mathrm{periodic},q_2}({\bm r}_3) \phi_{\mathrm{periodic},j}({\bm r}_3)$
			\State $\tilde{h}^{\sigma}_1({\bm G}) \gets$ {\sc FT}[$h^{\sigma}_1({\bm r}_3)$]
			\State $\tilde{\bm h}^{\sigma}_2({\bm G}) \gets i({\bm k} - {\bm q}_2 + {\bm G}) \tilde{u}^{\sigma, \sigma}({\bm k} - {\bm q}_2 + {\bm G})\tilde{h}^{\sigma}_1({\bm G})$
			\State ${\bm h}^{\sigma}_2({\bm r}_2) \gets$ {\sc FT}$^{-1}$[$\tilde{\bm h}^{\sigma}_2({\bm G})$]
			\State ${\bm h}^{\sigma}_3({\bm r}_2) \gets {\bm h}^{\sigma}_3({\bm r}_2) + {\bm h}^{\sigma}_3({\bm r}_2)\phi_{\mathrm{periodic},q_2}({\bm r}_2)f_{q_2}/N_{\bm k}$
		\EndFor
		\For {$q_1 = 1$ to $N$}
			\If{$\sigma_1\neq \sigma$}
				\State continue
			\EndIf
			\State ${\bm h}^{\sigma}_4({\bm r}_2) \gets {\bm h}^{\sigma}_3({\bm r}_2) \chi^*_{\mathrm{periodic},q_1}({\bm r}_2)$
			\State $\tilde{\bm h}^{\sigma}_4 ({\bm G})\gets$ {\sc FT}[${\bm h}^{\sigma}_4({\bm r}_2)$]
			\State $\tilde{h}^{\sigma}_5({\bm G}) \gets -i({\bm k} - {\bm q}_1 + {\bm G}) \tilde{u}^{\sigma, \sigma}({\bm k} - {\bm q}_1 + {\bm G})\cdot \tilde{\bm h}^{\sigma}_4({\bm G})$
			\State ${h}^{\sigma}_5({\bm r}_1) \gets$ {\sc FT}$^{-1}$[$\tilde{h}^{\sigma}_5({\bm G})$]
			\State $g_j({\bm r}_1) \gets g_j({\bm r}_1) - (1/2){h}^{\sigma}_5({\bm r}_1)\phi_{\mathrm{periodic},q_1}({\bm r}_1)f_{q_1}/N_{\bm k}$
		\EndFor
	\EndFor
   \end{algorithmic}
\end{algorithm}
\end{figure} \noindent
The divergence correction for {\bf 3b4} consists of the following two contributions.
One is that for ${\bm h}^{\sigma}_3({\bm r}_2)$ in Algorithm~\ref{algo:3b4}, which comes from the divergence correction for $\nabla_2 u_{23}$.
Since ${\bm h}^{\sigma}_3({\bm r}_2)$ is written as,
\begin{align}
{\bm h}^{\sigma}_3({\bm r}_2) = \frac{1}{N_{\bm k}}\sum_{q_2,{\bm G}} 
i({\bm k} - {\bm q}_{2} + {\bm G}) \tilde{u}^{\sigma, \sigma}({\bm k}- {\bm q}_{2} + {\bm G}) \notag \\
\times \mathrm{FT} [ \chi^*_{\mathrm{periodic},q_2} \phi_{\mathrm{periodic},j} ] ({\bm G}) 
 \phi_{\mathrm{periodic},q_2}({\bm r}_2) \mathrm{e}^{i{\bm G}\cdot {\bm r}_2} f_{q_2} \delta_{\sigma, \sigma_2}, \label{eq:div_corr_3b4}
\end{align}
the correction term for which is
\begin{align}
\frac{1}{N_{\bm k}}  \sum_{\mu_2}
\bigg[ -4\pi i A_{\sigma,\sigma} \sum_{\substack{{\bm q}, {\bm G}'\\ ({\bm q} + {\bm G}'\ne {\bm 0})}}  ({\bm q}+{\bm G}')A_{\mathrm{aux}}({\bm q}+{\bm G}') \bigg] \notag \\
\times \bigg[\int \mathrm{d}{\bm r}_3\ \chi^*_{\mathrm{periodic},q_{2k}} ({\bm r}_3) \phi_{\mathrm{periodic},j} ({\bm r}_3) \bigg]
 \phi_{\mathrm{periodic},q_{2k}}({\bm r}_2) f_{q_{2k}},  \label{eq:div_corr_3b4_2}
\end{align}
where $q_{2k} =  (\sigma, {\bm k},\mu_2)$ belongs to the same ${\bm k}$-point as that for $j = (\sigma, {\bm k},\mu_j)$, and
\begin{equation}
 \sum_{q_2,{\bm G}} 
({\bm k} - {\bm q}_{2} + {\bm G}) \tilde{u}^{\sigma, \sigma}({\bm k}- {\bm q}_{2} + {\bm G}) 
\end{equation}
in Eq.~(\ref{eq:div_corr_3b4}) is replaced with
\begin{equation}
-4\pi A_{\sigma,\sigma} \sum_{\mu_2} \sum_{\substack{{\bm q}, {\bm G}'\\ ({\bm q} + {\bm G}'\ne {\bm 0})}}  ({\bm q}+{\bm G}')A_{\mathrm{aux}}({\bm q}+{\bm G}')
\end{equation}
by taking a limit of ${\bm k} - {\bm q}_{2} + {\bm G} \to {\bm 0}$ and considering the correction term as in Eq.~(\ref{eq:correction_nabla}).
The other correction is required for $\nabla_2 u_{21}$ in $g_j({\bm r}_1)$ in Algorithm~\ref{algo:3b4}.
Since $g_j({\bm r}_1)$ in Algorithm~\ref{algo:3b4} can be written as,
\begin{align}
g_j({\bm r}_1) = \frac{1}{2N_{\bm k}}\sum_{q_1,{\bm G}} 
i({\bm k} - {\bm q}_{1} + {\bm G}) \tilde{u}^{\sigma, \sigma}({\bm k}- {\bm q}_{1} + {\bm G}) \notag \\
\cdot \mathrm{FT} [ \chi^*_{\mathrm{periodic},q_1} {\bm h}_3^{\sigma}] ({\bm G}) 
 \phi_{\mathrm{periodic},q_1}({\bm r}_1) \mathrm{e}^{i{\bm G}\cdot {\bm r}_1} f_{q_1} \delta_{\sigma, \sigma_1}, \label{eq:div_corr_3b4_3}
\end{align}
the correction term for which is
\begin{align}
\frac{1}{2N_{\bm k}}  \sum_{\mu_1}
\bigg[ -4\pi i A_{\sigma,\sigma} \sum_{\substack{{\bm q}, {\bm G}'\\ ({\bm q} + {\bm G}'\ne {\bm 0})}}  ({\bm q}+{\bm G}')A_{\mathrm{aux}}({\bm q}+{\bm G}') \bigg] \notag \\
\cdot \bigg[\int \mathrm{d}{\bm r}_3\ \chi^*_{\mathrm{periodic},q_{1k}} ({\bm r}_3){\bm h}_3^{\sigma} ({\bm r}_3) \bigg]
 \phi_{\mathrm{periodic},q_{1k}}({\bm r}_1) f_{q_{1k}},  \label{eq:div_corr_3b4_4}
\end{align}
where $q_{1k} =  (\sigma, {\bm k},\mu_2)$ belongs to the same ${\bm k}$-point as that for $j = (\sigma, {\bm k},\mu_j)$.
Note that ${\bm h}_3^{\sigma}$ in Eq.~(\ref{eq:div_corr_3b4_3}) is already corrected by adding Eq.~(\ref{eq:div_corr_3b4_2}).
These two corrections, Eqs.~(\ref{eq:div_corr_3b4_2}) and (\ref{eq:div_corr_3b4_4}), become zero when ${\bm k}$ is included in the SCF ${\bm k}$-mesh: Eq.~(\ref{eq:correction_nabla}) becomes zero due to the symmetry of the auxiliary function $A_{\mathrm{aux}}({\bm G})$.

A pseudocode for calculating the {\bf 3b5} term is shown in Algorithm~\ref{algo:3b5}.
\begin{figure}[h]
\begin{algorithm}[H]
  \caption{Calculate {\bf 3b5}: $g_j({\bm r}_1) = -\dfrac{1}{2}\sum_{q_1, q_2}^{\mathrm{occupied}} \langle *, q_1, q_2 | \nabla_2 u_{21} \cdot \nabla_2 u_{23} | q_2, j, q_1 \rangle \times \mathrm{exp}(-i{\bm k}\cdot {\bm r}_1)$ (except the divergence correction terms)}
  \label{algo:3b5}
   \begin{algorithmic}[1]
	\State $g_j({\bm r}_1)\gets 0$
	\State ${\bm h}^{\sigma_2}_3({\bm r}_2; q_2)$ ($q_2 = 1$ to $N$) $\gets 0$	
	\For {$q_{2;0} = 1$ to $N_{\mathrm{irred}}$}
		\For {$q_1 = 1$ to $N$ (MPI parallelized)}
			\If{$\sigma_1\neq \sigma_2$}
				\State continue
			\EndIf
			\State $h_1^{\sigma_2}({\bm r}_3) \gets  \chi^*_{\mathrm{periodic},q_{2;0}}({\bm r}_3) \phi_{\mathrm{periodic},q_1}({\bm r}_3)$
			\State $\tilde{h}^{\sigma_2}_1({\bm G}) \gets$ {\sc FT}[$h^{\sigma_2}_1({\bm r}_3)$]
			\State $\tilde{\bm h}^{\sigma_2}_2({\bm G}) \gets i({\bm q}_1 - {\bm q}_{2;0} + {\bm G}) \tilde{u}^{\sigma_2, \sigma_2}({\bm q}_1 - {\bm q}_{2;0} + {\bm G})\tilde{h}^{\sigma_2}_1({\bm G})$
			\State ${\bm h}^{\sigma_2}_2({\bm r}_2) \gets$  {\sc FT}$^{-1}$[$\tilde{\bm h}^{\sigma_2}_2({\bm G})$]
			\State ${\bm h}_3^{\sigma_2}({\bm r}_2; q_{2;0}) \gets {\bm h}_3^{\sigma_2}({\bm r}_2; q_{2;0}) + {\bm h}^{\sigma_2}_2({\bm r}_2) \chi^*_{\mathrm{periodic},q_1}({\bm r}_2) f_{q_1}/N_{\bm k}$
		\EndFor
		\State MPI Allreduce for $q_1$-parallelization
		\For {symmetry operation ($q_{2;0}\to q_2$)}
			\State make ${\bm h}^{\sigma_2}_3({\bm r}_2; q_2)$ from ${\bm h}^{\sigma_2}_3({\bm r}_2; q_{2;0})$ by symmetry operation
		    	\For {$j = 1$ to $N$ (MPI parallelized)}
				\If{$\sigma \neq \sigma_2$}
					\State continue
				\EndIf
				\State ${\bm h}_4^{\sigma_2}({\bm r}_2) \gets {\bm h}_3^{\sigma_2}({\bm r}_2; q_2) \phi_{\mathrm{periodic},j}({\bm r}_2)$	
				\State $\tilde{\bm h}^{\sigma_2}_4({\bm G}) \gets$ {\sc FT}[${\bm h}^{\sigma_2}_4({\bm r}_2)$]
				\State $\tilde{h}^{\sigma_2}_5({\bm G}) \gets -i({\bm k} - {\bm q}_{2} + {\bm G}) \tilde{u}^{\sigma_2, \sigma_2}({\bm k} - {\bm q}_{2} + {\bm G})\cdot \tilde{\bm h}^{\sigma_2}_4({\bm G})$
				\State ${h}^{\sigma_2}_5({\bm r}_1) \gets$  {\sc FT}$^{-1}$[$\tilde{h}^{\sigma_2}_5({\bm G})$]
				\State $g_j({\bm r}_1) \gets g_j({\bm r}_1) + (1/2){h}^{\sigma_2}_5({\bm r}_1) \phi_{\mathrm{periodic},q_2}({\bm r}_1) f_{q_2}/N_{\bm k}$
			\EndFor
		\EndFor
	\EndFor
   \end{algorithmic}
\end{algorithm}
\end{figure} \noindent
Since ${\bm h}^{\sigma_2}_3({\bm r}_1; q_2)$ in Algorithms~\ref{algo:3b2} and \ref{algo:3b5} are exactly the same, the symmetry operation for obtaining ${\bm h}^{\sigma_2}_3({\bm r}_1; q_2)$ from ${\bm h}^{\sigma_2}_3({\bm r}_1; q_{2;0})$ is also the same between them.
The divergence correction for {\bf 3b5} is calculated in the following way.
First, $g_j({\bm r}_1)$ in Algorithm~\ref{algo:3b5} can be written as
\begin{align}
&g_j({\bm r}_1) = \notag \\
&-\frac{1}{2N^2_{\bm k}}\sum_{q_1, q_2, {\bm G}, {\bm G}'} 
({\bm q}_1 - {\bm q}_2 + {\bm G}) \tilde{u}^{\sigma, \sigma}({\bm q}_1 - {\bm q}_2 + {\bm G})
\mathrm{FT} [ \chi^*_{\mathrm{periodic},q_{2}} \phi_{\mathrm{periodic},q_1} ] ({\bm G}) \notag \\
&\cdot ({\bm k} - {\bm q}_2 + {\bm G} + {\bm G}') \tilde{u}^{\sigma, \sigma}({\bm k} - {\bm q}_2 + {\bm G} + {\bm G}')
\mathrm{FT} [ \chi^*_{\mathrm{periodic},q_1} \phi_{\mathrm{periodic},j} ] ({\bm G}') \notag \\
&\times \phi_{\mathrm{periodic},q_2}({\bm r}_1)
\mathrm{e}^{i({\bm G}+{\bm G}')\cdot {\bm r}_1} f_{q_1} f_{q_2} \delta_{\sigma, \sigma_1} \delta_{\sigma, \sigma_2}. \label{eq:3b5_div} 
\end{align}
Here, ${\bm q}_1$-summation does not require the divergence correction since the divergence correction for $\sum_{q_1, {\bm G}} ({\bm q}_1 - {\bm q}_2 + {\bm G}) \tilde{u}^{\sigma, \sigma}({\bm q}_1 - {\bm q}_2 + {\bm G})$ becomes zero:  Eq.~(\ref{eq:correction_nabla}) becomes zero due to the symmetry of the auxiliary function $A_{\mathrm{aux}}({\bm G})$.
Thus, we concentrate on the divergence at ${\bm k} - {\bm q}_2 + {\bm G} + {\bm G}' \to {\bm 0}$ in the second Jastrow function in Eq.~(\ref{eq:3b5_div}).
By considering a limit of ${\bm q}_2 \to {\bm k}$ and ${\bm G}' \to -{\bm G}$ for Eq.~(\ref{eq:3b5_div}), we get the following correction term,
\begin{align}
&-\frac{1}{2N^2_{\bm k}}\sum_{q_1, \mu_2, {\bm G}} 
({\bm q}_1 - {\bm k} + {\bm G}) \tilde{u}^{\sigma, \sigma}({\bm q}_1 - {\bm k} + {\bm G})
\mathrm{FT} [ \chi^*_{\mathrm{periodic},q_{2k}} \phi_{\mathrm{periodic},q_1} ] ({\bm G}) \notag \\
&\cdot 
\bigg[ -4\pi A_{\sigma,\sigma} \sum_{\substack{{\bm q}, {\bm G}''\\ ({\bm q} + {\bm G}''\ne {\bm 0})}}  ({\bm q}+{\bm G}'')A_{\mathrm{aux}}({\bm q}+{\bm G}'') \bigg]
\mathrm{FT} [ \chi^*_{\mathrm{periodic},q_1} \phi_{\mathrm{periodic},j} ] (-{\bm G}) \notag \\
&\times \phi_{\mathrm{periodic},q_{2k}}({\bm r}_1) f_{q_1} f_{q_{2k}} \delta_{\sigma, \sigma_1}, \label{eq:3b5_div_2} 
\end{align}
where $q_{2k} =  (\sigma, {\bm k},\mu_2)$ belongs to the same ${\bm k}$-point as that for $j = (\sigma, {\bm k},\mu_j)$, and
\begin{equation}
 \sum_{q_2,{\bm G}'} 
({\bm k} - {\bm q}_2 + {\bm G} + {\bm G}') \tilde{u}^{\sigma, \sigma}({\bm k} - {\bm q}_2 + {\bm G} + {\bm G}')
\end{equation}
in Eq.~(\ref{eq:3b5_div}) is replaced with
\begin{equation}
-4\pi A_{\sigma,\sigma} \sum_{\mu_2} \sum_{\substack{{\bm q}, {\bm G}''\\ ({\bm q} + {\bm G}''\ne {\bm 0})}}  ({\bm q}+{\bm G}'')A_{\mathrm{aux}}({\bm q}+{\bm G}'').
\end{equation}
This correction term, Eq~(\ref{eq:3b5_div_2}), should also be corrected for the divergence of the Jastrow function therein.
This additional correction for Eq~(\ref{eq:3b5_div_2}) can be obtained by considering the divergence at ${\bm q}_1 - {\bm k} + {\bm G}\to {\bm 0}$ in Eq~(\ref{eq:3b5_div_2}):
\begin{align}
&-\frac{1}{2N^2_{\bm k}}\sum_{\mu_1, \mu_2, {\bm G}} 
\mathrm{FT} [ \chi^*_{\mathrm{periodic},q_{2k}} \phi_{\mathrm{periodic},q_{1k}} ] ({\bm 0})
\bigg[ -4\pi A_{\sigma,\sigma} \sum_{\substack{{\bm q}, {\bm G}''\\ ({\bm q} + {\bm G}''\ne {\bm 0})}}  ({\bm q}+{\bm G}'')A_{\mathrm{aux}}({\bm q}+{\bm G}'') \bigg]^2 \notag \\
&\times \mathrm{FT} [ \chi^*_{\mathrm{periodic},q_{1k}} \phi_{\mathrm{periodic},j} ] ({\bm 0}) 
\phi_{\mathrm{periodic},q_{2k}}({\bm r}_1) f_{q_{1k}} f_{q_{2k}} \notag \\
&= -\frac{1}{2N^2_{\bm k}}\sum_{\mu_1, {\bm G}} 
\bigg[ -4\pi A_{\sigma,\sigma} \sum_{\substack{{\bm q}, {\bm G}''\\ ({\bm q} + {\bm G}''\ne {\bm 0})}}  ({\bm q}+{\bm G}'')A_{\mathrm{aux}}({\bm q}+{\bm G}'') \bigg]^2 \notag \\
&\times \bigg[ \int \mathrm{d}{\bm r}_2\ \chi^*_{\mathrm{periodic},q_{1k}} ({\bm r}_2) \phi_{\mathrm{periodic},j}  ({\bm r}_2) \bigg]
\phi_{\mathrm{periodic},q_{1k}}({\bm r}_1) f_{q_{1k}}^2, \label{eq:3b5_div_3} 
\end{align}
where $q_{1k} =  (\sigma, {\bm k},\mu_1)$.

A pseudocode for calculating the {\bf 3b6} term is shown in Algorithm~\ref{algo:3b6}.
\begin{figure}[h]
\begin{algorithm}[H]
  \caption{Calculate {\bf 3b6}: $g_j({\bm r}_1) = \dfrac{1}{2}\sum_{q_1, q_2}^{\mathrm{occupied}} \langle *, q_1, q_2 | \nabla_2 u_{21} \cdot \nabla_2 u_{23} |  q_2, q_1, j \rangle \times \mathrm{exp}(-i{\bm k}\cdot {\bm r}_1)$ (except the divergence correction terms)}
 \label{algo:3b6}
   \begin{algorithmic}[1]
	\State $g_j({\bm r}_1) \gets {\bm 0}$
    	\For {$j = 1$ to $N$ (MPI parallelized)}
		\For {$q_2 = 1$ to $N$}
			\If{$\sigma_2\neq \sigma$}
				\State continue
			\EndIf
			\State $h_1^{\sigma}({\bm r}_3) \gets  \chi^*_{\mathrm{periodic},q_2}({\bm r}_3) \phi_{\mathrm{periodic},j}({\bm r}_3)$
			\State $\tilde{h}^{\sigma}_1({\bm G}) \gets$ {\sc FT}[$h^{\sigma}_1({\bm r}_3)$]
			\State $\tilde{\bm h}^{\sigma_1, \sigma}_2({\bm G}) \gets i({\bm k} - {\bm q}_2 + {\bm G}) \tilde{u}^{\sigma_1, \sigma}({\bm k} - {\bm q}_2 + {\bm G})\tilde{h}^{\sigma}_1({\bm G})$
			\State ${\bm h}^{\sigma_1, \sigma}_2({\bm r}_2) \gets$ {\sc FT}$^{-1}$[$\tilde{\bm h}^{\sigma_1, \sigma}_2({\bm G})$]
			\State ${\bm h}^{\sigma_1, \sigma}_3({\bm r}_2) \gets {\bm h}^{\sigma_1, \sigma}_2({\bm r}_2)n_{\sigma_1}({\bm r}_2)$
			\State $\tilde{\bm h}^{\sigma_1, \sigma}_3({\bm G}) \gets$ {\sc FT}[${\bm h}^{\sigma_1, \sigma}_3({\bm r}_2)$]
			\State $\tilde{h}^{\sigma}_4({\bm G}) \gets \sum_{\sigma_1} -i({\bm k} - {\bm q}_2 + {\bm G}) \tilde{u}^{\sigma_1, \sigma}({\bm k} - {\bm q}_2 + {\bm G})\cdot \tilde{\bm h}^{\sigma_1, \sigma}_3({\bm G})$
			\State ${h}^{\sigma}_4({\bm r}_1) \gets$ {\sc FT}$^{-1}$[$\tilde{h}^{\sigma}_4({\bm G})$]
			\State $g_j({\bm r}_1) \gets g_j({\bm r}_1)  + (1/2){h}^{\sigma}_4({\bm r}_1) \phi_{\mathrm{periodic},q_2}({\bm r}_1) f_{q_2}/N_{\bm k}$
		\EndFor
	\EndFor
   \end{algorithmic}
\end{algorithm}
\end{figure} \noindent
The divergence correction for {\bf 3b6} is the most complicated because $({\bm k} - {\bm q}_2 + {\bm G})$ divergence appears twice in Algorithm~\ref{algo:3b6}.
We shall see what correction terms are required.
$g_j({\bm r}_1)$ in Algorithm~\ref{algo:3b6} can be written as
\begin{align}
&g_j({\bm r}_1) = \notag \\
&\frac{1}{2N_{\bm k}}\sum_{\sigma_1, q_2, {\bm G}, {\bm G}'} 
({\bm k} - {\bm q}_2 + {\bm G}) \tilde{u}^{\sigma_1, \sigma}({\bm k} - {\bm q}_2 + {\bm G})
\mathrm{FT} [ \chi^*_{\mathrm{periodic},q_{2}} \phi_{\mathrm{periodic},j} ] ({\bm G}) \notag \\
&\cdot ({\bm k} - {\bm q}_2 + {\bm G} + {\bm G}') \tilde{u}^{\sigma_1, \sigma}({\bm k} - {\bm q}_2 + {\bm G} + {\bm G}')
\tilde{n}_{\sigma_1}({\bm G}') \notag \\
&\times \phi_{\mathrm{periodic},q_2}({\bm r}_1)
\mathrm{e}^{i({\bm G}+{\bm G}')\cdot {\bm r}_1} f_{q_2} \delta_{\sigma, \sigma_2}. \label{eq:3b6_div} 
\end{align}

First, we consider the case where ${\bm k}$ is included in the SCF ${\bm k}$-mesh.
For this case, the divergence correction can be considered by a similar way to that for {\bf 3a2}.
The divergence correction term for the ${\bm G}={\bm G}'={\bm 0}, {\bm q}_2\to {\bm k}$ component in Eq.~(\ref{eq:3b6_div}) is
\begin{align}
&\sum_{\sigma_1, \mu_2}2\pi A_{\sigma_1,\sigma} 
\bigg[ \frac{A_{\sigma_1,\sigma}\Omega}{\sqrt{\pi \alpha}} -
\frac{4\pi A_{\sigma_1,\sigma}}{N_{\bm k}} \bigg( -\alpha + \sum_{\substack{{\bm q}, {\bm G}''\\ ({\bm k} - {\bm q} + {\bm G}''\ne {\bm 0})}}  A_{\mathrm{aux}}({\bm k} - {\bm q}+{\bm G}'') \bigg)
+ \frac{\tilde{u}^{\sigma,\sigma_2}_{\mathrm{short}}({\bm 0})}{N_{\bm k}}
 \bigg] \notag \\
& \times \bigg[ \int \mathrm{d}{\bm r}_2\ \chi^*_{\mathrm{periodic},q_{2k}} ({\bm r}_3) \phi_{\mathrm{periodic},j}  ({\bm r}_3) \bigg] 
\tilde{n}_{\sigma_1}({\bm 0}) \phi_{\mathrm{periodic},q_{2k}}({\bm r}_1)
f_{q_{2k}}, \label{eq:3b6_div_scf1}
\end{align}
where $q_{2k} =  (\sigma, {\bm k},\mu_2)$ belongs to the same ${\bm k}$-point as that for $j = (\sigma, {\bm k},\mu_j)$, and we use
\begin{equation}
\lim_{{\bm k}- {\bm q}_{2},{\bm G},{\bm G}' \to {\bm 0}}  ({\bm k} - {\bm q}_2 + {\bm G})\cdot ({\bm k} - {\bm q}_{2} + {\bm G} + {\bm G}')  \tilde{u}^{\sigma_1, \sigma}({\bm k} - {\bm q}_{2} +{\bm G} +  {\bm G}') = 4\pi A_{\sigma_1,\sigma}
\end{equation}
and Eqs.~(\ref{eq:correction2}) and (\ref{eq:u_short}).
The divergence correction for ${\bm G}={\bm 0}, {\bm G}'\ne {\bm 0}, {\bm q}_2\to {\bm k}$ in Eq.~(\ref{eq:3b6_div}) is
\begin{align}
&\sum_{\sigma_1, \mu_2,{\bm G}'\ne {\bm 0}}  \frac{2\pi A_{\sigma_1,\sigma}}{N_{\bm k}}
 \bigg[ \int \mathrm{d}{\bm r}_2\ \chi^*_{\mathrm{periodic},q_{2k}} ({\bm r}_3) \phi_{\mathrm{periodic},j}  ({\bm r}_3) \bigg] \notag \\
&\times \tilde{u}^{\sigma_1, \sigma}({\bm G}')
\tilde{n}_{\sigma_1}({\bm G}') \phi_{\mathrm{periodic},q_{2k}}({\bm r}_1)
\mathrm{e}^{i {\bm G}' \cdot {\bm r}_1} f_{q_{2k}}, \label{eq:3b6_div_scf2} 
\end{align}
where we only consider the first term of the right-hand side in
\begin{equation}
({\bm k} - {\bm q}_2 + {\bm G}) \cdot ({\bm k} - {\bm q}_2 + {\bm G} + {\bm G}')  = ({\bm k} - {\bm q}_2 + {\bm G})^2 + ({\bm k} - {\bm q}_2 + {\bm G}) \cdot {\bm G}'
\end{equation}
and the second term is not considered because the divergence correction for $({\bm k} - {\bm q}_2 + {\bm G})  \tilde{u}^{\sigma_1, \sigma}({\bm k} - {\bm q}_2 + {\bm G})$ becomes zero when ${\bm k}$ is included in the SCF ${\bm k}$-mesh, as we have seen for several cases in this paper.
In the same manner, the divergence correction for ${\bm G}\ne{\bm 0}, {\bm G}+{\bm G}'={\bm 0}, {\bm q}_2\to {\bm k}$ in Eq.~(\ref{eq:3b6_div}) is
\begin{align}
&\sum_{\sigma_1, \mu_2, {\bm G}\ne {\bm 0}}  \frac{2\pi A_{\sigma_1,\sigma}}{N_{\bm k}}
\tilde{u}^{\sigma_1, \sigma}({\bm G})
\mathrm{FT} [ \chi^*_{\mathrm{periodic},q_{2k}} \phi_{\mathrm{periodic},j} ] ({\bm G})  \tilde{n}_{\sigma_1}(-{\bm G})\notag \\
&\times \phi_{\mathrm{periodic},q_{2k}}({\bm r}_1) f_{q_{2k}}. \label{eq:3b6_div_scf3} 
\end{align}

Second, we consider the case where ${\bm k}$ is not included in the SCF ${\bm k}$-mesh.
The divergence correction term for the ${\bm G}={\bm G}'={\bm 0}, {\bm q}_2\to {\bm k}$ component in Eq.~(\ref{eq:3b6_div}) is
\begin{align}
&\sum_{\mu_2, \sigma_1} 2\pi A_{\sigma_1,\sigma}^2 
\bigg[ \frac{\Omega}{\sqrt{\pi \alpha}} -
\frac{4\pi}{N_{\bm k}} \sum_{{\bm q}, {\bm G}''}  A_{\mathrm{aux}}({\bm k} - {\bm q}+{\bm G}'') 
 \bigg] \notag \\
& \times \bigg[ \int \mathrm{d}{\bm r}_2\ \chi^*_{\mathrm{periodic},q_{2k}} ({\bm r}_3) \phi_{\mathrm{periodic},j}  ({\bm r}_3) \bigg] 
\tilde{n}_{\sigma_1}({\bm 0}) \phi_{\mathrm{periodic},q_{2k}}({\bm r}_1)
f_{q_{2k}}, \label{eq:3b6_div_band1}
\end{align}
where $\alpha$ and $\tilde{u}^{\sigma,\sigma_2}_{\mathrm{short}}$ in Eq.~(\ref{eq:3b6_div_scf1}) are removed (see Sec.~\ref{sec:divcorr}).
For the divergence correction for ${\bm G}={\bm 0}, {\bm G}'\ne {\bm 0}, {\bm q}_2\to {\bm k}$ in Eq.~(\ref{eq:3b6_div}), we consider the divergence correction for $\nabla u$.
Namely, the divergence correction is
\begin{align}
&-\sum_{\sigma_1, \mu_2, {\bm G}'\ne{\bm 0}}  \frac{2\pi A_{\sigma_1,\sigma}}{N_{\bm k}}
\bigg[  \sum_{{\bm q}, {\bm G}''}  ({\bm k} - {\bm q}+{\bm G}'')A_{\mathrm{aux}}({\bm k} - {\bm q}+{\bm G}'') \bigg] \notag \\
&\cdot \bigg[ \int \mathrm{d}{\bm r}_2\ \chi^*_{\mathrm{periodic},q_{2k}} ({\bm r}_3) \phi_{\mathrm{periodic},j}  ({\bm r}_3) \bigg]
{\bm G}'\tilde{u}^{\sigma_1, \sigma}({\bm G}')
\tilde{n}_{\sigma_1}({\bm G}') \phi_{\mathrm{periodic},q_{2k}}({\bm r}_1)
\mathrm{e}^{i {\bm G}'\cdot {\bm r}_1} f_{q_{2k}}, \label{eq:3b6_div_band2} 
\end{align}
by using Eq.~(\ref{eq:correction_nabla}).
In the same manner, the divergence correction for ${\bm G}\ne{\bm 0}, {\bm G}+{\bm G}'={\bm 0}, {\bm q}_2\to {\bm k}$ in Eq.~(\ref{eq:3b6_div}) is
\begin{align}
&-\sum_{\sigma_1, \mu_2, {\bm G}\ne{\bm 0}}  \frac{2\pi A_{\sigma_1,\sigma}}{N_{\bm k}}
{\bm G} \tilde{u}^{\sigma_1, \sigma}({\bm G})
\mathrm{FT} [ \chi^*_{\mathrm{periodic},q_{2k}} \phi_{\mathrm{periodic},j} ] ({\bm G}) \notag \\
&\cdot \bigg[  \sum_{{\bm q}, {\bm G}''}  ({\bm k} - {\bm q}+{\bm G}'')A_{\mathrm{aux}}({\bm k} - {\bm q}+{\bm G}'') \bigg] 
\tilde{n}_{\sigma_1}(-{\bm G}) \phi_{\mathrm{periodic},q_{2k}}({\bm r}_1) f_{q_{2k}}. \label{eq:3b6_div_band3} 
\end{align}

\subsubsection{Equations used for calculating the two-body and three-body terms}

As a short summary, we show a list of equations used for calculating the two-body and three-body terms in the (BI)TC method as follows.
\begin{description}
\item[2ah] Algorithm~\ref{algo:2ah}
\item[2bh1, 2bh2] in the same way as Algorithm~\ref{algo:2ah} (not shown)
\item[2ax] Algorithm~\ref{algo:2ax} with the divergence correction, Eq.~(\ref{eq:divcorr_2ax_scf}) (for ${\bm k}$ included in the SCF ${\bm k}$-mesh) or Eq.~(\ref{eq:divcorr_2ax_band}) (otherwise)
\item[2bx1] in the same way as Algorithm~\ref{algo:2ax} (not shown) with the divergence correction, Eq.~(\ref{eq:divcorr_2bx1}) (for ${\bm k}$ not included in the SCF ${\bm k}$-mesh) 
\item[2bx2] in the same way as Algorithm~\ref{algo:2ax} (not shown) with the divergence correction, Eq.~(\ref{eq:divcorr_2bx2}) (for ${\bm k}$ not included in the SCF ${\bm k}$-mesh) 
\item[3a1] Algorithm~\ref{algo:3a1}
\item[3a2] Algorithm~\ref{algo:3a2} with the divergence correction for $h^{\sigma, \sigma_2}_5({\bm r}_1; q_{2;0})$, Eqs.~(\ref{eq:divcor_3a2_first}) and (\ref{eq:divcor_3a2_second}). When the density-matrix mixing is used, these equations are replaced with Eqs.~(\ref{eq:divcor_3a2_first_density_matrix_mixing}) and (\ref{eq:divcor_3a2_second_density_matrix_mixing}).
\item[3a3] Algorithm~\ref{algo:3a3} with the divergence correction for ${\bm h}^{\sigma}_4({\bm r}_1)$, Eq.~(\ref{eq:divcorr_3a3}) (for ${\bm k}$ not included in the SCF ${\bm k}$-mesh) 
\item[3a4] Algorithm~\ref{algo:3a4} with the divergence correction, Eqs.~(\ref{eq:div_corr_3a4_2}) and (\ref{eq:div_corr_3a4_3}) (for ${\bm k}$ not included in the SCF ${\bm k}$-mesh) 
\item[3a5] equivalent to {\bf 3a4}
\item[3a6] equivalent to {\bf 3a3}
\item[3b1] Algorithm~\ref{algo:3b1}
\item[3b2] Algorithm~\ref{algo:3b2}
\item[3b3] Algorithm~\ref{algo:3b3} with the divergence correction, Eq.~(\ref{eq:div_corr_3b3_2})  (for ${\bm k}$ not included in the SCF ${\bm k}$-mesh) 
\item[3b4] Algorithm~\ref{algo:3b4} with the divergence correction for ${\bm h}^{\sigma}_3({\bm r}_2)$, Eq.~(\ref{eq:div_corr_3b4_2}), and that for $g_j({\bm r}_1)$, Eq.~(\ref{eq:div_corr_3b4_4}) (for ${\bm k}$ not included in the SCF ${\bm k}$-mesh)
\item[3b5] Algorithm~\ref{algo:3b5} with the divergence correction, Eqs.~(\ref{eq:3b5_div_2}) and (\ref{eq:3b5_div_3}) (for ${\bm k}$ not included in the SCF ${\bm k}$-mesh)
\item[3b6] Algorithm~\ref{algo:3b6} with the divergence correction, Eqs.~(\ref{eq:3b6_div_scf1}), (\ref{eq:3b6_div_scf2}), and (\ref{eq:3b6_div_scf3}) (for ${\bm k}$ included in the SCF ${\bm k}$-mesh) or Eqs.~(\ref{eq:3b6_div_band1}), (\ref{eq:3b6_div_band2}), and (\ref{eq:3b6_div_band3}) (otherwise)
\item[3c*] equivalent to {\bf 3b*}
\end{description}

\section{How to use TC++}
\label{howto}

\subsection{Requirements}

TC\verb!++! requires an MPI C\verb!++! compiler that supports C\verb!++!11, a Fortran90 compiler, and the following libraries: FFTW3~\cite{FFTW3}, Eigen (Eigen 3)~\cite{Eigen}, and Boost~\cite{boost}.
Quantum ESPRESSO (QE) ver.6.2 or newer is also required for performing calculation in advance of the TC calculation.

\subsection{Download and install}

Download the source files from https://github.com/masaochi/TC and unzip it. Then,
\begin{verbatim}
cd src
\end{verbatim}
and edit Makefile to specify compilers and libraries. Finally, typing
\begin{verbatim}
make
\end{verbatim}
will create an execution file named {\bf tc\verb!++!} in {\bf src}.
As an alternative way for installation, cmake is also available in our code (from ver.1.2). Typing
\begin{verbatim}
mkdir build && cd build
cmake ..
make
make install
\end{verbatim}
will also create the execution file {\bf tc\verb!++!}. Several options for cmake that might be required to specify the compilers and libraries are listed in the online users' guide.

After compilation, it is recommended to perform test calculation to verify that your installation was successfully done. A test suite is provided in {\bf test} folder (from ver.1.2). Please type
\begin{verbatim}
cd test
\end{verbatim}
and copy the execution file {\bf tc\verb!++!} to the {\bf test} directory. Finally, you can perform test calculation by typing
\begin{verbatim}
python3 test.py
\end{verbatim}
and its result will be shown in your screen.

\subsection{Functionalities}

TC\verb!++! is a free/libre open-source software of the TC method for first-principles calculation of solids.
Supported functionalities are listed below.
\begin{itemize}
\item Method: free-electron mode (FREE), HF, TC, BITC
\item Mode: SCF and band calculations
\item Solid-state calculation under the periodic boundary condition. Homogeneous-electron-gas calculation using a periodic cell is also possible by ignoring pseudopotentials.
\item Plane-wave basis set
\item Norm-conserving pseudopotentials without partial core correction
\item Non-spin-polarized calculation or spin-polarized calculation with the following conditions satisfied: spin-collinear state without spin-orbit coupling, {\bf no\_t\_rev} and {\bf noinv} should be true in QE
\item Monkhorst-Pack ${\bm k}$-grid~\cite{MPkgrid} with/without a shift. A ${\bm k}$-grid should not break any crystal symmetry (e.g., a $2\times 3\times 4\ {\bm k}$-grid for the simple-cubic lattice is not allowed). $\Gamma$-only calculation is at present not supported. 
\end{itemize}

\subsection{How to use}
\subsubsection{Precalculation using QE}

Before performing TC\verb!++! calculation, one should perform calculation using QE to get the crystal-symmetry information, initial estimate of one-electron orbitals, and so on.
Any calculation method in QE, such as DFT and HF, is acceptable as long as one can get one-electron orbitals.

Note that acceptable pseudopotentials are a bit limited: norm-conserving pseudopotentials without partial core correction.
You can get them, e.g., in Pseudopotential Library~\cite{pseudopotential_library}.

It is recommended to perform QE calculation in the same environment for Fortran90 as TC\verb!++! because TC\verb!++! reads binary files containing wave-function data dumped by QE.
In TC\verb!++!, Fortran90 is used only for this purpose.

\subsubsection{Input files for TC++}

Three inputs are required for TC\verb!++!. 
One is the {\bf save} directory obtained by QE calculation, which includes {\bf data-file-schema.xml} and wave-function files such as {\bf wfc1.dat}.
Another one is pseudopotential files that should be the same as those used in the QE precalculation.
Also TC\verb!++! requires {\bf input.in} containing several input keywords for running TC\verb!++!.
The example is shown below.
\begin{verbatim}
calc_method  TC  # comment can be added like this
calc_mode  SCF
pseudo_dir  /home/user/where_pseudo_potentials_are_placed
qe_save_dir  /home/user/where_QEcalc_was_performed/prefix.save
\end{verbatim}
A complete list of keywords in {\bf input.in} is shown in Tables~\ref{tab:input-keywords-mandatory} and \ref{tab:input-keywords-optional}.
For restarting SCF calculation or performing band calculation after SCF, TC\verb!++! requires some other input files dumped by TC\verb!++!. Please see Sec.~\ref{sec:output_files}.

\begin{table}
\caption{Mandatory Keywords in {\bf input.in}. Optional keywords are shown in Table~\ref{tab:input-keywords-optional}.}
\label{tab:input-keywords-mandatory}
\centering
\begin{tabular}{cp{2.5 cm}p{8 cm}}
\hline
keyword & type & description \\
\hline
{\bf calc\_method} & string & [available values: FREE, HF, TC, BITC] Calculation method. No electron-electron interaction is considered for FREE, i.e., the kinetic energy and pseudopotentials are only considered.\\
{\bf calc\_mode} & string & [available values: SCF, BAND] Calculation mode. BAND calculation should be performed after SCF calculation.\\
{\bf pseudo\_dir} & string & A directory where pseudopotential files are placed, e.g., /home/user/where\_pseudopot\_are\_placed\\
{\bf qe\_save\_dir} & string & A {\bf save} directory created by QE, e.g., /home/user/where\_QEcalc\_was\_performed/prefix.save\\
\hline
\end{tabular}
\end{table}

\begin{table}
\caption{Optional keywords in {\bf input.in}. Mandatory keywords are shown in Table~\ref{tab:input-keywords-mandatory}.}
\label{tab:input-keywords-optional}
\centering
\begin{tabular}{cp{2.4 cm}p{8.1 cm}}
\hline
keyword & type & description \\
\hline
{\bf A\_up\_up} & real & [default: 1.0] $A_{\uparrow, \uparrow}$ in Eq.~(\ref{eq:Jastrow}), normalized by Eq.~(\ref{eq:A_normalize}). Namely, 1.0 means the value shown in Eq.~(\ref{eq:A_normalize}). 
$C_{\uparrow, \uparrow}$ in Eq.~(\ref{eq:Jastrow}) is set so as to satisfy the cusp condition. Not used for {\bf calc\_method} = FREE or HF.\\
{\bf A\_up\_dn} & real & [default: 1.0] $A_{\uparrow, \downarrow} = A_{\downarrow, \uparrow}$, same as above. \\
{\bf A\_dn\_dn} & real & [default: 1.0] $A_{\downarrow, \downarrow}$, same as above. Users cannot specify different values for {\bf A\_up\_up} and {\bf A\_dn\_dn} in non-spin-polarized calculation.\\
{\bf num\_bands\_tc} & integer ($\geq 1$, $\leq$ {\bf nbnd} in QE) & [default: {\bf nbnd} in QE] The number of bands, which can be smaller than {\bf nbnd} in QE.\\
{\bf smearing\_mode} & string & [default: gaussian] [available values: fixed, gaussian] We recommend {\bf smearing\_mode} = fixed and gaussian for insulators and metals, respectively.\\
{\bf smearing\_width} & real ($\geq 0$) & [default: 0.01] In Hartree unit. Not used for {\bf smearing\_mode} = fixed. A negative value will be ignored.\\
{\bf restarts} & boolean & [default: false] When {\bf restarts} = true, TC\verb!++! restarts calculation from a previous run.\\
{\bf includes\_div\_correction} & boolean & [default: true] Whether the divergence correction described in this paper is included.\\
{\bf energy\_tolerance} & real ($\geq 0$) & [default: 1e-5] In Hartree unit. Convergence criteria for the total energy ({\bf calc\_mode} = SCF) or a sum of eigenvalues ({\bf calc\_mode} = BAND).\\
{\bf charge\_tolerance} & real ($\geq 0$) & [default: 1e-4] In $e^-$. Convergence criteria for the charge density, used only for {\bf calc\_mode} = SCF.\\
{\bf max\_num\_iterations} & integer ($\geq 0$) & [default: 30 for {\bf calc\_mode} $=$ SCF, 15 for {\bf calc\_mode} $=$ BAND] Maximum number of iterations for the self-consistent-field loop.\\
{\bf mixes\_density\_matrix} & boolean & [default: false] The density matrix (true) or the density (false) is used for mixing.\\
{\bf mixing\_beta} & real ($> 0$) & [default: 0.7] Mixing ratio for simple density mixing: new density = {\bf mixing\_beta} $\times$ new density + ($1-${\bf mixing\_beta}) $\times$ old density,
 used only for {\bf calc\_mode} = SCF.\\
{\bf num\_refresh\_david} & integer ($\geq 1$) & [default: 1] Trial vectors are updated by {\bf num\_refresh\_david} times for each update of the Fock operator in Davidson diagonalization.\\
{\bf max\_num\_blocks\_david} & integer ($\geq 2$) & [default: 2] This keyword determines a size of subspace dimension: subspace dimension = {\bf max\_num\_blocks\_david} $\times$ {\bf num\_bands\_tc} (see {\bf diago\_david\_ndim} in QE). Increasing this value can improve convergence while computational time is proportional to it.\\
{\bf is\_heg} & boolean & [default: false] Switches on the homogeneous-electron-gas mode where pseudopotentials and the Ewald energy are ignored (i.e., a lattice is ignored).\\
\hline
\end{tabular}
\end{table}

\subsubsection{How to run TC++}

An example command to run TC\verb!++! is as follows.
\begin{verbatim}
mpirun -np 4 $HOME/TC++/ver.1.0/src/tc++
\end{verbatim}
Since TC\verb!++! does not use OpenMP parallelization, please set {\bf OMP\_NUM\_THREADS} to be 1.

\subsubsection{Output files for TC++ \label{sec:output_files}}

The following outputs are obtained by TC\verb!++! calculation:
\begin{itemize}
\item Standard output shows error messages. Please check it when calculation unexpectedly stops.
\item {\bf output.out} shows much information including a list of ${\bm k}$-points and symmetries, total energy, eigenvalues, computation time, and convergence information.
\item {\bf tc\_bandplot.dat} shows band eigenvalues obtained by BAND calculation. Users can plot the band dispersion using this file. For example, ``plot 'tc\_bandplot.dat' u 4:5 w l'' in {\bf gnuplot} will show the band structure. The Fermi energy obtained by SCF calculation is also shown in the second line of this file.
\end{itemize}

The following binary files are dumped and required for subsequent TC\verb!++! calculation:
\begin{itemize}
\item {\bf tc\_energy\_scf.dat} contains SCF energy eigenvalues that are used for restarting SCF calculation or performing subsequent BAND calculation. Dumped in SCF calculation.
\item {\bf tc\_energy\_band.dat} contains BAND energy eigenvalues that are used for restarting BAND calculation. Dumped in BAND calculation.
\item {\bf tc\_wfc\_scf.dat} contains SCF wave functions that are used for restarting SCF calculation or performing subsequent BAND calculation. Dumped in SCF calculation.
\item {\bf tc\_wfc\_band.dat} contains BAND wave functions that are used for restarting BAND calculation. Dumped in BAND calculation.
\item {\bf tc\_scfinfo.dat} contains several information of SCF calculation that are used for subsequent BAND calculation. Dumped in SCF calculation.
\end{itemize}
Here, {\bf tc\_energy\_*.dat} and {\bf tc\_wfc\_*.dat} are dumped in each self-consistent iteration so that users can restart calculation when calculation stops.

\section{Results}
\label{results}

\subsection{bulk silicon}

As the first example, we show how to run the band-structure calculation of bulk silicon using TC\verb!++!.
First, we performed SCF calculation using QE by the following input file, 
\begin{verbatim}
&control
  prefix = 'prefix'
  calculation = 'scf'
  pseudo_dir = '/home/user/QE/pseudo_potential/'
  outdir = './'
  verbosity = 'high'
  disk_io = 'low'
/
&system
  ibrav = 2
  celldm(1) = 10.26
  nat = 2
  ntyp = 1
  nbnd = 10
  ecutwfc = 20.0
  occupations = 'fixed'
/
&electrons
  conv_thr = 1.0d-8
/
ATOMIC_SPECIES
 Si 1.0 Si.upf
ATOMIC_POSITIONS {alat}
 Si 0.00 0.00 0.00
 Si 0.25 0.25 0.25
K_POINTS {automatic}
 8 8 8 0 0 0
\end{verbatim}
Here, we used the Ne-core pseudopotential of silicon~\cite{Si_pp} taken from Pseudopotential Library~\cite{pseudopotential_library}.
Any calculation method in QE, such as DFT and HF, is acceptable as long as one can get one-electron orbitals.
To obtain a band structure, we also performed the band calculation using QE by the following input file,
\begin{verbatim}
&control
  prefix = 'prefix'
  calculation = 'bands'
  pseudo_dir = '/home/user/QE/pseudo_potential/'
  outdir = './'
  verbosity = 'high'
  disk_io = 'low'
/
&system
  ibrav = 2
  celldm(1) = 10.26
  nat = 2
  ntyp = 1
  nbnd = 10
  ecutwfc = 20.0
  occupations = 'fixed'
/
&electrons
  conv_thr = 1.0d-8
/
ATOMIC_SPECIES
 Si 1.0 Si.upf
ATOMIC_POSITIONS {alat}
 Si 0.00 0.00 0.00
 Si 0.25 0.25 0.25
K_POINTS {crystal_b}
3
 0.5 0.5 0.0 20
 0.0 0.0 0.0 20
 0.5 0.0 0.0 0
\end{verbatim}
Here, we copied the directory for SCF calculation and performed band calculation there.
Namely, we performed band calculation in a different directory from that for SCF calculation.

Next, we performed SCF calculation with HF or TC or BITC using the following input file, {\bf input.in},
\begin{verbatim}
calc_method  HF # change here (TC, BITC)
calc_mode  SCF
pseudo_dir  /home/user/QE/pseudo_potential
qe_save_dir  /home/user/where_QE_SCFcalc_was_performed/prefix.save
smearing_mode  fixed
\end{verbatim}
where {\bf pseudo\_dir} and {\bf qe\_save\_dir} should be appropriately specified.
After SCF calculation, we should check whether ``convergence is achieved!'' is shown in {\bf output.out}.
If the convergence is not achieved, we can restart calculation using {\bf input.in} with the following line added:
\begin{verbatim}
restarts  true
\end{verbatim}
However, it is often difficult to achieve convergence in BITC calculations (see Sec.~\ref{sec:comment}).
While convergence can be improved by increasing the number of ${\bm k}$-points and/or {\bf max\_num\_blocks\_david} (e.g., to 5),
we did not do so in this tutorial calculation since it is often not necessary to get convergence with a default value of convergence criteria, {\bf energy\_tolerance} and {\bf charge\_tolerance}.
To improve the convergence, it is also effective to reduce {\bf mixing\_beta} with {\bf mixes\_density\_matrix} $=$ true. 
The band structures shown later were obtained without taking these ways or restarting calculation.
Finally, we performed the band calculation using the following input file, {\bf input.in},
\begin{verbatim}
calc_method  HF # change here (TC, BITC)
calc_mode  BAND
pseudo_dir  /home/user/QE/pseudo_potential
qe_save_dir  /home/user/where_QE_BANDcalc_was_performed/prefix.save
smearing_mode  fixed
\end{verbatim}
Note that {\bf qe\_save\_dir} is different from that used in SCF calculation.
Users can apply {\bf restarts = true} also for BAND calculation if necessary.
A small error will remain in these tutorial calculations of the TC and BITC methods, which can be reduced by increasing the number of ${\bm k}$-points and/or changing the choice of the band ${\bm k}$-points (see Sec.~\ref{sec:comment}).

\begin{figure}
\begin{center}
\includegraphics[width=10 cm]{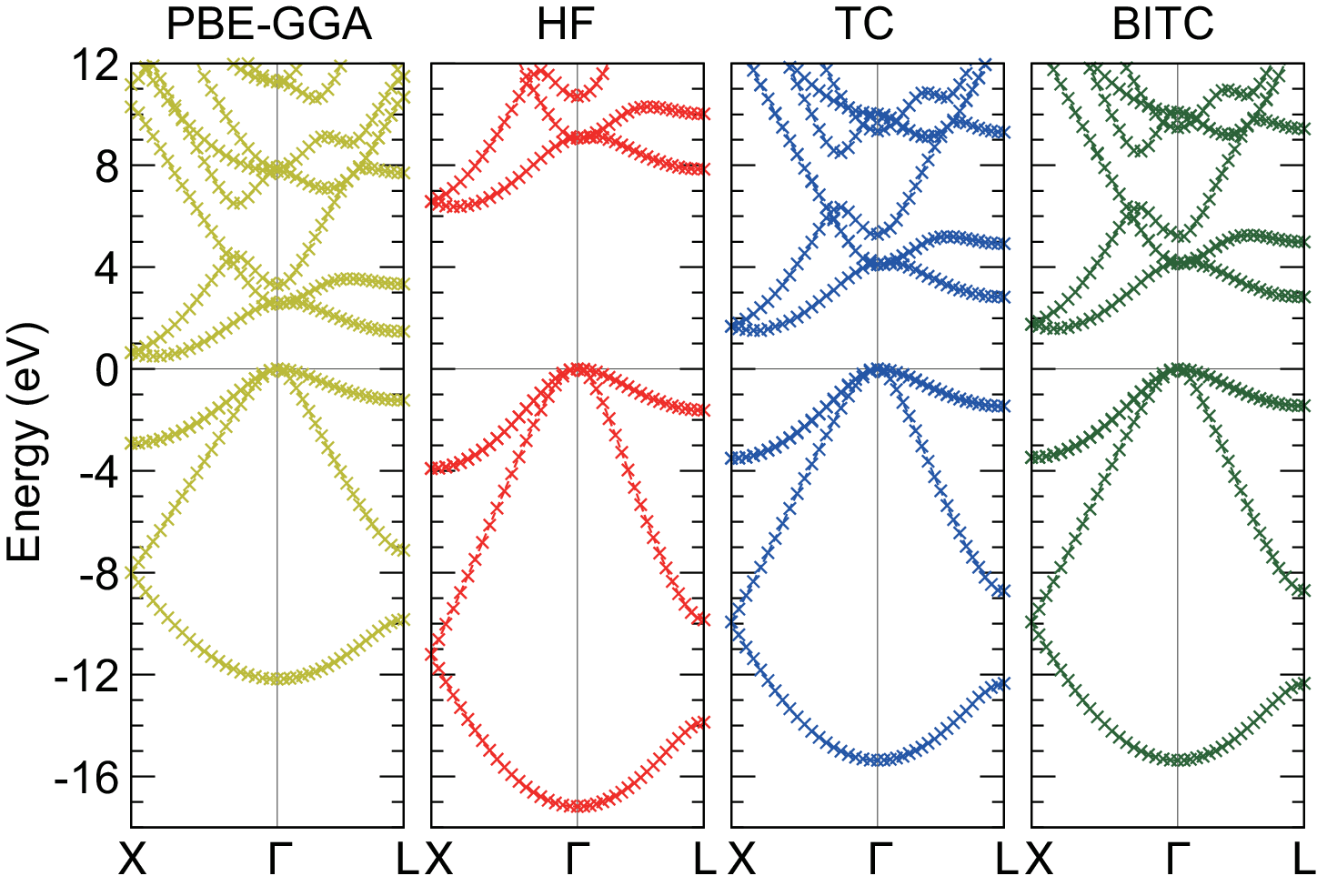}
\caption{Band structures of bulk silicon calculated with PBE-GGA~\cite{PBE} (obtained by using QE), HF, TC, and BITC methods.}
\label{fig:si}
\end{center}
\end{figure}

The calculated band structures are shown in Fig.~\ref{fig:si}, which were plotted using the fourth and fifth columns in {\bf tc\_bandplot.dat}.
The indirect band gap is 0.5 eV, 6.4 eV, 1.5 eV, 1.6 eV for PBE-GGA, HF, TC, and BITC methods, respectively.
Since the experimental band gap of bulk silicon is 1.17 eV~\cite{Si_gap}, the accuracy of the band gap is improved in TC calculation as reported in our previous study~\cite{TCaccel}.
On the other hand, the valence bandwidth is overestimated in the TC method ($\simeq$ 15 eV) compared with the experimental value, 12.5$\pm$0.6 eV~\cite{Si_vbw}.
We reported in the previous study that the overestimation of the valence bandwidth is much improved by using a He-core pseudopotential where $2s,2p$ orbitals are treated as the valence orbitals~\cite{TCPW}.

In TC\verb!++!, so-called {\it fake}-SCF calculation is also possible, where SCF and band calculations are simultaneously performed by specifying the ${\bm k}$-points with an appropriate weight.
Namely, users can use the following input file for QE when using a $4\times 4\times 4$ ${\bm k}$-mesh,
\begin{verbatim}
&control
  prefix = 'prefix'
  calculation = 'scf'
  pseudo_dir = '/home/user/QE/pseudo_potential/'
  outdir = './'
  verbosity = 'high'
  disk_io = 'low'
/
&system
  ibrav = 2
  celldm(1) = 10.26
  nat = 2
  ntyp = 1
  nbnd = 10
  ecutwfc = 20.0
  occupations = 'fixed'
/
&electrons
  conv_thr = 1.0d-8
/
ATOMIC_SPECIES
 Si 1.0 Si.upf
ATOMIC_POSITIONS {alat}
 Si 0.00 0.00 0.00
 Si 0.25 0.25 0.25
K_POINTS {crystal}
19
 0.0 0.0 0.0 0.03125
 0.0 0.0 0.25 0.25
 0.0 0.0 -0.5 0.125
 0.0 0.25 0.25 0.1875
 0.0 0.25 -0.5 0.75
 0.0 0.25 -0.25 0.375
 0.0 -0.5 -0.5 0.09375
 0.25 -0.5 -0.25 0.1875
 0.0 0.0 0.0 0.0
 0.05 0.0 0.0 0.0
 0.1 0.0 0.0 0.0
 0.15 0.0 0.0 0.0
 0.2 0.0 0.0 0.0
 0.25 0.0 0.0 0.0
 0.3 0.0 0.0 0.0
 0.35 0.0 0.0 0.0
 0.4 0.0 0.0 0.0
 0.45 0.0 0.0 0.0
 0.5 0.0 0.0 0.0
\end{verbatim}
and perform SCF calculation with TC\verb!++!, which gives the SCF and BAND eigenvalues simultaneously.
However, we do not recommend this way by the following reasons: band eigenvalues are not checked for convergence (see {\bf energy\_tolerance} in Table~\ref{tab:input-keywords-optional}), and computational cost becomes expensive because the computation time is proportional to $N_{\bm k}^2$ in the TC method.
Note that {\bf tc\_bandplot.dat} is not dumped in {\it fake}-SCF calculation since {\bf calc\_mode} = SCF. If users would like to perform band calculation in this way, they should read band eigenvalues from {\bf output.out}.

\subsection{homogeneous electron gas}

TC\verb!++! also supports calculation of homogeneous electron gas. First, we performed SCF calculation using QE with the following input file, 
\begin{verbatim}
&control
  prefix = 'prefix'
  calculation = 'scf'
  pseudo_dir = '/home/user/QE/pseudo_potential/'
  outdir = './'
  verbosity = 'high'
  disk_io = 'low'
/
&system
  ibrav = 1
  celldm(1) = 7.67663317071 ! Bohr
  nat = 1
  ntyp = 1
  nbnd = 20
  ecutwfc = 20.0
  occupations = 'smearing'
  smearing = 'gauss'
  degauss = 0.03 ! Ry
/
&electrons
  conv_thr = 1.0d-8
/
ATOMIC_SPECIES
 Si 1.0 Si.upf
ATOMIC_POSITIONS {alat}
 Si 0.00 0.00 0.00
K_POINTS {automatic}
 12 12 12 0 0 0
\end{verbatim}
where the pseudopotential file, Si.upf, placed in {\bf pseudo\_dir} is used because calculation of homogeneous electron gas is not implemented in QE.
Four valence electrons in the simple-cubic lattice with this lattice constant correspond to the $r_s$ parameter of 3 Bohr in electron gas.
For a band-structure plot, we also performed the band calculation using QE with the following input file,
\begin{verbatim}
&control
  prefix = 'prefix'
  calculation = 'bands'
  pseudo_dir = '/home/user/QE/pseudo_potential/'
  outdir = './'
  verbosity = 'high'
  disk_io = 'low'
/
&system
  ibrav = 1
  celldm(1) = 7.67663317071 ! Bohr
  nat = 1
  ntyp = 1
  nbnd = 20
  ecutwfc = 20.0
  occupations = 'smearing'
  smearing = 'gauss'
  degauss = 0.03 ! Ry
/
&electrons
  conv_thr = 1.0d-8
/
ATOMIC_SPECIES
 Si 1.0 Si.upf
ATOMIC_POSITIONS {alat}
 Si 0.00 0.00 0.00
K_POINTS {tpiba_b}
3
-0.5 -0.5 -0.5 20
0.0 0.0 0.0 20
0.5 0.0 0.0 0
\end{verbatim}
Here, we copied the directory for SCF calculation and performed band calculation there.
Namely, we performed band calculation in a different directory from that for SCF calculation.

\begin{figure}
\begin{center}
\includegraphics[width=9 cm]{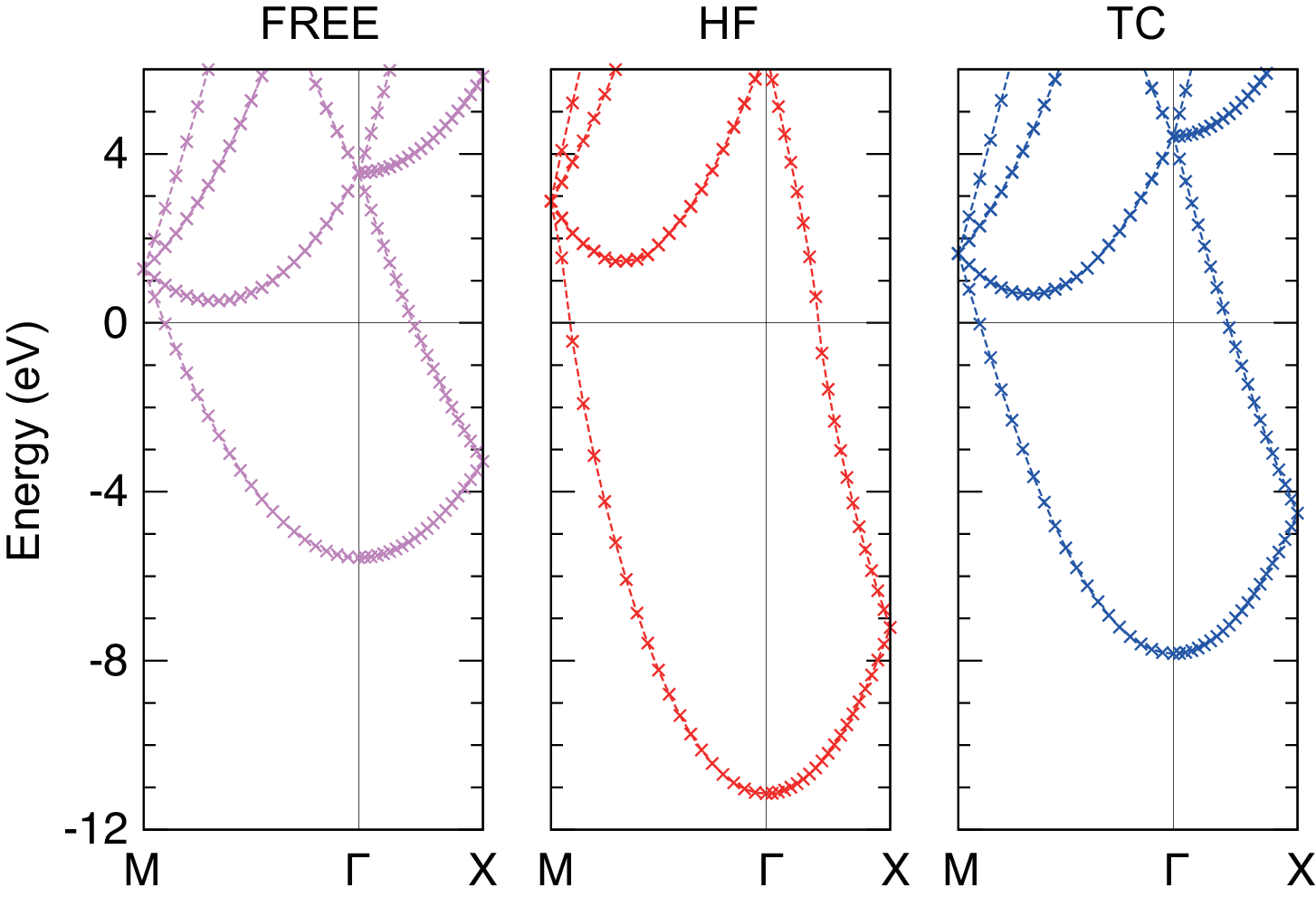}
\caption{Band structures of homogeneous electron gas using {\bf calc\_method} $=$ FREE, HF, and TC. The Brillouin zone for the simple cubic lattice is used: X$=(\pi /a, 0, 0)$ and M$=(\pi /a, \pi /a, 0)$ where $a$ is the lattice constant.}
\label{fig:heg}
\end{center}
\end{figure}

Next, we performed SCF calculation with FREE (free-electron mode) or HF or TC using the following input file, {\bf input.in},
\begin{verbatim}
calc_method  FREE # change here (HF, TC)
calc_mode  SCF
pseudo_dir  /home/user/QE/pseudo_potential
qe_save_dir  /home/user/where_QE_SCFcalc_was_performed/prefix.save
smearing_mode  gaussian
smearing_width  0.02 # in Ht.
is_heg  true
\end{verbatim}
where {\bf qe\_save\_dir} and {\bf pseudo\_dir} should be appropriately specified.
Finally, we performed the band calculation using the following input file, {\bf input.in},
\begin{verbatim}
calc_method  FREE # change here (HF, TC)
calc_mode  BAND
pseudo_dir  /home/user/QE/pseudo_potential
qe_save_dir  /home/user/where_QE_BANDcalc_was_performed/prefix.save
smearing_mode  gaussian
smearing_width  0.02 # in Ht.
is_heg  true
\end{verbatim}

The calculated band structures are shown in Fig.~\ref{fig:heg}.
One notable feature is that the HF band structure has a well-known singularity at the Fermi energy: the density of states becomes zero at the Fermi energy with a logarithmic singularity.
This is due to a lack of the screening effect of the electron-electron interaction in the Hartree-Fock theory. As a result, the HF band structure is quite dispersive near the Fermi energy.
On the other hand, the TC band structure does not have this kind of unphysical behavior thanks to the Jastrow factor that includes the screening effect.
These are consistent with those reported in \cite{Sakuma}.
Note that BITC should offer the same result as TC because left and right one-electron orbitals are the same plane waves for homogeneous electron gas.

Users can use a different value for the lattice type, the atomic species, and the lattice constant. The subsequent TC\verb!++! run only uses the number of electrons and the periodic cell.
Since TC\verb!++! can use crystal symmetries existing in the QE input, high-symmetry structure is preferable for efficient computation.

\subsection{Other comments and tips for calculation~\label{sec:comment}}

Computational time of the HF and (BI)TC methods is $\mathcal{O}(N^2_{\bm k}N^2_{\mathrm{bands}}N_{\mathrm{pw}}\ln N_{\mathrm{pw}})$,
and required memory size is $\mathcal{O}(N_{\bm k}N_{\mathrm{bands}}N_{\mathrm{pw}})$.
Such a relatively low scaling compared with other post-HF methods is one of the great advantages of the (BI)TC method.

When the convergence of the TC calculation is difficult, it might be effective to (i) increase the number of ${\bm k}$-points, (ii) increase {\bf max\_num\_blocks\_david} (e.g., to 5), and (iii) increase the number of bands, and (iv) reduce {\bf mixing\_beta} with {\bf mixes\_density\_matrix} $=$ true.
In particular, (i) is the most effective in many cases because the divergence of the interaction terms in the reciprocal space can make a large error in the (BI)TC and HF calculation, while it is partially alleviated by the divergence correction.
This error can be regarded as a sampling error of a rapidly changing function in the reciprocal space, and thus is resolved by using a fine ${\bm k}$-mesh.
Using a fine ${\bm k}$-mesh is also effective to make the band structure smooth.
(ii) and (iii) increase a subspace dimension for diagonalization.
For (iv), please note that computational time becomes longer by around a factor of two when using {\bf mixes\_density\_matrix} $=$ true.
Because the TC method handles the non-Hermitian Hamiltonian, it seems that achieving the convergence in calculation is more difficult that other methods.
This tendency is more conspicuous in BITC calculations.

Related to the above-mentioned convergence issue, it is often difficult to get the smooth band dispersion.
To get a smooth band dispersion, users should not take a band ${\bm k}$-point that is very close to (but different from) the SCF ${\bm k}$-points.
This is because the interaction terms include the divergence such as $1/|{\bm k}-{\bm q}+{\bm G}|^2$, where ${\bm k}, {\bm q}, {\bm G}$ are the band ${\bm k}$-point, the SCF ${\bm k}$-point, and the reciprocal vector, respectively. This divergence is problematic when ${\bm k} \simeq {\bm q}$.

For a large system, a large memory consumption can be problematic in the TC calculation.
Increasing the number of MPI processes can alleviate this issue, by distributing large arrays to many MPI processes.

 One of the important features of the TC method is that one can optimize one-electron orbitals in the presence of the Jastrow factor, which can improve VMC and DMC results where HF or DFT orbitals are usually used. From this perspective, it might be important to investigate several types of the Jastrow factors, e.g., for obtaining a highly accurate nodal structure of the many-body wave function, which is a key for improving the accuracy of DMC. These are important ongoing issues for a future release.

\section{Summary}
\label{summary}

In this paper, we present our implementation of TC\verb!++!, a free/libre open-source software of the TC method for first-principles calculation of solids.
We describe our calculation algorithm in detail, including the way to handle the divergence of the effective potentials in the reciprocal space.
Our computational code enables ones to easily perform first-principles calculation of solids based on the wave-function theory.
Some application results of TC\verb!++! are promising.
We believe that TC\verb!++! will make an important contribution for the development of the wave-function theory in solids.

\section*{Declaration of competing interest}
The authors declare that they have no known competing financial interests or personal relationships that could have appeared to influence the work reported in this paper.

\section*{Acknowledgements}
This work was supported by JST FOREST Program (Grant Number JPMJFR212P), Japan.
We thank fruitful discussion with Dr. Rei Sakuma, Dr. Naoto Umezawa, Dr. Keitaro Sodeyama, and Prof. Shinji Tsuneyuki.

\section*{Appendix A: Invariance of the diagonal element of the eigenvalue matrix in the TC SCF equation}

The SCF equation, Eq.~(\ref{eq:SCF}), can be written as
\begin{equation}
\hat{h}\phi_i({\bm r}) = \sum_{j=1}^N \epsilon_{ij} \phi_j({\bm r}),
\end{equation}
where the orthonormal condition $\langle \phi_i | \phi_j \rangle = \delta_{i,j}$ is satisfied. We also write the eigenvalue equation of $\hat{h}$ as
\begin{equation}
\hat{h}\tilde{\phi}_i({\bm r}) = \tilde{\epsilon}_{i} \tilde{\phi}_i({\bm r}),
\end{equation}
where $\tilde{\phi}$ and $\tilde{\epsilon}$ are the eigenvector and the eigenvalue of $\hat{h}$, respectively.
Note that, $\tilde{\phi}$ does not mean the Fourier transform of $\phi$ in this Appendix.
Because we get $\phi$ by the Gram--Schmidt orthonormalization of the eigenvectors $\tilde{\phi}$,
$\phi_j$ can be represented as a linear combination of $\tilde{\phi}_i$ ($i\leq j$) and vice versa.
In the following proof, $V_i$ is the subspace spanned by $\phi_1, \phi_2, \dots, \phi_i$ (or $\tilde{\phi}_1, \tilde{\phi}_2, \dots, \tilde{\phi}_i$).

It is obvious that $f \in V_i \Rightarrow \hat{h}f \in V_i$ since $f$ can be expanded with $\tilde{\phi}_1, \tilde{\phi}_2, \dots, \tilde{\phi}_i$.
Therefore, $\epsilon_{ij} = \langle \phi_i|\hat{h}|\phi_j \rangle = 0$ holds for $i>j$ because $\hat{h} \phi_j \in V_j$, which is orthogonal to $\phi_i$.
By defining coefficients $c_i$ as
\begin{equation}
\phi_i = c_i \tilde{\phi}_i + f,\ \ (f \in V_{i-1})
\end{equation}
the diagonal element of $\epsilon$ can be calculated as
\begin{equation}
\epsilon_{ii} = \frac{\langle \phi_i | \hat{h} | \phi_i \rangle}{\langle \phi_i | \phi_i \rangle} = \frac{c_i \tilde{\epsilon}_i \langle \phi_i | \tilde{\phi}_i \rangle}{c_i \langle \phi_i | \tilde{\phi}_i \rangle} = \tilde{\epsilon_i}.
\end{equation}
Thus, the diagonal element of the eigenvalue matrix is invariant against the Gram--Schmidt orthonormalization.

\section*{Appendix B: Fourier transform of $(\nabla u)^2$}

For the Jastrow function shown in Eq.~(\ref{eq:Jastrow}), $(\nabla u)^2$ is calculated as
\begin{equation}
(\nabla u)^2(r) = \frac{A^2}{r^2}\left( \frac{1}{C}\mathrm{e}^{-r/C} + \frac{1}{r}(\mathrm{e}^{-r/C}-1) \right)^2.
\end{equation}
Therefore, we get
\begin{align}
\widetilde{(\nabla u)^2}(G) 
&= \int \mathrm{d}{\bm r}\ (\nabla u)^2 \mathrm{e}^{-i{\bm G}\cdot {\bm r}} \\
&= \frac{4\pi}{G} \mathrm{Im}\bigg[ \int_0^{\infty} \mathrm{d}r\ (\nabla u)^2 r\mathrm{e}^{iGr} \bigg] \\
&= \frac{4\pi A^2}{G} \mathrm{Im}\bigg[ \int_0^{\infty} \mathrm{d}r\ 
\left( \frac{1}{C}\mathrm{e}^{-r/C} + \frac{1}{r}(\mathrm{e}^{-r/C}-1) \right)^2 \frac{1}{r} \mathrm{e}^{iGr} \bigg] \\
&= \frac{4\pi A^2}{C g} \mathrm{Im}\bigg[ \int_0^{\infty} \mathrm{d}r\ 
\left( \mathrm{e}^{-r} + \frac{1}{r}(\mathrm{e}^{-r}-1) \right)^2 \frac{1}{r} \mathrm{e}^{igr} \bigg] \ \ \ (g = CG)\\
&=  \frac{4\pi A^2}{C g} \mathrm{Im} [ F(g)]. \label{eq:appendix1}
\end{align}
Here, $\mathrm{Im} [ F(g=0)] =0$ because the integrand in $F(g=0)$ is real. Therefore, instead of calculating $F(g)$ directly,
we first calculate $F'(g) (=\mathrm{d}F/\mathrm{d}g)$ then integrate it again, to avoid divergence.
For this purpose, we define
\begin{equation}
H(g; \alpha) = \int_0^{\infty} \mathrm{d}r\ 
\left( \alpha \mathrm{e}^{-\alpha r} + \frac{1}{r}(\mathrm{e}^{-\alpha r}-1) \right)^2 \mathrm{e}^{igr}
\end{equation}
for $\alpha >0$, and we can see $F'(g)=i H(g; \alpha=1)$. Because $\lim_{\alpha \to 0} H(g; \alpha)=0$, we can get $H(g; \alpha)$ by 
\begin{equation}
H(g; \alpha) = \int_{0}^{\alpha} \mathrm{d}\tilde{\alpha} \ \frac{\mathrm{d}H}{\mathrm{d}\tilde{\alpha}}.
\end{equation}
This integrand can be calculated as follows:
\begin{align}
\frac{\mathrm{d}H}{\mathrm{d}\tilde{\alpha}}
&= -2\tilde{\alpha}\int_0^{\infty} \mathrm{d}r\ 
\left( \alpha r \mathrm{e}^{-2\tilde{\alpha} r} + \mathrm{e}^{-2\tilde{\alpha} r} - \mathrm{e}^{-\tilde{\alpha} r} \right) \mathrm{e}^{igr} \\
&= -2\tilde{\alpha} \left( \frac{\tilde{\alpha}}{(-2\tilde{\alpha} + ig)^2} - \frac{1}{-2\tilde{\alpha} + ig} + \frac{1}{-\tilde{\alpha} + ig} \right) \\
&= \frac{1}{2} + \frac{g^2}{2}\frac{1}{(2\tilde{\alpha} - ig)^2} - \frac{2ig}{2\tilde{\alpha} - ig} + \frac{2 ig}{\tilde{\alpha} - ig}.
\end{align}
By integrating it with respect to $\tilde{\alpha}$, we get
\begin{align}
H(g; \alpha)
&= \bigg[  \frac{\tilde{\alpha}}{2} - \frac{g^2}{4}\frac{1}{2\tilde{\alpha} - ig} - ig\ln (2\tilde{\alpha} -ig) + 2 ig \ln (\tilde{\alpha} -ig) \bigg]_0^{\alpha}  \\
&= \frac{\alpha}{2} - \frac{g^2}{4}\frac{1}{2\alpha - ig} + \frac{i}{4g} - ig\ln (2\alpha -ig) + 2 ig \ln (\alpha -ig) -ig \ln (-ig).
\end{align}
Therefore, we get
\begin{align}
F'(g) &= iH(g; \alpha=1) \\
&= \frac{1}{4}g + \frac{i}{2-ig} - \frac{1}{4g}  + g\ln (2-ig) - 2g \ln (1-ig) + g \ln (-ig).
\end{align}
By integrating $F'(g)$ (i.e., $\int_0^g \mathrm{d}{\tilde g} \ F'({\tilde g} )$), we get
\begin{equation}
\mathrm{Im} [F(g)] = \mathrm{Im}\bigg[ \left( 1 + \frac{g^2}{2}\right) \ln (2-ig) - (1+g^2) \ln (1-ig) + \frac{g^2}{2}\ln (-ig) \bigg],
\end{equation}
where we remove some real terms from $F(g)$ that are irrelevant to $\mathrm{Im} [F(g)]$.
By using Eq.~(\ref{eq:appendix1}) and
\begin{equation}
\mathrm{Im}[ \ln (2-ig) ] = -\arctan \frac{g}{2},\ \ \mathrm{Im}[ \ln (1-ig) ] = -\arctan g,\ \ \mathrm{Im}[ \ln (-ig) ] = -\frac{\pi}{2},
\end{equation}
we get Eq.~(\ref{eq:fourier_transform_nabla2u}).


\begin{thebibliography}{999}

\bibitem{QMC} R. J. Needs, M. D. Towler, N. D. Drummond, and P. L{\'o}pez R{\'ios}, J. Phys.: Condens. Matter {\bf 22}, 023201 (2009).

\bibitem{FCI1} G. H. Booth, A. J. W. Thom, and A. Alavi, J. Chem. Phys. {\bf 131}, 054106 (2009).
\bibitem{FCI2} D. Cleland, G. H. Booth, and A. Alavi, J. Chem. Phys. {\bf 132}, 041103 (2010).
\bibitem{FCI3} W. Dobrautz, S. D. Smart, and A. Alavi, J. Chem. Phys. {\bf 151}, 094104 (2019).

\bibitem{BoysHandy} S. F. Boys and N. C. Handy, Proc. R. Soc. London Ser. A \textbf{309}, 209 (1969); {\it ibid.} \textbf{310}, 43 (1969); {\it ibid.} \textbf{310}, 63 (1969); {\it ibid.} \textbf{311}, 309 (1969).
\bibitem{Handy} N. C. Handy, Mol. Phys. \textbf{21}, 817 (1971).

\bibitem{Ten-no1} S. Ten-no, Chem. Phys. Lett. \textbf{330}, 169 (2000); {\it ibid.} 175 (2000). 
\bibitem{Ten-no2} O. Hino, Y. Tanimura, S. Ten-no, J. Chem. Phys. \textbf{115}, 7865 (2001). 
\bibitem{Umezawa} N. Umezawa and S. Tsuneyuki, J. Chem. Phys. \textbf{119}, 10015 (2003). 

\bibitem{elgas_Armour} E. A. G. Armour, J. Phys. C: Solid State Phys. {\bf 13}, 343 (1980).
\bibitem{Umezawa_elgas} N. Umezawa and S. Tsuneyuki, Phys. Rev. B {\bf 69}, 165102 (2004).
\bibitem{Sakuma} R. Sakuma and S. Tsuneyuki, J. Phys. Soc. Jpn. \textbf{75}, 103705 (2006).
\bibitem{Luo_elgas} H. Luo, J. Chem. Phys. {\bf 136}, 224111 (2012).
\bibitem{FCI_TC_elgas} H. Luo and A. Alavi, J. Chem. Theory. Comput. {\bf 14}, 1403 (2018). 
\bibitem{perturbation_elgas} H. Luo and A. Alavi, J. Chem. Phys. {\bf 157}, 074105 (2022).

\bibitem{TCaccel} M. Ochi, K. Sodeyama, R. Sakuma, and S. Tsuneyuki, J. Chem. Phys. \textbf{136}, 094108 (2012).
\bibitem{TCjfo} M. Ochi, K. Sodeyama, and S. Tsuneyuki, J. Chem. Phys. \textbf{140}, 074112 (2014).
\bibitem{TCPW} M. Ochi, Y. Yamamoto, R. Arita, and S. Tsuneyuki, J. Chem. Phys. \textbf{144}, 104109 (2016).
\bibitem{TCZnO} M. Ochi, R. Arita, and S. Tsuneyuki, Phys. Rev. Lett. \textbf{118}, 026402 (2017).

\bibitem{cusp} T. Kato, Commun. Pure Appl. Math. \textbf{10}, 151 (1957).
\bibitem{cusp2} R. T. Pack and W. B. Brown, J. Chem. Phys. \textbf{45}, 556 (1966).

\bibitem{Ten-no3} O. Hino, Y. Tanimura, S. Ten-no, Chem. Phys. Lett. \textbf{353}, 317 (2002). 
\bibitem{TCCC2021} T. Schraivogel, A. J. Cohen, A. Alavi, and D. Kats, J. Chem. Phys. {\bf 155}, 191101 (2021). 

\bibitem{Umezawa_CIS} N. Umezawa and S. Tsuneyuki, J. Chem. Phys. {\bf 121}, 7070 (2004). 
\bibitem{LuoVTC} H. Luo, J. Chem. Phys. \textbf{133}, 154109 (2010). 
\bibitem{Luo_multiconf} H. Luo, J. Chem. Phys. {\bf 135}, 024109 (2011). 
\bibitem{Giner_He} E. Giner, J. Chem Phys. {\bf 154}, 084119 (2021). 
\bibitem{SCI} A. Ammar, A. Scemama, and E. Giner, arXiv:2207.08399 (2022).

\bibitem{TCCIS} M. Ochi and S. Tsuneyuki, J. Chem. Theory Comput. \textbf{10}, 4098 (2014).
\bibitem{TCMP2} M. Ochi and S. Tsuneyuki, Chem. Phys. Lett. \textbf{621}, 177 (2015).

\bibitem{Umezawa_beta} R. Prasad, N. Umezawa, D. Domin, R. Salomon-Ferrer, and W. A. Lester, Jr., J. Chem. Phys. \textbf{126}, 164109 (2007). 
\bibitem{LuoTC} H. Luo, W. Hackbusch, and H.-J. Flad, Mol. Phys. \textbf{108}, 425 (2010). 
\bibitem{TCatoms_HFJastrow} A. J. Cohen, H. Luo, K. Guther, W. Dobrautz, D. P. Tew, and A. Alavi, J. Chem. Phys. {\bf 151}, 061101 (2019). 
\bibitem{TCatoms_oneparam} W. Dobrautz, A. J. Cohen, A. Alavi, and E. Giner, J. Chem. Phys. {\bf 156}, 234108 (2022). 
\bibitem{TCatom} M. Ochi, arXiv: 2109.05803 (2021).

\bibitem{FCI_canoTC_DMRG} S. Sharma, T. Yanai, G. H. Booth, C. J. Umrigar, and G. K.-L. Chan, J. Chem. Phys. {\bf 140}, 104112 (2014). 
\bibitem{FCI_canoTC} J. A. F. Kersten, G. H. Booth, and A. Alavi, J. Chem. Phys. {\bf 145}, 054117 (2016). 
\bibitem{FCI_Bedimer} K. Guther, A. J. Cohen, H. Luo, and A. Alavi, J. Chem. Phys. {\bf 155}, 011102 (2021). 
\bibitem{FCI_largeCI} A. Ammar, E. Giner, and A. Scemama, J. Chem. Theory Comput. {\bf 18}, 5325 (2022).

\bibitem{FCI_Hubbard} W. Dobrautz, H. Luo, and A. Alavi, Phys. Rev. B {\bf 99}, 075119 (2019).

\bibitem{FCI_1dgas} P. Jeszenszki, H. Luo, A. Alavi, and J. Brand, Phys. Rev. A {\bf 98}, 053627 (2018). 

\bibitem{FCI_cold} P. Jeszenszki, U. Ebling, H. Luo, A. Alavi, and J. Brand, Phys. Rev. Res. {\bf 2}, 043270 (2020). 


\bibitem{CanonicalTC} T. Yanai and T. Shiozaki, J. Chem. Phys. \textbf{136}, 084107 (2012).
\bibitem{CanonicalTC_qCCSD} M. Motta, T. P. Gujarati, J. E. Rice, A. Kumar, C. Masteran, J. A. Latone, E. Lee, E. F. Valeev, and T. Y. Takeshita, Phys. Chem. Chem. Phys. {\bf 22}, 24270 (2020). 

\bibitem{TCHubbard} S. Tsuneyuki, Prog. Theor. Phys. Suppl. \textbf{176}, 134 (2008).
\bibitem{TCDMRG} A. Baiardi and M. Reiher, J. Chem. Phys. {\bf 153}, 164115 (2020).
\bibitem{LieAlgebra} J. M. Wahlen-Strothman, C. A. Jim{\' e}nez-Hoyos, T. M. Henderson, and G. E. Scuseria, Phys. Rev. B \textbf{91}, 041114(R) (2015).

\bibitem{McArdle} S. McArdle and D. P. Tew, arXiv:2006.11181. 
\bibitem{quantum_simulation} A. Kumar, A. Asthana, C. Masteran, E. F. Valeev, Y. Zhang, L. Cincio, S. Tretiak, and P. A. Dub, J. Chem. Theory Comput. {\bf 18}, 5312 (2022).

\bibitem{Umezawa_Exc} N. Umezawa and T. Chikyow, Phys. Rev. A {\bf 73}, 062116 (2006).
\bibitem{Umezawa_Exc2} N. Umezawa, J. Chem Phys. {\bf 147}, 104104 (2017).

\bibitem{github_TC} https://github.com/masaochi/TC

\bibitem{QMCreview} W. M. C. Foulkes, L. Mitas, R. J. Needs, and G. Rajagopal, Rev. Mod. Phys. \textbf{73}, 33 (2001).
\bibitem{Ceperley} D. M. Ceperley, Phys. Rev. B \textbf{18}, 3126 (1978).
\bibitem{CeperleyAlder} D. M. Ceperley and B. J. Alder, Phys. Rev. Lett. \textbf{45}, 566 (1980).


\bibitem{Ten-nocusp} S. Ten-no, J. Chem. Phys. \textbf{121}, 117 (2004).
\bibitem{cuspUmrigar} C.-J. Huang, C. Filippi, C. J. Umrigar, J. Chem. Phys. {\bf 108}, 8838 (1998).

\bibitem{BohmPines} D. Bohm and D. Pines, Phys. Rev. \textbf{92}, 609 (1953).

\bibitem{Davidson1} E. R. Davidson, J. Comput. Phys. \textbf{17}, 87 (1975).
\bibitem{Davidson2} M. Crouzeix, B. Philippe, M. Sadkane, SIAM J. Sci. Comput. \textbf{15}, 62 (1994).
\bibitem{Payne_precon} M. C. Payne, M. P. Teter, D. C. Allan, T. A. Arias, and J. D. Joannopoulos, Rev. Mod. Phys. \textbf{64}, 1045 (1992).
\bibitem{Hirao_Nakatsuji} K. Hirao and H. Nakatsuji, J. Comput. Phys. \textbf{45}, 246 (1982).

\bibitem{memo_density} While we call $n({\bm x})$ the density, this function does not correspond to the electron density of the many-body state unlike the density functional theory, owing to the existence of the Jastrow factor.

\bibitem{Gygi} F. Gygi and A. Baldereschi, Phys. Rev. B {\bf 34}, 4405(R) (1986).
\bibitem{div_corr1} S. Massidda, M. Posternak, and A. Baldereschi, Phys. Rev. B {\bf 48}, 5058 (1993).

\bibitem{note_symmetry} While $\phi_j({\bm r}) \to \phi_j(S^{-1}{\bm r} - {\bm t})$ is a consequence of the symmetry operation in regular notation, we instead use $\phi_j(S({\bm r} + {\bm t}))$ in this paper that is consistent with our implementation. In fact, this change makes no problem because the inverse of a symmetry operation is always included in the space group.

\bibitem{memo_ewald} In TC\verb!++!, the Ewald energy is simply taken from QE output and reused.
\bibitem{KBpp} L. Kleinman and D. M. Bylander, Phys. Rev. Lett. {\bf 48}, 1425 (1982).

\bibitem{div_corr2} P. Carrier, S. Rohra, and A. G{\"o}rling, Phys. Rev. B {\bf 75}, 205126 (2007).


\bibitem{FFTW3} https://www.fftw.org/
\bibitem{Eigen} https://eigen.tuxfamily.org/
\bibitem{boost} https://www.boost.org/

\bibitem{MPkgrid} H. J. Monkhorst and J. D. Pack, Phys. Rev. B {\bf 13}, 5188 (1976).

\bibitem{pseudopotential_library} https://pseudopotentiallibrary.org/


\bibitem{Si_pp} M. Chandler Bennett, G. Wang, A. Annaberdiyev, C. A. Melton, L. Schulenburger, and L. Mitas, J. Chem. Phys. {\bf 149}, 104108 (2018).
\bibitem{PBE} J. P. Perdew, K. Burke, and M. Ernzerhof, Phys. Rev. Lett. \textbf{77}, 3865 (1996).
\bibitem{Si_gap} C. Kittel, Introduction to Solid State Physics, 6th ed., Wiley, New York, 1986.
\bibitem{Si_vbw} M. Rohlfing, P. Kr{\"u}ger, and J. Pollmann, Phys. Rev. B {\bf 48}, 17791 (1993).

\end{thebibliography}
\end{document}